\documentclass[aps]{revtex4-1}
\usepackage{amsmath,amssymb}
\usepackage{color,graphicx}
\usepackage{multirow}
\usepackage{amsmath,amssymb}
\usepackage{rotating}
\usepackage{slashbox}
\newcommand{\be}{\begin{equation}}
\newcommand{\ee}{\end{equation}}
\newcommand{\calkan}{\int\limits_0^\infty dr \int\limits_0^{2\pi} d\phi \int\limits_{0}^{\pi} d\theta}
\newcommand{\calkap}{\int\limits_0^1 dr \int\limits_0^{2\pi} d\phi \int\limits_{0}^{\pi} d\theta}
\newcommand{\iksy}{x_1^2+x_2^2+x_3^2}
\newcommand{\uki}{u_1^2+u_2^2+u_3^2}
\newcommand{\vki}{v_1^2+v_2^2+v_3^2}
\begin{document}
\title{Ultrarelativistic bound states in the  spherical well.}
\author{Mariusz  \.{Z}aba  and Piotr Garbaczewski}
\affiliation{Institute of Physics, University of Opole, 45-052
Opole, Poland}
\date{\today }
\begin{abstract}
We  address  an eigenvalue problem for  the ultrarelativistic
(Cauchy) operator  $(-\Delta )^{1/2}$, whose action is  restricted
to functions that vanish   beyond the interior of a unit sphere in
three spatial dimensions.  We provide high accuracy spectral data
for lowest eigenvalues and eigenfunctions  of this infinite
spherical   well problem.  Our focus is on radial  and orbital  shapes of eigenfunctions.
 The spectrum  consists of an ordered set of  strictly positive eigenvalues  which naturally  splits
  into  non-overlapping,    orbitally  labelled $E_{(k,l)}$   series.  For each   orbital label  $l=0,1,2,...$
   the label  $k =1,2,...$  enumerates consecutive $l$-th series eigenvalues.
 Each of them is $2l+1$-degenerate.  The $l=0$   eigenvalues  series   $E_{(k,0)}$
  are  identical with the set of  even labeled   eigenvalues  for the $d=1$ Cauchy well:  $E_{(k,0)}(d=3)=E_{2 k}(d=1)$.
  Likewise,  the  eigenfunctions  $\psi _{(k,0 )}(d=3)$  and   $\psi _{2k }(d=1)$  show  affinity.  We have identified the
   generic functional form of eigenfunctions of the spherical well which  appear to be  composed of  a product of a  solid
   harmonic and  of a suitable  purely radial function.  The method to evaluate (approximately) the latter
   has been found to follow the  universal pattern  which  effectively  allows to  skip all, sometimes involved,  intermediate
   calculations (those were in usage, while  computing the eigenvalues for  $l \leq 3$).

\end{abstract}
 \maketitle

\section{Motivation.}

A classical relativistic  Hamiltonian  $H= \sqrt{  m^2c^4 + c^2p^2} - mc^2$,
  where $c$ stands for the velocity of light, upon a standard canonical quantization recipe ($p \rightarrow -i\hbar \nabla $)  gives rise to the
   energy operator   $\hat{H} = \sqrt{-\hbar ^2c^2 \Delta + m^2c^4} - mc^2$. Its ultrarelativistic version (often interpreted
   as the  mass zero limit of the former)
   reads   $\hat{H} =  \hbar c \sqrt{- \Delta }$.

 Both operators are spatially nonlocal. The meaning of symbolic   expressions (like e.g. the square root of the minus Laplacian)
  is   well established,  see e.g. \cite{GS} and  references there in.  Compare e.g. also \cite{KKM}-\cite{KGSZ}.

Spectral problems for  $d=3$  quasi-relativistic  quantum   systems  in the presence  of  harmonic or  Coulomb potentials
 have  received an ample coverage in the literature, in  the context
of high-energy physics  (mostly computer-assisted  spectral  outcomes), \cite{lucha1,lucha2}, mathematical physics
 \cite{herbst},   and stability of matter  problems  (an enormous literature on the high level of mathematical rigor),
 \cite{lieb}.
 The $d=1$  quasi-relativistic  oscillator and  the finite well problems (the latter has  never been elevated to $d=3$)
 were   elaborated in detail   in \cite{acta}, see also  \cite{KKM}  where the  infinite well problem
 has been analyzed in  some depth.

  The  $d=3$ ultrarelativistic  operator  can  be given a  physical interpretation  within  the photon
  wave mechanics framework,  \cite{IBB}, see also \cite{GS,ZG}. As  well  it  may serve  as a  natural  approximation  of the  "true" generator in the
  quasi-relativistic quantum  mechanics  of nearly massless  particles.
     Another view  is to give the  ultrarelativistic operator  a status of one specific   (Cauchy)  example in an infinite
    family  of fractional (strictly speaking, L\'{e}vy stable, admitting a conceptual extension  from $d=1$ to $d=3$)
     energy operators. Each member of the family  gives   rise to  the  legitimate Schr\"{o}dinger-type
    evolution equation and   various  Schr\"{o}dinger-type spectral (eigenvalue) problems in the presence of
    external potentials.  We recall \cite{GS}, that such fractional quantum mechanics framework appears to be
     devoid of any natural  massive particle content,  which is the case in all pedestrian discussions  of the standard Schr\"{o}dinger picture
   quantum mechanics.

Various $d=1$ spectral problems for fractional  operators (Cauchy in this number), like e.g.  those with  the harmonic or anharmonic
 potentials,  have been widely  studied in  (mostly mathematical)  literature. The  essential progress has been
  made just recently, \cite{gar,lorinczi,lorinczi1}.   The infinite fractional  well in $d=1$  has been studied primarily
  by mathematicians   and   preliminary  attempts were made to attack  the fully-fledged   $d=3$ spectral problem,
    \cite{KKMS}-\cite{DKK2}.

    In the mathematically oriented research, the  main objective for the eigenfunctions was to deduce
    approximate formulas,   next  monotonicity, concavity and norm estimates, plus the decay rates at the boundary.
With respect to the eigenvalues, the focus was on  results concerning properties  of the spectrum,
 like e.g. multiplicity and approximation   of eigenvalues, with suitable upper and lower accuracy bounds.

 We follow   a bit more pragmatic line of research  in Refs. \cite{ZG,KGSZ},  with the aim to deduce most accurate to date
 shapes  of eigenfunctions and  possibly most accurate  approximate eigenvalues for the ultrarelativistic case
 proper,  with a focus on would-be   simplest models   of the  finite and infinite   Cauchy wells.
 Our previous   analysis \cite{ZG}  has been restricted to $d=1$, like  in  the   mathematical references mentioned above.

Interestingly, the  $d=3$ investigation of the  spherical well   analog of the  $d=1$  fractional (and thus also Cauchy)
 infinite well problem  has been  initiated only recently, \cite{D,DKK,DKK2}. The existence of solutions to the eigenvalue problem
  has been demonstrated, together with that of a non-decreasing unbounded   sequence of   eigenvalues,
  the lowest eigenvalue being positive and simple, \cite{DKK}.
      An analysis  has been focused  on finding two-sided bounds  for the  eigenvalues of the
 fractional Laplace operator in the unit ball.  An efficient numerical scheme has been proposed and few   exemplary eigenvalues were
  obtained in the $d=3$ case.

    Some general properties of the  fractional   unit ball   spectrum were established, including links of  lowest  $d=1$ eigenvalues
    (specifically, the least one)  with these  related to the $d=3$ problem.
In the derivations, the Authors have  employed so-called solid harmonics, hence worked with  a definite orbital
 (angular momentum) input.   However, consequences of the  orbital dependence,
 except for mentioning  the  trivial  orbital  label  $l=0$ case,   have been  basically  left aside.

   The methods of Ref. \cite{DKK} do  not   give access to explicit  eigenfunctions,  and thence to   their
    approximate shapes.     We are vitally interested in the   orbital $l=0,1,2...$    dependence and
     the expected  $|m|\leq l$
    degeneracy   of the spectrum for each value of  $l$.
It is instructive to  note   that the   only   spectral  solution  in existence, with the $d=3$ generator involved, is that
of the   Cauchy oscillator, \cite{remb1}. It has been solved   exclusively in  the orbital $l=0$ sector and
so far no data are available about   $l\geq 1$ sectors of this specific model system.

  The major purpose of the present paper is  to overcome the  above mentioned (orbital)  shortcomings of
  the existing   $d=3$    formalism for a fractional  infinite well, \cite{DKK,DKK2}.
We are mostly interested in the ultrarelativistic   spectral   problem    $d=3$.
  Therefore, instead of addressing the whole one-parameter family of fractional energy operators,
  we restrict considerations  to the Cauchy  operator.
  This  entails an exploration of  affinities  of the $d=3$ problem  with the   previously
  resolved $d=1$ Cauchy  case, \cite{ZG}, which  go deeper than predicted in Ref. \cite{DKK}.

  Since calculation methods involving fractional operators (with a possible exception of  so-called   fractional derivatives, that share a number
of  shortcomings with the  Fourier multiplier methods, c.f. \cite{ZG} and \cite{GS}) are not bread and butter in the physics-oriented research,
 we pay  attention to  a number of  essential details.
   Our methodology can be  extended   to other fractional spectral problems as well,  but the
    Cauchy (ultrarelativistic) case  is a  perfect playground, where  analytic and numerical intricacies  related to nonlocal operators can be
      efficiently  kept under control.
Additionally, among all fractional generators, it is the Cauchy one  which  remains  close enough to traditional physicists'
 intuitions about what the quantum theory is about, \cite{GS}.

\section{Infinite  spherical  Cauchy well.}

We depart  from a formal  eigenvalue problem for a nonlocal $\alpha \in (0,2)$  fractional operator
\be
(-\Delta)^{\alpha/2}f(x)=E f(x),
\ee
  in a bounded open domain $D  \subset \mathbb{R}^d $,  $d=1,2,3...$ ,
 with a zero condition  in the complement of $D$  (exterior Dirichlet boundary data), meaning that  there holds
   $f(x)= 0$ for $x \notin D$.

Before imposing  the boundary data,   let us recall that for all $x\in\mathbb{R}^d$  the
nonlocal  operator $(-\Delta)^{\alpha /2}$  is defined as follows, \cite{GS,DKK}:
\be
(-\Delta)^{\alpha/2}f(x)=\mathcal{A}_{\alpha,d}\lim\limits_{\varepsilon\to 0^+} \int\limits_{\mathbb{R}^d\cap\{|y-x|>\varepsilon\}}
\frac{f(x)-f(y)}{|x-y|^{\alpha+d}}dy, \quad 0<\alpha<2,
\ee
where the (L\'{e}vy measure) normalisation coefficient $\mathcal{A}_{\alpha,d}$ reads
\be
\mathcal{A}_{\alpha,d}=\frac{2^\alpha\Gamma(\frac{\alpha+d}{2})}{\pi^{d/2}|\Gamma(-\frac{\alpha}{2})|}.
\ee
Since we are interested in the ultrarelativistic (Cauchy) operator, we shall ultimately set
  $\alpha=1, d=3$ and accordingly  $\mathcal{A}_{1,3}=\pi^{-2}$.\\

The  implementation of  exterior Dirichlet boundary data  upon a nonlocal operator,   which is a priori  defined everywhere in  $\mathbb{R}^d$,   is not a trivial affair, c.f.
the $d=1$ analysis of this issue in Refs. \cite{acta,ZG,KGSZ}.   Our $d=3$ solution  of the spectral problem for the infinite Cauchy well   will  rely  in part  on  $d=1$  intuitions   of  Ref. \cite{ZG}.
It  is  possible   due to  the    radial symmetry of the Cauchy generator  in $d=3$, c.f. also \cite{D,DKK,DKK2}  which  enforces a "natural" topology  of the infinite  well  in $d=3$ as
 that of  the spherical well  (actually the unit ball).

 We point out  that  in conjunction with the standard Laplacian,  a typical well shape,  considered in the literature,   is  that of  a cube.  Nonetheless,  the spherical well, both finite and infinite,
  has received some attention in  the  quantum theory tetxtbooks  and in the nuclear physics literature \cite{G}.\\

For clarity of discussion  and further usage in the present paper,  we find instructive to  set a link with a discussion of Refs. \cite{ZG,KGSZ}, on how to reconcile the $d=1$
  spatial nonlocality of the generator with the exterior   Dirichlet boundary data.    Namely, in the  $d=1$  case,  the general expression Eq.  (2) takes the form of the Cauchy principal value
   (relative to $0$) of the   integral
 $(-\Delta )^{1/2} f(x)=\frac{1}{\pi}\int\limits_{-\infty}^\infty\frac{f(x)- f(t)}{(t-x)^2}dt$.   Let us assume that
 $x\in D=(-1,1)\subset \mathbb{R}$ and demand $f(x) $ to vanish on
  the complement of $D$.
Keeping in mind the integration  singularities
(their  impact  has been made explicit in  Eq. (5) of Ref. \cite{KGSZ}),   we can pass to another   form  of  $ (-\Delta )^{1/2}f  $:
\be
(-\Delta )^{1/2} f(x)=   (p.v.) \left[  \frac{1}{\pi}\int\limits_{-\infty}^\infty\frac{f(x)}{(t-x)^2}dt-
\frac{1}{\pi}\int\limits_{-1}^1\frac{f(t)}{(t-x)^2}dt \right],
\ee
where the Cauchy principal value symbol  $(p.v)$  appears in the self-explanatory notation.
Although   Eq. (4)  looks  excessively formal, since both integrals are  hypersingular \cite{KGSZ}, the adopted  $(p.v.)$  recipe
(given $x$,  execute integrations over  $|t-x|>0$,  subtract  two   finite integrals,  ultimately  take the $\epsilon \rightarrow 0$ limit), allows to  handle all  obstacles.\\

We shall elevate this $d=1$ observation to $d=3$. Effectively, in three dimensions,   the fractional operator while acting on functions $f$  that vanish everywhere,
except  for an  open set $D \subset \mathbb{R}^3$  (i.e.   vanish for $|\textbf{r}|=r \geq 1$),   may be considered as the   $(p.v.)$-regularized difference of two singular integrals,
 in close  affinity with Eq. (4).    Namely,  in view of Eq. (2)  we have:
\be
(-\Delta )^{1/2}  f(\textbf{r})=
 (p.v.) \left[  \frac{1}{\pi ^2}\int_{\mathbb{R}^3}\frac{f(\textbf{r})}{(\textbf{u}  -
  \textbf{r})^4}d^3u - \frac{1}{\pi ^2}\int_{D} \frac{f(\textbf{u})}{(\textbf{u} - \textbf{r})^4}d^3u \right]
\equiv  I_1(\textbf{r}) - I_2(\textbf{r}),
\ee
where the notation $(p.v.)$ indicates that, given $\textbf{r}=(x,y,z) \in D \subset \mathbb{R}^3$,  integrations are carried out over $\textbf{u}=(p,t,s) \in \mathbb{R}^3$ such that  $|\textbf{u}-\textbf{r}|>\epsilon $,
and  subsequently  the $\epsilon \rightarrow 0$
limit is to  follow.

In computations  to be carried out in below, we shall simplify the notation by skipping the $(p.v.)$ symbol  and  passing to a formal difference   $I_1(\textbf{r}) - I_2(\textbf{r})$  of  singular  integrals.
 We shall make   explicit the    divergent contributions that are   cancelled away in the $(p.v.)$ procedure.    We note, that  in spherical coordinates, $I_1(\textbf{r}) $    involves an integration with respect to the radial
 parameter  $r \in (0,\infty)$, while   $I_2(\textbf{r})$  refers to the radial integration over   $r\in (0,1)$ (the unit ball assumption).  \\

\section{Ground state  and other purely radial eigenfunctions.}

In the present section we shall use a notation $D= \{\textbf{r}= (x,y,z)\in\mathbb{R}^3:\> x^2+y^2+z^2 < 1\}$.
Upon  assuming that the eigenfunction  shows up the radial dependence only  $f(\textbf{r}) = \psi (r)$, with $r=\sqrt{x^2+y^2+z^2}$,  we may
 consider the eigenvalue problem in a simpler form.

  Namely, since  for a purely radial function  we have an identity
  $f(\textbf{r})= f(0,0,|z|)= \psi (|z|)$, it suffices to investigate   $f(\textbf{r})$
  along the  $z$-semiaxis,   for all $(0,0,|z|)\in D$,
    i.e.   $\psi (|z|)$ for   $|z|<1$.   Quite analogously we may proceed with  $|x|<1$, and likewise with  $|y|<1$.

Guided by intuitions coming from our previous analysis of the infinite Cauchy well in $d=1$, \cite{ZG}, we seek the ground state
 function   of the $d=3$   (infinite) spherical well problem in the form of  power  series:
 \be
f(\textbf{r}) =\psi(r) =  C \sqrt{1-r^2}\sum_{n=0}^\infty\alpha_{2n}r^{2n},\qquad \alpha_0=1,
\ee
where  $C$  is the normalization constant defined through  $|C|^2\int_{4\pi} d\Omega\int_0^1 r^2\psi^*(r)\psi(r)dr=1$.
In view of $\psi (r)= \psi (|z|)$,  instead of the fully-fledged eigenvalue problem (1), with   (5) implicit,
 we  shall seek solutions of
\be
(-\Delta)^{1/2}\psi (|z|)=E  \psi (|z|),\qquad |z|<1,
\ee
for  $\textbf{r}=(0,0, |z|), \, |z|<1$    hence  effectively  along  the interval $z\in (-1,1)$ on the $z$-semiaxis. We recall that
$\psi (|z|)$ needs to  vanish identically  for $|z|\geq 1$.

First we shall establish  what is the output of the action  (5)   of the  Cauchy operator upon radial functions of the
 form $r^{2n} \sqrt{1-r^2}$, with $n=0,1,2,...$, while  evaluated at  $r=|z|$, for rationale  c.f. Eq. (6).\\

\subsection{ $(-\Delta)^{1/2} [r^{2n} \sqrt{1-r^2}](|z|)$.}

  Let us begin from the $n=0$ case.  By  direct computation, one    arrives at:
\be
\left((-\Delta)^{1/2}\sqrt{1-x^2-y^2-z^2}\right)(0,0,|z|)=2.
\ee
We shall take an opportunity to  perform  computations in  detail for this exemplary case, to indicate how potentially divergent
 terms are $(p.v.)$-handled.  Integrations will  be  carried out in spherical coordinates:
\be
\left\{
  \begin{array}{ll}
    u=r\cos\phi\sin\theta,  \\
    v=r\sin\phi\sin\theta,  \\
    w=r\cos\theta,
  \end{array}
\right.
\qquad |J|=r^2\sin\theta,\qquad
\left\{
  \begin{array}{ll}
    r\geqslant 0,  \\
    0\leqslant \phi <2\pi,  \\
    0 \leqslant \theta <\pi.
  \end{array}
\right.
\ee

We interpret the left-hand-side   of Eq. (8) as a  (controlled)   subtraction of two  singular  integrals $I_1(r) - I_2(r)$,  with
$r=\sqrt{x^2+y^2+z^2}$,  c.f.    Eq. (5).  Keeping in mind the  $(p.v.)$ recipe, we shall evaluate  each of these integrals separately with
 divergent terms clearly isolated.  We know that they are to be  cancelled  away in the subtraction procedure.

Let us   consider   $I_1(|z|)$  and   $I_2(|z|)$   at the origin, specified by the value   $|z|=0$. We have:
\be
I_1(0) =\frac{1}{\pi^2}\calkan \frac{r^2\sin\theta}{r^4}=
\frac{4}{\pi}\int\limits_0^\infty\frac{dr}{r^2}=\lim\limits_{r\to 0} \frac{4}{\pi r},
\ee
while for $I_2(|z|)$ there holds
\be
I_2(0) =\frac{1}{\pi^2}\calkap \frac{r^2\sin\theta\sqrt{1-r^2} }{r^4}=\frac{4}{\pi}\int\limits_0^1 dr\frac{\sqrt{1-r^2}}{r^2}=\frac{4}{\pi}\left(-\frac{\pi}{2}+\lim\limits_{r\to 0} \frac{1}{r}\right).
\ee
Since we effectively follow the $(p.v.)$ recipe, the divergent terms cancel each other  and  we   arrive  at  $I_1(0) -I_2(0)=2$.
 Let us consider  $1>|z|\neq 0$. Accordingly:
\be
I_1(|z|) =\frac{\sqrt{1-|z|^2}}{\pi^2}\calkan \frac{r^2\sin\theta}{(r^2-2r|z|\cos\theta+|z|^2)^2}.
\ee
The $\phi$  integration produces a $2\pi$  factor.  By  employing
\be
\int\frac{\sin x\, dx}{(A-B\cos x)^2}=- \frac{1}{B(A-B\cos x)},
\ee
we  get
\be
I_1(|z|)=\frac{\sqrt{1-|z|^2}}{\pi |z|}\int\limits_0^\infty  r\left(\frac{1}{(r-|z|)^2}-\frac{1}{(r+|z|)^2}\right)dr.
\ee
 $I_1(|z|)$ can be rewritten as a difference of two integrals, the first of which is singular.
 In view of the implicit $(p.v.)$ recipe,  the first integration is carried over intervals  $(0,|z|-\varepsilon)$ and
  $(|z|+\varepsilon,\infty)$, where  $\varepsilon>0$   and the ultimate limiting procedure $\varepsilon\to 0$ is implicit
   while computing $I_1(|z|) - I_2(|z|)$.
Because of
\be
\int \frac{r\,dr}{(r\pm |z|)^2}=\pm\frac{|z|}{r\pm |z|}+\ln\left|r\pm|z|\right|,
\ee
we have
\be
I_1(|z|)=\frac{2\sqrt{1-|z|^2}}{\pi}\lim\limits_{\varepsilon\to 0}\frac{1}{\varepsilon}.
\ee
The second entry  in $I_1(|z|) - I_2(|z|)$ reads
\be
I_2 (|z|) =\frac{1}{\pi^2}\calkap \frac{r^2\sin\theta \sqrt{1-r^2}}{(r^2-2r|z|\cos\theta+|z|^2)^2}=\frac{1}{\pi |z|}\int\limits_0^1 r\sqrt{1-r^2}\left(\frac{1}{(r-|z|)^2}-\frac{1}{(r+|z|)^2}\right)dr.
\ee
Indefinite integrals
\be
\int \frac{r\sqrt{1-r^2}}{(r\pm|z|)^2}dr=\frac{(r\pm 2|z|)\sqrt{1-r^2}}{r\pm |z|}\pm2|z|\arcsin(r)-\frac{(-1+2|z|^2)\ln\left|r\pm |z|\right|}{\sqrt{1-|z|^2}}+\frac{(-1+2|z|^2)\ln(1\pm r|z|+\sqrt{1-r^2}\sqrt{1-|z|^2})}{\sqrt{1-|z|^2}},
\ee
need some care concerning the integration intervals (c.f. the previous singular case Eq. (16)) and  keeping in mind  an ultimate
  $\varepsilon \to 0$ limit.  The final result is:
\be
I_2(|z|) =-2+\frac{1}{\pi}\lim\limits_{\varepsilon\to 0}\left(\frac{\sqrt{1-(|z|-\varepsilon)^2}}{\varepsilon}+\frac{\sqrt{1-(|z|+\varepsilon)^2}}{\varepsilon}\right).
\ee
The difference $I_1(|z|)  -I_2(|z|)$,  if carried out in the $(p.v.)$ manner, involves a well defined  limiting  expression
\be
\lim\limits_{\varepsilon\to 0}\left(\frac{2\sqrt{1-|z|^2}}{\varepsilon}-\frac{\sqrt{1-(|z|-\varepsilon)^2}}{\varepsilon}-\frac{\sqrt{1-(|z|+\varepsilon)^2}}{\varepsilon}\right)=0,
\ee
hence for all  $0<|z|<1$,  there holds  $I_1(|z|) -I_2(|z|)=  2$ as  anticipated in Eq. (8).\\

Since $r=\sqrt{x^2+y^2+z^2}$, we can proceed  analogously  to  evaluate  $(-\Delta )^{1/2} ( r^{2n}\sqrt{1-r^2})$, $n\in \mathbb{N}$
at $ (0,0,|z|)\in D$. We get:
\be
\left((-\Delta)^{1/2}r^2\sqrt{1-r^2}\right)(0,0,|z|)=-1+4|z|^2=\left(2\left(-\frac{1}{2}\right)+4\cdot 1\cdot r^2\right)(0,0,|z|),
\ee
\be
\left((-\Delta)^{1/2}r^4\sqrt{1-r^2}\right)(0,0,|z|)=-\frac{1}{4}-2|z|^2+6|z|^4=\left(2\left(-\frac{1}{8}\right)+4\left(-\frac{1}{2}\right)r^2+6\cdot 1\cdot r^4\right)(0,0,|z|),
\ee
\be
\left((-\Delta)^{1/2}r^6\sqrt{1-r^2}\right)(0,0,|z|)=-\frac{1}{8}-\frac{1}{2}|z|^2-3|z|^4+8|z|^6=\left(2\left(-\frac{1}{16}\right)+4\left(-\frac{1}{8}\right)r^2+6\left(-\frac{1}{2}\right) r^4+8\cdot 1\cdot r^6\right)(0,0,|z|).
\ee
and more generally:
\be
(-\Delta)^{1/2}r^{2n}\sqrt{1-r^2}(0,0,|z|)=\left(2c_{2n}+4c_{2n-2}r^2+\ldots+(2n+2)c_0r^{2n}\right)(0,0,|z|),
\ee
where  $c_{2n}$   are  coefficients of the Taylor expansion of $\sqrt{1-r^2}$, with  $r<1$:
\be
\sqrt{1-r^2}=\sum_{n=0}^{\infty}c_{2n}r^{2n}=\sum_{n=0}^{\infty}\frac{(2n)!}{(1-2n)(n!)^24^n}r^{2n}.
\ee

Our major observation, to be employed in below, is that if we act $(-\Delta)^{1/2}$ upon $\sqrt{1-r^2}\, w_{2n}(r)$ where  $w_{2n}(r) = \sum_{k=0}^{n} \alpha_{2k}r^{2k}$
 is a polynomial of the $2n$-th degree,  the outcome is the sole (no $\sqrt{1-r^2}$ factor)  polynomial of the $2n$-th degree, compare e.g.  Eq. (24),  see  also \cite{ZG,D}.

 On the other hand, we   have assumed that the ground state   $\psi (r)=\psi (|z|)$  should have  a functional form
$\psi (|z|) = C \sqrt{1-r^2}\sum_{k=0}^\infty\alpha_{2k}\, r^{2k}$, $\alpha_0=1$,  where $C$ is the $L^2(D)$ normalization constant.
Under these premises, the  validity of  the eigenvalue equation   $(-\Delta)^{1/2}\psi(|z|)=E \psi(|z|)$,  for all    $|z|<1$, is far from being obvious.

\subsection{Approximate ground state function.}

Further procedure follows the main idea of Ref. \cite{ZG}.
Expansions coefficients $\alpha_{2k}$  of $\psi (|z|)$ and the would be eigenvalue $E$,  at the moment remain unknown.
 Nonetheless, presuming all necessary convergence properties, upon  inserting $\psi (|z|)$ to the eigenvalue equation  (7),
we  formally  get
\be
\sum_{k=0}^\infty\alpha_{2k}\sum_{j=0}^k  a_{j,k}\, |z|^{2k-2j}=E\sum_{j=0}^\infty
\sum_{k=0}^\infty  c_{2j} \alpha_{2k}|z|^{2k +2j}\, ,
\ee
where coefficients of the generating matrix  read:
\be
a_{j,k}= 2 (k-j +1) c_{2j}.
\ee

Since we do not see any prospect to solve the above equation (27) analytically with respect to  $E$  and all $\alpha _{2k}, k=1,2,...$
($\alpha _0=1$ being presumed), following the idea of  Ref. \cite{ZG}   (c.f. specifically Section III there in),
 we reiterate to approximate  solution methods  which are based on a  suitable  truncation of  the infinite series
  on both right and left-hand-sides of Eq. (26). We shall discuss truncations  to polynomial expressions of   degrees ranging
   up to $2n=500$.   \\

 (i)  We deliberately insert a  truncated   test  function     (remember about  $r=|z|<1$)
  \be
  \psi ^{(2n)} (r)= C^{(2n)}   \sqrt{1-r^2}\,  w_{2n} (r)=   C^{(2n)}   \sqrt{1-r^2}   \sum_{k=0}^{n}  \alpha_{2k}\, r^{2k}
  \ee
  into the eigenvalue equation  $ (-\Delta)^{1/2}\psi (r)=E  \psi (r),\, 0<r<1$,  compare e.g. Eq. (7).   Clearly, in view of (24),
  the left-hand-side  of the eigenvalue equation  (26)  becomes  a polynomial of the degree $2n$.  $C^{(2n)}$ stands for
   the corresponding normalisation coeeficient.\\

 (ii) The right-hand-side  series      of Eq. (26)  needs to be  truncated   carefully
  to yield a polynomial of the degree $(2n)$  as well,
  so that  we ultimately  get  $n$ equations involving the   unknown energy eigenvalue $E$  and  $n$ coefficients
  $\alpha _{2k}, k=1,2,...,n$,  (we assume   $\alpha _0=1$). \\

(iii) Our approximate function obeys the boundary condition (e.g. vanishes at $r=1$). We extend this boundary condition to the output
of  $(-\Delta )^{1/2} \psi ^{(2n)}(r)$, i.e. we demand
  \be
\lim\limits_{r\to 1}(-\Delta)^{1/2}\psi ^{(2n)}(r)=0.
\ee
which completes the system of $n$ equations  for $n+1$ unknowns (mentioned in (ii))
 by  a supplementary  $n+1$-st   constraint. \\

To derive the  system  of  linear equations   resulting  from our assumptions (ii) and (iii), let us
 rewrite the left-hand-side of the identity (26) as follows
\begin{eqnarray*}
(-\Delta )^{1/2} \, \left(\sum\limits_{k=0}^\infty\alpha_{2k}x^{2k}\sqrt{1-x^2}\right)=
\sum\limits_{k=0}^\infty\alpha_{2k}\left( 2c_{2k}+4c_{2k-2}x^2+\ldots (2k+2)c_0x^{2k}\right)=\\
(\alpha_0\cdot 2c_0+\alpha_2\cdot 2 c_2+\ldots)+(\alpha_2\cdot 4c_0+\alpha_4\cdot 4c_2+\ldots)x^2+(\alpha_4\cdot 6c_0+
\alpha_6\cdot 6c_2+\ldots)x^4+\ldots=\\
(\alpha_0a_{0,0}+\alpha_2a_{1,1}+\ldots)+(\alpha_2a_{0,1}+\alpha_4a_{1,2}+\ldots)x^2+(\alpha_4a_{0,2}+\alpha_6a_{1,3}+\ldots)x^4+\ldots,
\end{eqnarray*}
where the  definition (27) of expansion coefficients  $a_{j,k}$  has been employed.
The  right-hand side of (26) reads:
\begin{eqnarray*}
E\sum\limits_{k=0}^\infty\alpha_{2k}x^{2k}\sqrt{1-x^2}=E(\alpha_0+\alpha_2x^2+\ldots)(c_0+c_2x^2+\ldots)=\\
E(\alpha_0c_0+(\alpha_0c_2+\alpha_2c_0)x^2+(\alpha_0c_4+\alpha_2c_2+\alpha_4c_0)x^4+\ldots).
\end{eqnarray*}
We truncate the power series  in (26)  at the order $2n$ and compare  coefficients  staying at consecutive powers of $x^{2k}$ up
to $k=n$. The  result comprises $n$ equations
\be
\begin{split}
\alpha_0a_{0,0}+\alpha_2a_{1,1}+\ldots+\alpha_{2n}a_{n,n}&=E\alpha_0c_0\\
\alpha_2a_{0,1}+\alpha_4a_{1,2}+\ldots+\alpha_{2n}a_{n-1,n}&=E(\alpha_0c_2+\alpha_2c_0)\\
\alpha_4a_{0,2}+\alpha_6a_{1,3}+\ldots+\alpha_{2n}a_{n-2,n}&=E(\alpha_0c_4+\alpha_2c_2+\alpha_4c_0)\\
&\vdots\\
\alpha_{2n-2}a_{0,n-1}+\alpha_{2n}a_{1,n}&=E(\alpha_0c_{2n-2}+\alpha_2c_{2n-4}+\ldots+\alpha_{2n-2}c_0).\nonumber
\end{split}
\end{equation}

Accordingly, the linear system of equations  with unknown  $E$  and  $\alpha_{2k}, 1\leq k \leq n$, associated with
a truncation of (26) to finite polynomial expressions of degree  $2n$  receives  the final  form
\begin{eqnarray}
\sum\limits_{k=i}^n \alpha_{2k}a_{k-i,k}=E\sum\limits_{k=0}^i \alpha_{2k}c_{2(i-k)},\qquad i=0,1,\ldots,n-1,\nonumber\\
\sum\limits_{m=0}^n \left(\alpha_{2m}\sum\limits_{k=0}^m a_{k,m}\right)=0,
\end{eqnarray}
that is amenable to computer assisted solution methods.  The  last  identity  comes  from the boundary condition (iii). \\

We solve the  linear  system  (30)  by means of  the  Wolfram  Mathematica routines. These  provide  a perfect tool
to solve  large  systems of linear  equations.
One needs to realize that    (30)   has more than one solution.
To select the solution which yields the best approximation of the ground state,  we  seek the lowest eigenvalue $E$
 in the set of  all   (approximate) energy values obtained, compare e.g. also \cite{ZG}. \\

\begin{table}
\begin{center}
\begin{tabular}{|c||c|c|c|c|c|c|c|c|c|c|}
  \hline
  - & C & \textbf{E} & $\alpha_2$ & $\alpha_4$ & $\alpha_6$ & $\alpha_8$ & $\alpha_{10}$ & $\alpha_{12}$ & $\alpha_{14}$ & $\alpha_{16}$\\
  \hline
  $w_2$ & 1.056807 & 2.666667 & -0.666667 & - & - & - & - & - & - & - \\
  $w_4$ & 1.140012 & 2.863894 & -0.913200 & 0.197227 & - & - & - & - & - & -\\
  $w_6$ & 1.106161 & 2.799020 & -0.843785 & 0.207082 & -0.056047 & - & - & - & - & -\\
  $w_8$ & 1.099255 & 2.786553 & -0.831059 & 0.205789 & -0.045361 & -0.016270 & - & - & - & -\\
  $w_{10}$ & 1.094163 & 2.777689 & -0.822196 & 0.204434 & -0.041563 & -0.007494 & -0.015020 & - & - & -\\
  $w_{12}$ & 1.090862 & 2.772063 & -0.816638 & 0.203467 & -0.039622 & -0.005198 & -0.008023 & -0.012058 & - & -\\
  $w_{14}$ & 1.088597 & 2.768252 & -0.812904 & 0.202774 & -0.038442 & -0.004068 & -0.006161 & -0.006311 & -0.010044 & -\\
  $w_{16}$ & 1.086983 & 2.765561 & -0.810281 & 0.202268 & -0.037661 & -0.003395 & -0.005232 & -0.004792 & -0.005218 & -0.008531\\
  $w_{18}$ & 1.085796 & 2.763594 & -0.808372 & 0.201890 & -0.037115 & -0.002954 & -0.004670 & -0.004031 & -0.003948 & -0.004406\\
  $w_{20}$ & 1.084900 & 2.762114 & -0.806941 & 0.201601 & -0.036717 & -0.002645 & -0.004295 & -0.003567 & -0.003309 & -0.003323\\
  $w_{30}$ & 1.082578 & 2.758299 & -0.803271 & 0.200837 & -0.035739 & -0.001932 & -0.003475 & -0.002638 & -0.002207 & -0.001922\\
  $w_{40}$ & 1.081679 & 2.756826 & -0.801863 & 0.200534 & -0.035380 & -0.001686 & -0.003205 & -0.002353 & -0.001901 & -0.001589\\
  $w_{50}$ & 1.081242 & 2.756110 & -0.801180 & 0.200385 & -0.035210 & -0.001573 & -0.003082 & -0.002227 & -0.001770 & -0.001451\\
  $w_{60}$ & 1.080999 & 2.755709 & -0.800799 & 0.200300 & -0.035116 & -0.001511 & -0.003016 & -0.002160 & -0.001701 & -0.001380\\
  $w_{70}$ & 1.080849 & 2.755463 & -0.800565 & 0.200248 & -0.035059 & -0.001474 & -0.002977 & -0.002120 & -0.001660 & -0.001338\\
  $w_{80}$ & 1.080751 & 2.755301 & -0.800411 & 0.200214 & -0.035022 & -0.001450 & -0.002951 & -0.002094 & -0.001634 & -0.001311\\
  $w_{90}$ & 1.080683 & 2.755188 & -0.800305 & 0.200190 & -0.034996 & -0.001436 & -0.002933 & -0.002077 & -0.001616 & -0.001293\\
  $w_{100}$& 1.080634 & 2.755107 & -0.800228 & 0.200173 & -0.034978 & -0.001422 & -0.002921 & -0.002064 & -0.001604 & -0.001281\\
  $w_{150}$& 1.080517 & 2.754913 & -0.800044 & 0.200131 & -0.034934 & -0.001394 & -0.002892 & -0.002035 & -0.001574 & -0.001251\\
  $w_{200}$& 1.080476 & 2.754844 & -0.799979 & 0.200116 & -0.034918 & -0.001384 & -0.002881 & -0.002025 & -0.001564 & -0.001241\\
  $w_{300}$& 1.080446 & 2.754795 & -0.799932 & 0.200105 & -0.034907 & -0.001377 & -0.002874 & -0.002017 & -0.001557 & -0.001234\\
  $w_{400}$& 1.080436 & 2.754777 & -0.799916 & 0.200102 & -0.034903 & -0.001375 & -0.002871 & -0.002015 & -0.001555 & -0.001231\\
  $w_{500}$& 1.080431 & \textbf{2.754769} & -0.799908 & 0.200100 & -0.034901 & -0.001374 & -0.002870 & -0.002014 & -0.001553 & -0.001230\\
  \hline
\end{tabular}
\end{center}
\caption{Computed  exemplary  expansion coefficients  $\alpha_{2k}, k\leq  n$    for polynomial  entries  $w_{2n}= \sum_{k=0}^{n}  \alpha_{2k}\, r^{2k} , n=1, 2 ,....,250$  in the approximate ground state function
 expression $\psi ^{(2n)}(r) = C^{(2n)}   \sqrt{1-r^2} w_{2n}(r) $.  We have a clear  picture of the convergence properties (and stabilization tendency)  of $E^{(2n)}$ and $C^{(2n)}$, together with that of displayed coefficients,
 with the growth of $n$ towards $250$.  We indicate that  our  spectral  result  $E^{(500)}= 2.754769$ may  be set in comparison with $E=  2.75476$ computed independently  in Ref. \cite{D}.}
 \end{table}

\begin{figure}[h]
\begin{center}
\centering
\includegraphics[width=110mm,height=110mm]{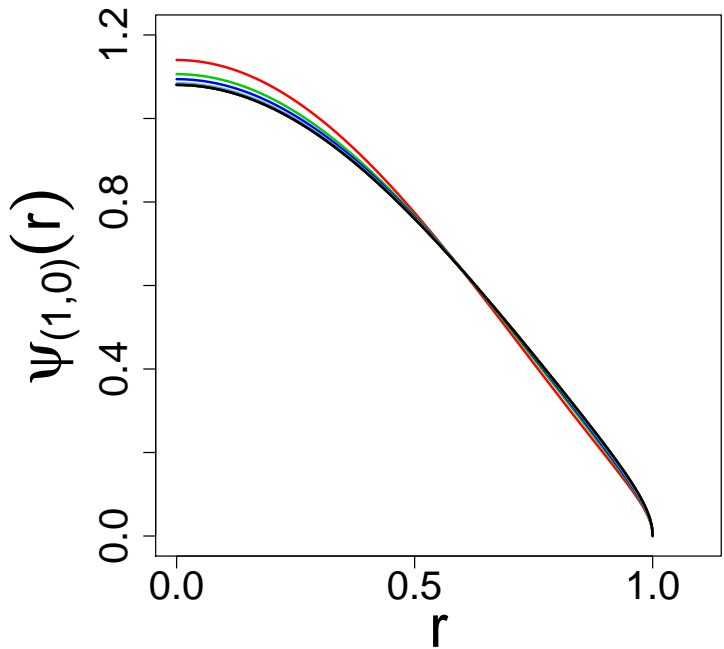}
\caption{A comparative display of approximate ground state functions  $\psi _{(1,0)}^{(2n)}(r)$  for polynomial (approximation)
 degrees   $2n= 4,6,10,20,30,50,70,100,150,200,500$. For each consecutive  curve, the maximum location  drops down with the growth
 of  $2n$. The $2n=500$ curve is depicted in black. }
\end{center}
\end{figure}
\begin{figure}[h]
\begin{center}
\centering
\includegraphics[width=70mm,height=70mm]{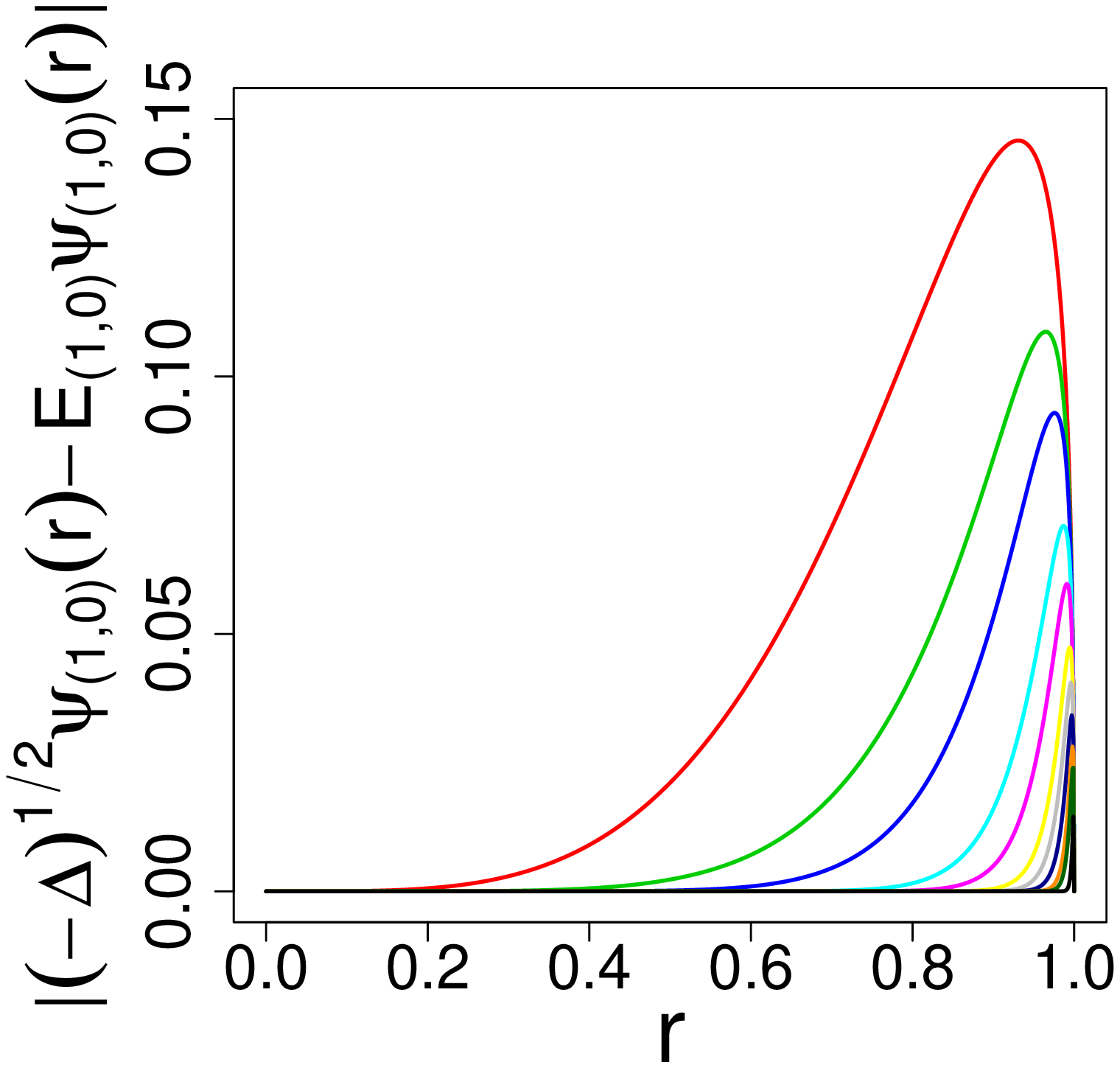}
\includegraphics[width=70mm,height=70mm]{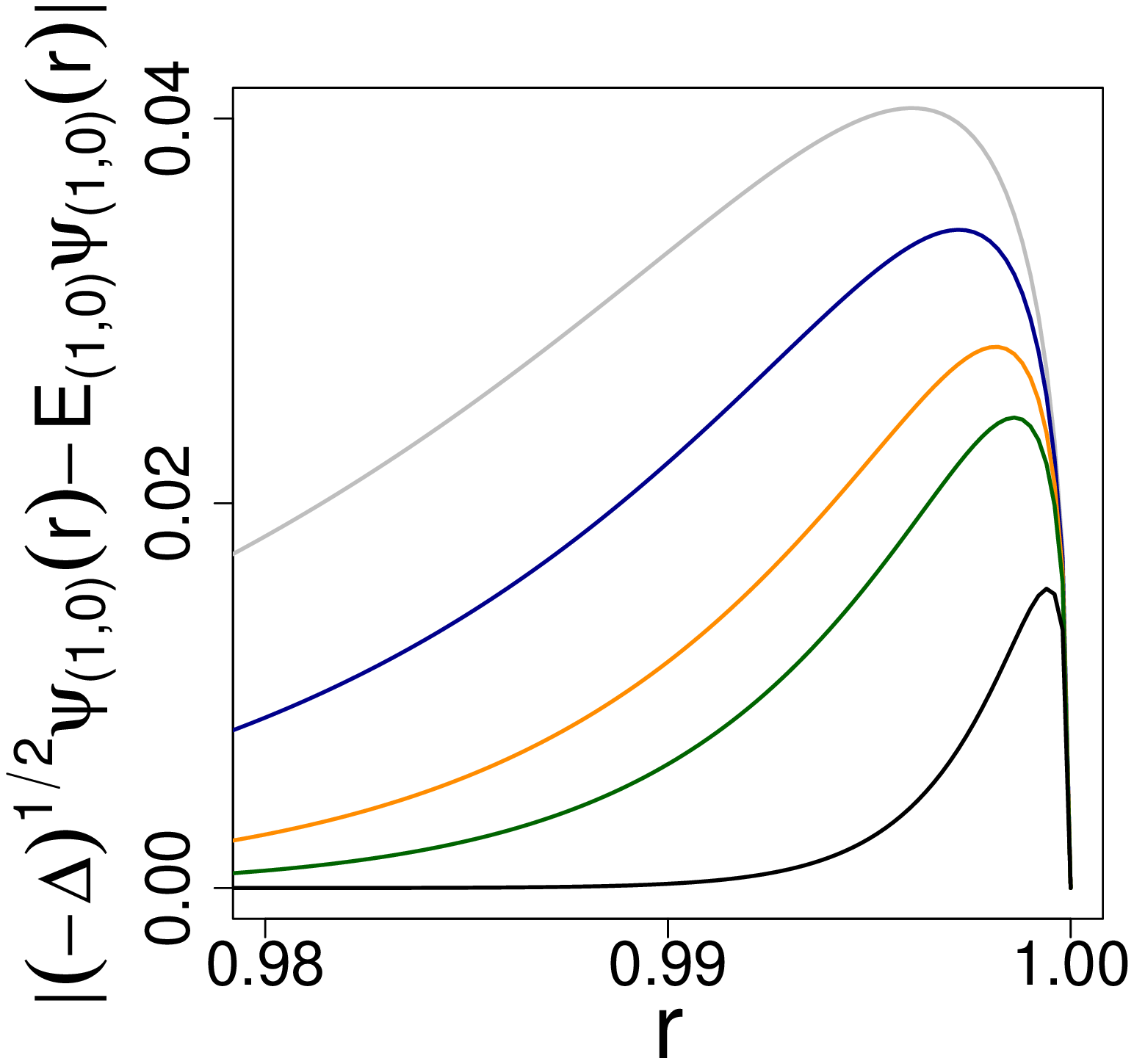}
\caption{Detuning between the polynomial    $(-\Delta )^{1/2} \psi ^{(2n)}_{(1,0)}(r) $  of the degree $2n$ and the non-polynomial
 approximate expression $E^{(2n)}_{(1,0)}  \psi ^{(2n)}_{(1,0)}(r)$.   Left panel:  a comparative display for
  $2n=4,6,10,20,30,50,70,100,150,200,500$.  Right panel: the detuning display for  $2n=70,100,150,200,500$. The $2n=500$ curve is
  depicted in black.}
\end{center}
\end{figure}

\begin{figure}[h]
\begin{center}
\centering
\includegraphics[width=50mm ,height=50mm]{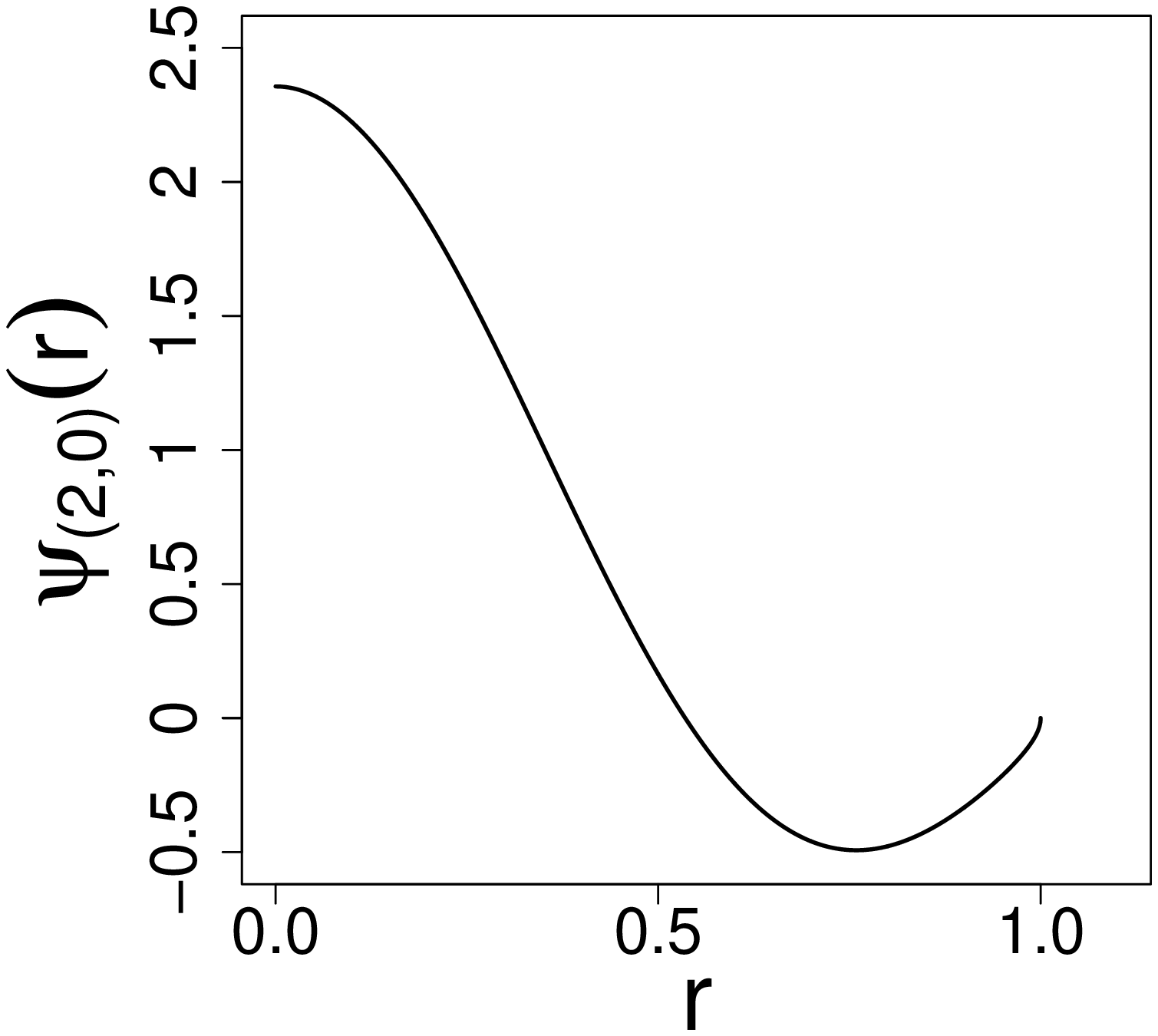}
\includegraphics[width=50mm,height=50mm]{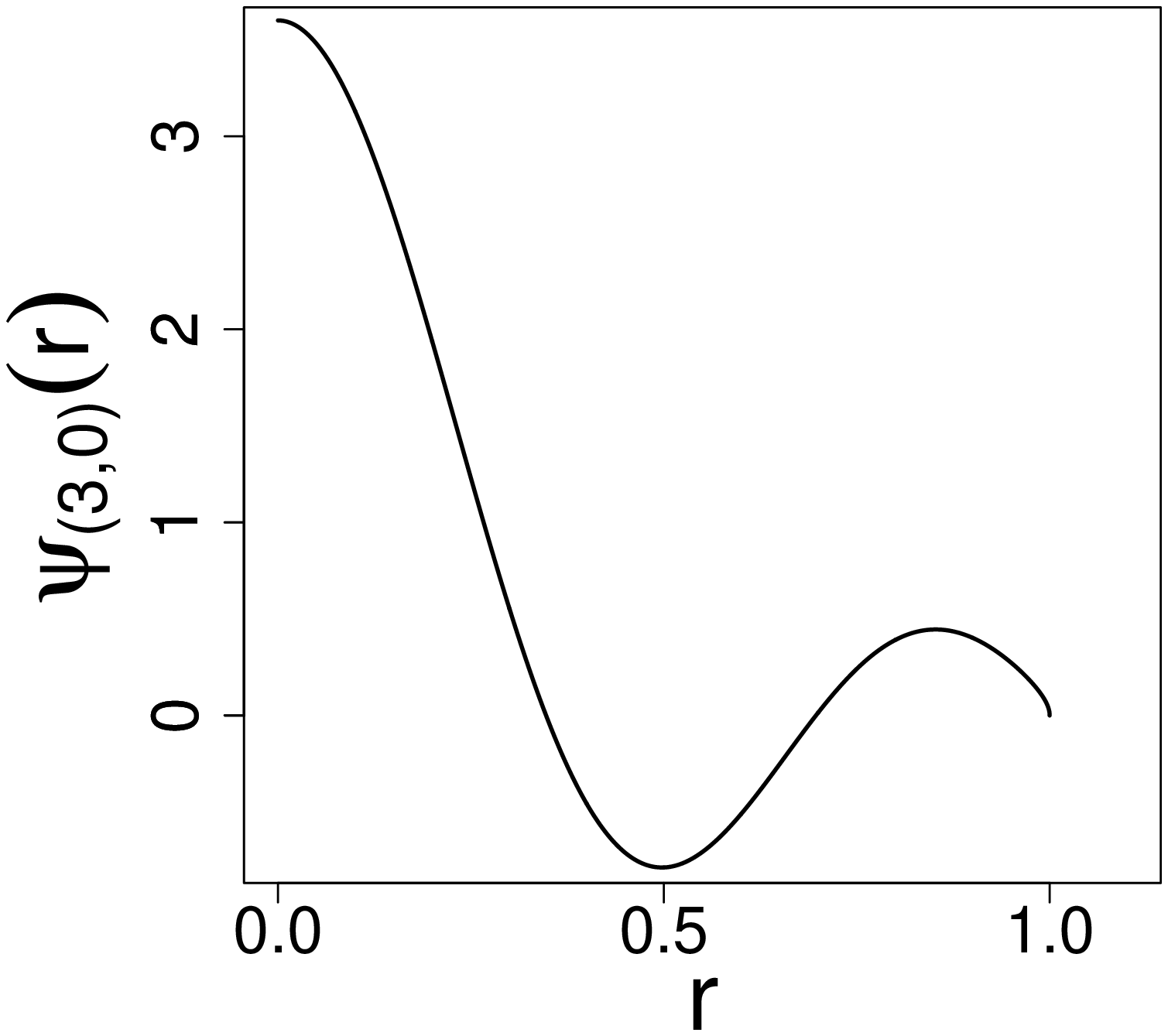}
\includegraphics[width=50mm,height=50mm]{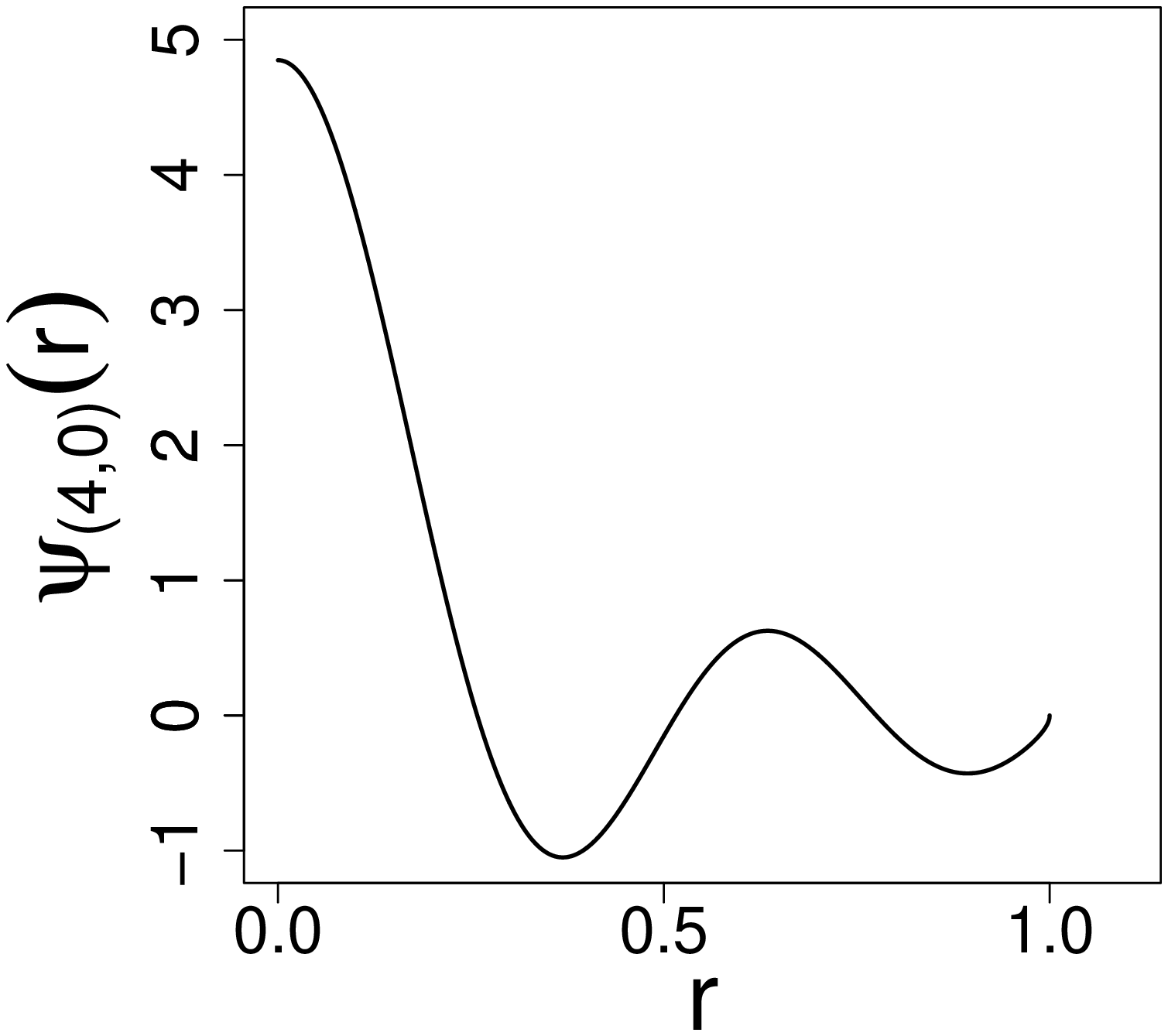}
\caption{Contour shapes of approximate   purely radial  eigenfunctions  $\psi ^{(500)}_{(k,0)}(r)$  for $k=2, 3, 4$.}
\end{center}
\end{figure}
\begin{figure}[h]
\begin{center}
\centering
\includegraphics[width=60mm,height=60mm]{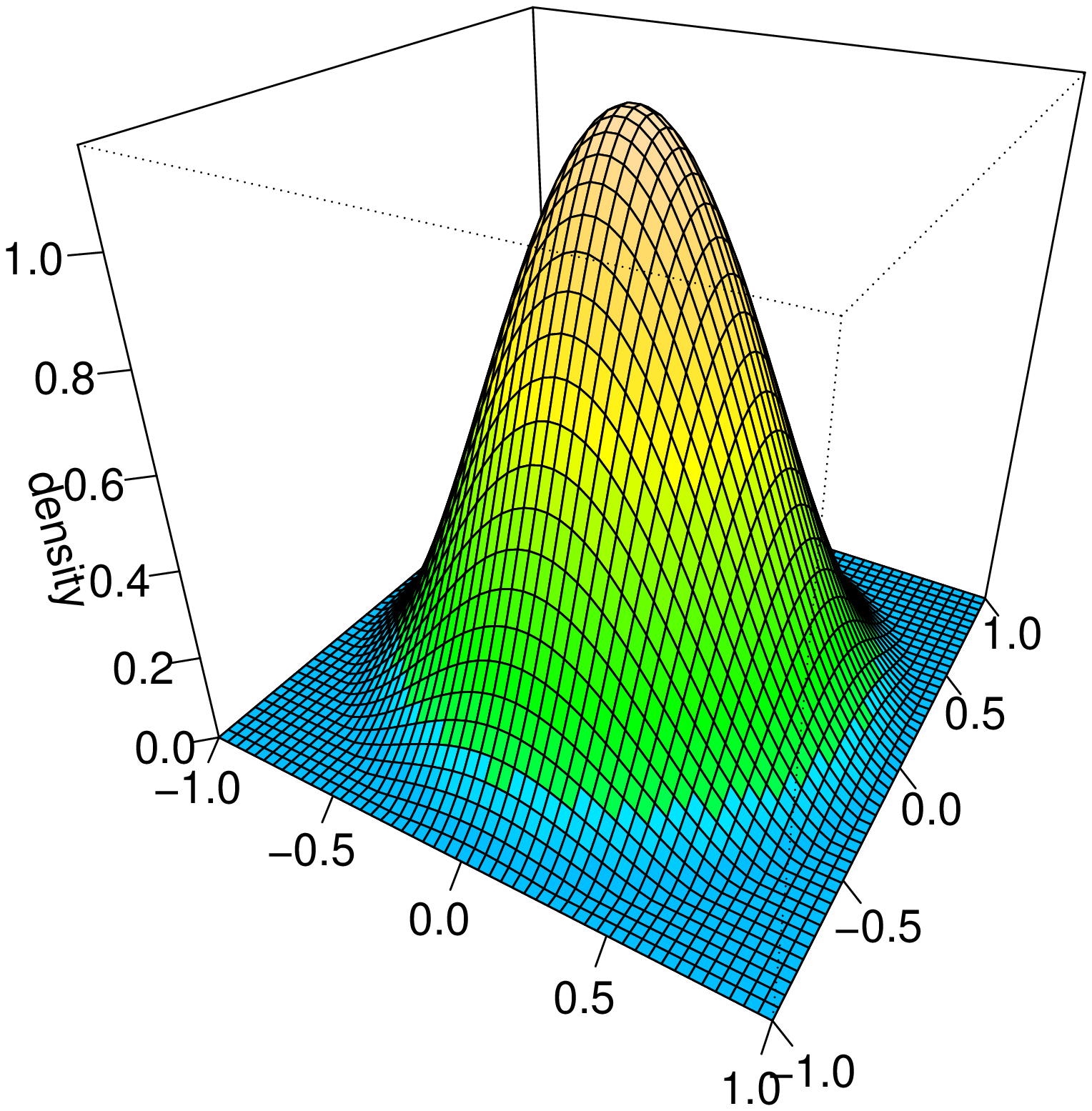}
\includegraphics[width=60mm,height=60mm]{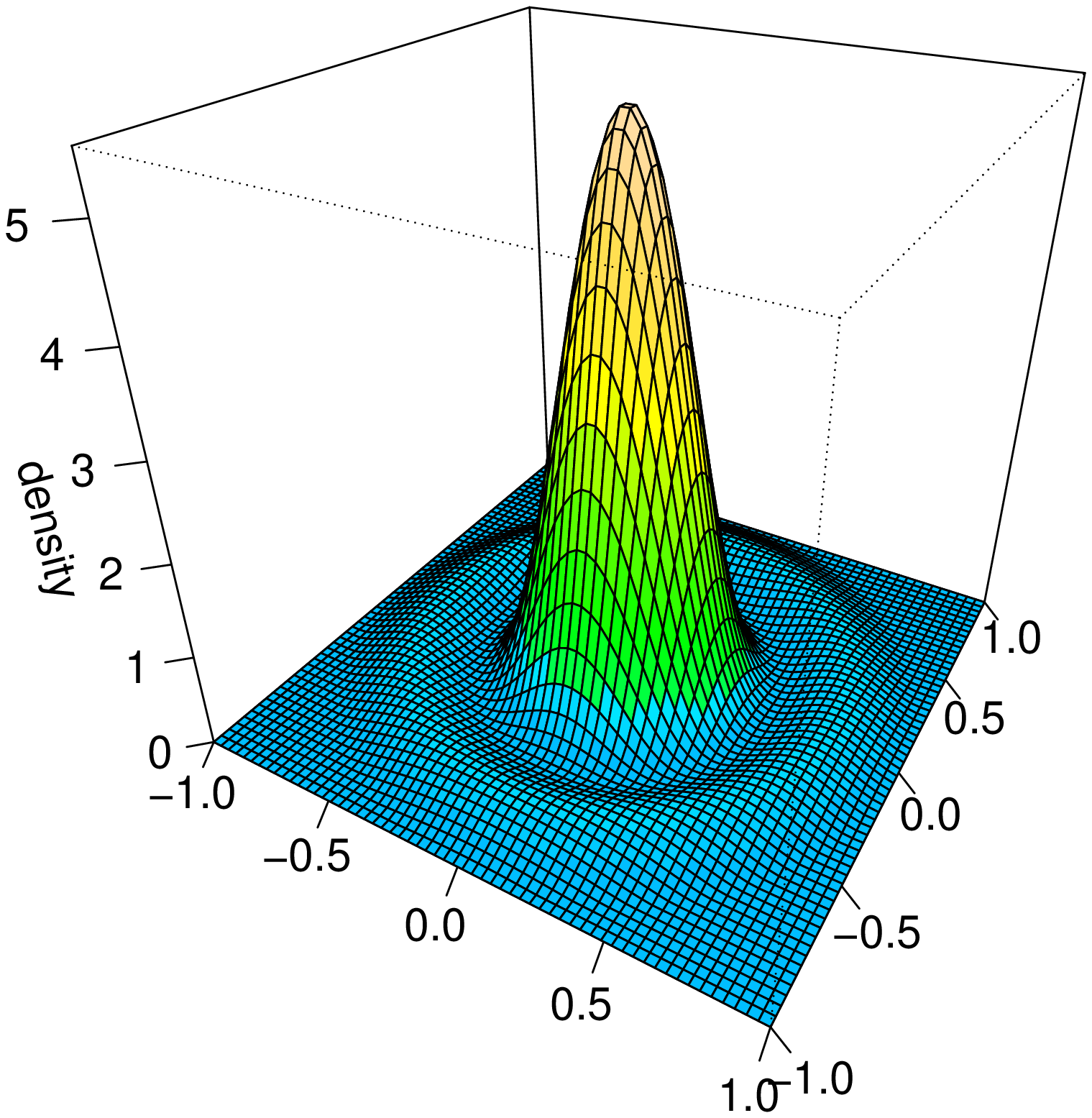}
\includegraphics[width=60mm,height=60mm]{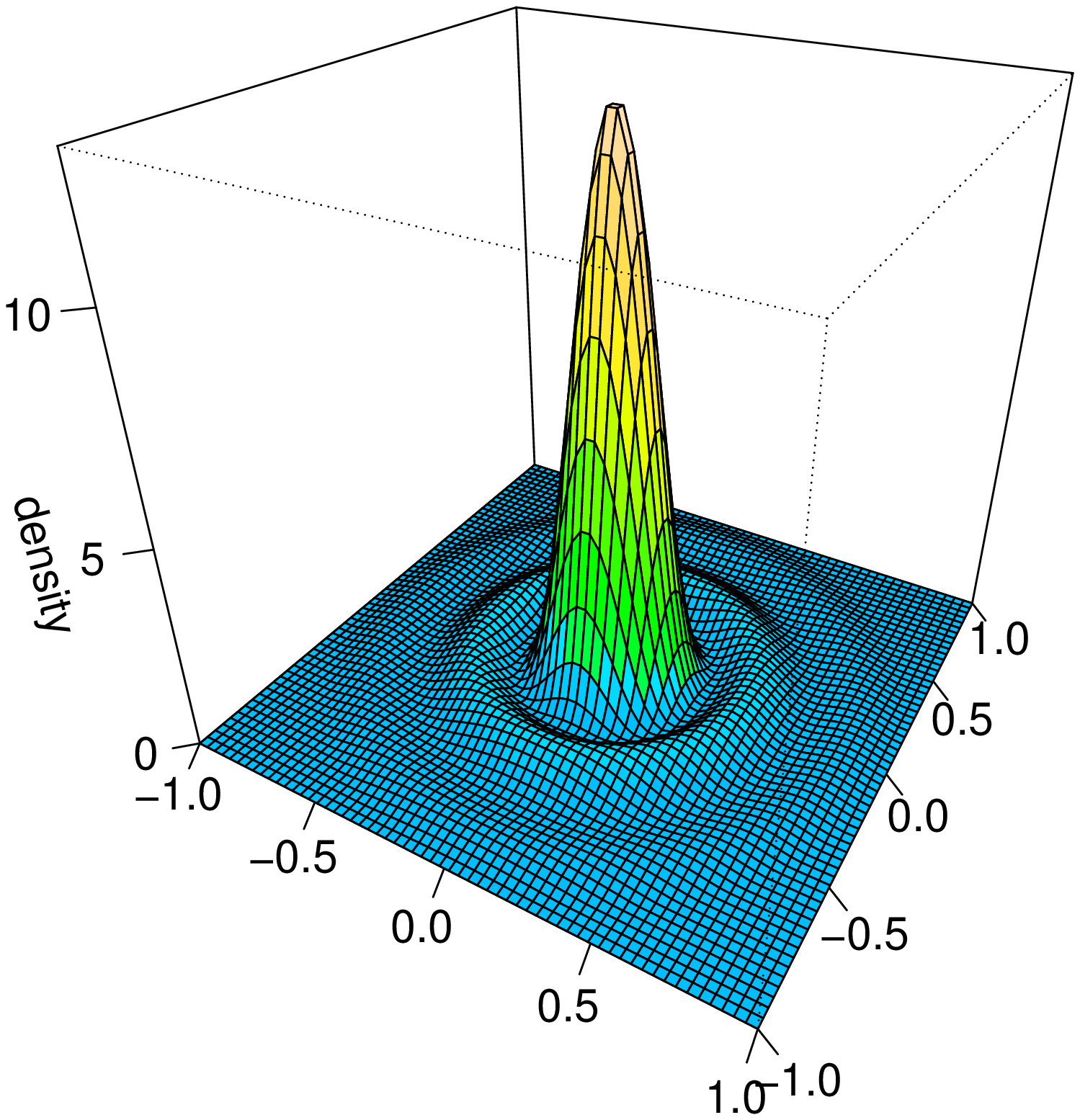}
\includegraphics[width=60mm,height=60mm]{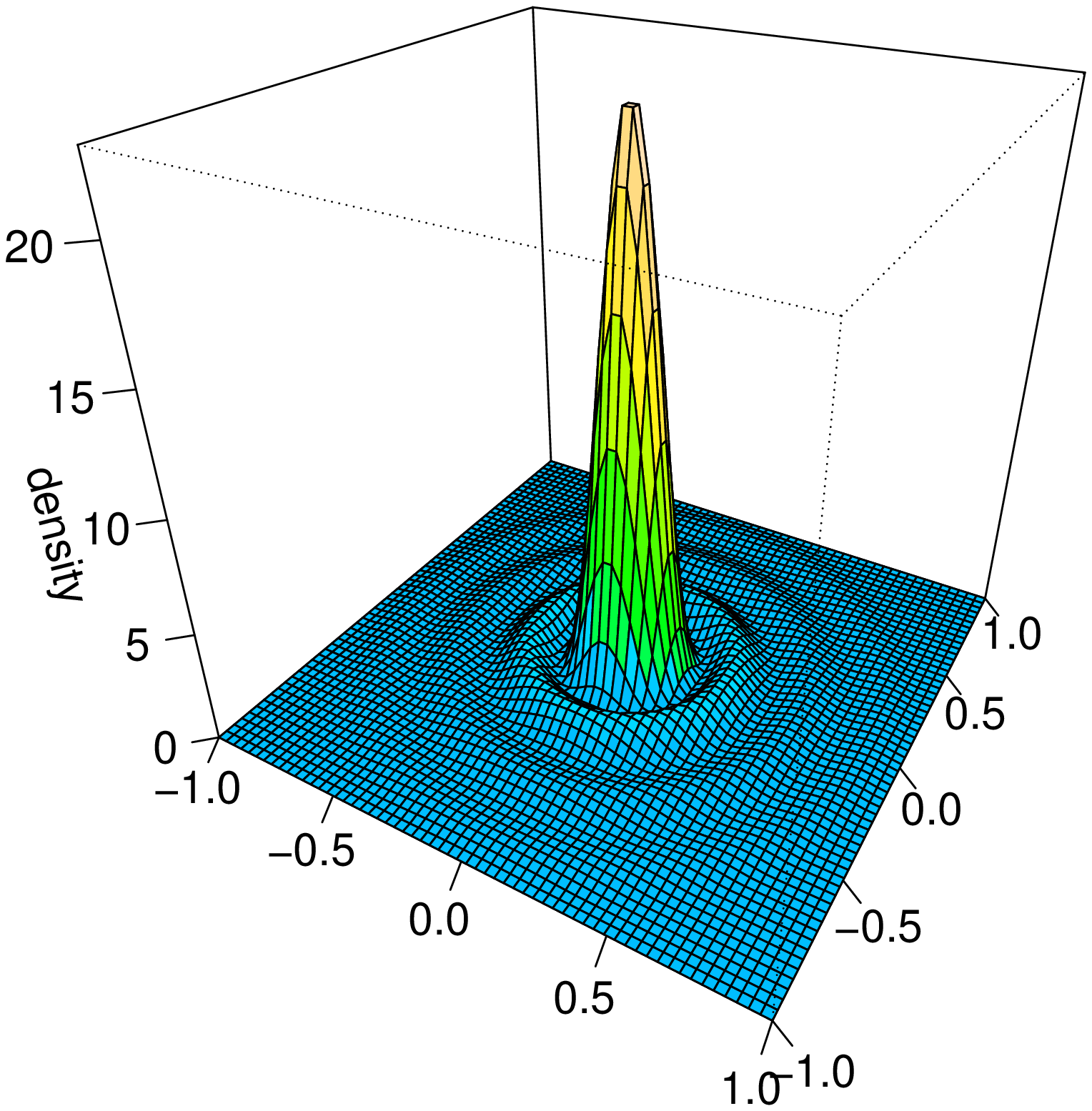}

\caption{Comparative display of  polar $(r,\theta )$  plots for  probability densities  $|\psi^{(500)}_{(k,0)}(r)|^2$  with  $k=1, 2, 3, 4$.
 Note  scale changes along  the vertical (density values) axis, necessary  to fit the location of maxima. }
\end{center}
\end{figure}

We  have  explicitly computed  solution  values of $E^{(2n)}$ and $C^{(2n)}$, together with all expansion coefficients $\alpha^{(2n)}_{2k}, k=1,...,n$
for each polynomial $w_{2n}$ appearing as a building block of an approximate
ground state function   $\psi ^{(2n)}(r) = C^{(2n)}   \sqrt{1-r^2} w_{2n}(r) $.  In Table I, we have displayed  selected coefficients only. The convergence properties  of the data associated with $\psi _{2n}$,
as $n$ approaches $250$,   set a solid ground for further computational analysis of excited states.

The data displayed in Table I clearly demonstrate that the approximate ground state eigenvalue  $E^{(2n)}$   drops down, showing a distinctive stabilization tendency.
Our computed  (approximate) ground state eigenvalue  $E^{(500)} =2.754769$, up to the fifth decimal digit coincides with the  value   obtained independently in \cite{D} (see e.g. Table 4 on page 552).\\

Since   we have in hands   all coefficients $\alpha ^{(2n)}_{2k}, k\geq n$ (not displayed in the present paper),
that determine  consecutive  polynomials $w_{2n}$ from n=1 up to $n=250$, it is possible to make a comparative display of
various  curves   $\psi ^{(2n)}(r)$  with $n\leq 250$.
The data in Fig. 1 show convincingly  how close to the true (limiting) ground state of the  (ultrarelativistic)   infinite spherical well  we actually are,
even for relatively small values of $n$.

In Fig. 1   we   employ   the notation $\psi _{(1,0)} (r)$ for
 the   ground state function and  its  approximations.   We tentatively mention that the   $2l+1$-fold    ($|m| \leq  l$)
  degeneracy of eigenvalues in each  $l>0 $  spectral  series  will enforce the usage of the third  index $m$.
  We  anticipate as well   the splitting of the spherical well spectrum into  the family of  independent
   $l=0,1,2,...$  eigenvalue    series $E_{(k,l)}$, $k=1,2,3,...$.

       We have displayed curves   $\psi^{(2n)}_{(1,0)}(r)$  for   $2n=4,6,10,20,30,50,70,100,150,200,500$.
 The best approximation ($2n=500$) of the ground state function is depicted in black.  We point out that maxima of approximating curves consecutively
  drop down with the growth of $n$.

All coefficients $\alpha _{2k}, k\leq n$, together with  $E^{(2n)}$,  were explicitly computed  after completing a severe truncation
 of the resultant polynomial expressions on  both  sides of the  identity  (26), down to the degree  $2n$.
Accordingly, the right-hand-side of (26)  seems to have not much in common with  $E^{(2n)}\, \psi ^{(2n)}(r)$, where merely $w_{2n}(r)$
factor obeys the truncation restriction, while   $\sqrt{1-r^2}$  remains untouched (is not truncated at all).   That is not so.

    In Fig. 2, for  each value of $2n=4,6,10,20,30,50,70,100,150,200,500$,  we  compare directly the computed polynomial expression
    of the $2n$-th degree $(-\Delta  )^{1/2} \psi ^{(2n)} (r)$  with the complete (openly non-polynomial)
    expression $E^{(2n)} \, \psi ^{(2n)}  (r) = E^{(2n)}  C^{(2n)} \sqrt{1-r^2} w_{2n}(r)$.  That is accomplished by means of the   point-wise detuning  measure
 $|(-\Delta  )^{1/2} \psi ^{(2n)} (r) -  E^{(2n)} \, \psi ^{(2n)}  (r)|$ which quantifies a difference (actually its modulus)  between  the two pertinent expressions.

The  detuning  proves to be  fairly small ($< 0.017$  for $2n=500$), remains sharply concentrated in a  close vicinity
of the $r=1$ boundary (negligible for $r< 0.993$),
and quickly decays to $0$  with the growth of $n$.  In   Fig, 2  we have  displayed
a  convincing graphical proof of   both the  reliability of  our approximation method   and of  the  conspicuous    convergence (in fact that of the detuning)
of   $\psi ^{(2n)}(r)$  towards an ultimate ground state $\psi _{(1,0)}(r)$, as $n\rightarrow \infty $.\\

\subsection{$l=0$ series.}

We point out that the  system of equations  (31) allows to deduce approximate radial forms of higher eigenfunctions and eigenvalues
in the infinite  spherical  well problem.  Clearly, there are many other  solutions available (including
the complex ones, which we discard).
After selecting the lowest eigenvalue $E=E_{(1,0)}$ (associated with the ground state $\psi_ {(1,0)}(r)$),
in the increasingly ordered set of $E$'s, we  select the least one with the property   $E_{(2,0)}< E_{(3,0)}$.

The approximate value of $E^{(2n)}_{(2,0)}$ for the  $2n=500$-th  excited state
\be
\psi^{(2n)}_{(2,0)}(r,\phi,\theta)=C^{(2n)} \sqrt{1-r^2}\sum_{k=0}^{n}\alpha_{2k}r^{2k},\qquad \alpha_0=1
\ee
 reads $E_{(2,0)}^{(500)}=5.8922138$.

 Obviously, in the course of the  computation (according to (30))
 we have recovered not only the approximate
 eigenvalue, but the approximate eigenfunction as well.  We recall that  expansion  coefficients
 $\alpha_{2k}$ with $k\leq 250$ come out as solutions of Eq. (31)  together with the value of  $E^{(2k)}$.
  We do not reproduce the detailed computation data  (available upon request),
    we also  abstain from presenting   the  detuning estimates.
     Results are similar to those obtained for the ground state function.

Selecting other solutions of Eq. (30) associated with  with  consecutive eigenvalues in an increasing sequence
$E_{(1,0)} <E_{(2,0)}<E_{(3,0)}<...$   we are able to deduce the   functional forms $\psi _{(k,0)}^{(2n)}(r)$  of
  higher  (approximate)
 excited eigenfunctions of the purely radial form.   These eigenvalues correspond to other purely radial
  solutions $\psi^{(2n)}_{(k,0)}(r)$, $k=2,3,4$  of Eqs. (30).

In Fig. 3  we  depict the (contour)  shapes of lowest radial
eigenfunctions, while in Fig. 4   polar $(r,\theta )$ diagrams of
 related  probability densities, with the $z$-axis directed perpendicular and
inwards, relative to the  picture frame.

 \subsection{Link between $d=1$ and $d=3$ infinite well  ($l=0$) spectral series.}

\begin{table}[t]
\begin{tabular}{|c||c|c|c|c|c|c|}
  \hline
  $E_{(k,0)}(d=3)=E_{2k}(d=1)$ & 1 & 2 & 3 & 4& 5 & 6   \\
  \hline\hline
  2n=500  & 2.754769 & 5.892214 & 9.033009 & 12.174403&  15.316005 &   18.457716 \\
  Ref. \cite{KGSZ} & 2.754795 & 5.892233 & 9.032984 & 12.174295 & 15.315777 & 18.457329 \\
  Ref. \cite{K} & 2.748894 & 5.890486 & 9.032079 & 12.173672& 15.315554 & *   \\
  \hline
\end{tabular}
\caption{A comparison of approximate eigenvalues  $E_{(k,0)}$ of the infinite spherical Cauchy well with approximate eigenvalues
$E_{2k}$  of the $d=1$ infinite Cauchy well, as reported in  \cite{KGSZ,K}.}
\end{table}

   For a particular  choice   $2n=500$ of the  polynomial $w_{(2n)}$ degree,  we have computed  few   lowest  eigenvalues.
   They read:   $E^{(500)}_{(2,0)}= 5.892214$, $ E^{(500)}_{(3,0)}= 9.033009$ and $E^{(500)}_{(4,0)}= 12.174403$. Interestingly,
 the  obtained   $d=3$ eigenvalues, at least up to five decimal digits,   coincide with   independently derived
   even-labelled eigenvalues  $E_2, E_4, E_6, E_8$  of the $d=1$ infinite Cauchy well  spectral problem, \cite{ZG,KGSZ}.

 In Table II we have collected  comparatively the pertinent  computed  $d=3$   eigenvalues   $E_{(k,0)}$ with $d=1$  results taken from
  \cite{KGSZ,K}.  We point out that  eigenvalues   computed in \cite{K} and  \cite{KGSZ}   originally  were set against the
  (asymptotically valid in $d=1$) formula  $E_k \sim k\pi/2-\pi/8$,  where  $k\in 2\mathbb{N}$.

 We note that the computation fidelity of higher eigenvalues is quite sensitive on the sufficiently large degree $2n$
  of the polynomial approximation involved. Accordingly,  we    need very large $2n$ to infer a reliable
  approximation  of  $E_{(k,0)}$ if $k=100$ for example.

We hereby identify  the  generic feature  of  the   spectrum  of the Cauchy spherical well, that
 a subset  $E_{(k,0)}$   of all  $l=0$   eigenvalues    is identical with that of  {\it even} labeled
  eigenvalues  of  the $d=1$  Cauchy well:  $E_{(k,0)}(d=3)=E_{2k} (d=1)$.   Its asymptotic (large $k$)
   behavior is controlled by the $d=1$  formula,  \cite{K,KKMS,ZG}
\be
   E_{2k} \sim k\pi -\pi/8.
\ee
In the literature we have found  some hints   in this  connection, \cite{D,DKK}, but the above conjecture has never been spelled out.
The pertinent discussion has been focused on
relating  the ground state eigenvalue of the Cauchy well in $d>2$  with a suitable lower dimensional even-labeled eigenvalue.

Indeed, our computed ground state eigenvalue appears to coincide (fapp - for all practical purposes) with the first excited state eigenvalue
of the $d=1$ Cauchy well, c.f. \cite{ZG,KGSZ}.
  This observation finds support in earlier theoretical results \cite{D}. Namely, if  $E_*$
  is the lowest eigenvalue for which there exists    the   odd eigenfunction in the  $d$-dimensional space,   $E_*$ coincides
  with the eigenvalue  of the radial ground state in the  $(d+2)$ - dimensional space. This observation holds true for any $d$
  (cf. Theorem 2 in Ref. \cite{D}).

Let us notice that in accordance with our analysis of the $d=3$ case,
 polynomial  expansion  coefficients of the  ground state function are identical with $d=1$ polynomial coefficients
 of the first excited state. To this end one should simply compare Eqs. (58)-(62)  of Ref. \cite{ZG}  with our present formulas
  (26)-(31). Indeed, we have:
  \be
\begin{array}{ccc}
\psi_2(x)=C\sqrt{1-x^2}\sum\limits_{n=0}^\infty \beta_{2n+1}x^{2n+1},& \qquad & \psi_{(1,0)}(r)=C\sqrt{1-r^2}\sum\limits_{n=0}^\infty \alpha_{2n}r^{2n},\\
d=1, & \qquad & d=3,
\end{array}
\end{equation}
and clearly   $\beta_{2n+1}=\alpha_{2n}$,  for any  $n\in \mathbb{N}$.
Because other radial solutions  (excited states) come out the same way from (30), our  ($d=3$ versus $d=1$)
  conjecture seems to be unquestionably valid. \\

{\bf Remark  1:}
This  peculiar interplay between $d=1$ and $d=3$ spectral data may justified directly by investigating the  properties
of   respective  Cauchy operators.
Let $|z|\neq 0$.  Assuming that we deal with the purely radial $d=3$ eigenfunction we have
\be
\left((-\Delta )^{1/2} \psi(x,y,z)\right)(0,0,|z|)=\left(I_1-I_2\right)=E\psi(|z|),\nonumber
\ee
where
\be
I_1=\frac{1}{\pi^2}\left(\calkan\frac{\psi(|z|)r^2\sin\theta}{(r^2-2r|z|\cos\theta+|z|^2)^2}\right)=\frac{\psi(|z|)}{\pi |z|}\int\limits_0^\infty  r\left(\frac{1}{(r-|z|)^2}-\frac{1}{(r+|z|)^2}\right)dr,\nonumber
\ee
and
\be
I_2=\frac{1}{\pi^2}\left(\calkap\frac{\psi(r)r^2\sin\theta}{(r^2-2r|z|\cos\theta+|z|^2)^2}\right)=\frac{1}{\pi |z|}\int\limits_0^1 r\psi(r)\left(\frac{1}{(r-|z|)^2}-\frac{1}{(r+|z|)^2}\right)dr.\nonumber
\ee
Let us assume that actually  $\psi (r)$  is an even function i.e. $\psi (r)=\psi (-r)$. Then,  presuming    $0<z<1$  we can  make a formal
 change of   the integration variable  $r \rightarrow -r$ in  the second integrand (and related integral). We get:
 \be
I_1=\frac{\psi(z)}{\pi z}\int\limits_{-\infty}^\infty  \frac{r}{(r-z)^2}dr,\nonumber
\ee
and accordingly ($r$ is interpreted to belong to the integration  interval $[-1,1]$)
\be
I_2=\frac{1}{\pi z}\int\limits_{-1}^1  \frac{r\psi(r)}{(r-z)^2}dr.\nonumber
\ee
Presuming  $-1<z<0$  we get similar  results (except that we  make a  change of  the  integration variable
 $r \rightarrow -r$  in the first integrand). That extends the validity of the previous two identities (for $I_1$ and $I_2$)
  to any $z\in(-1,0)\cup(0,1)$.\\
Remembering that all integrations are carried out in the sense of the Cauchy principal values, we have also:
\be
\int\limits_{-\infty}^\infty\frac{r}{(r-z)^2}dr=\int\limits_{-\infty}^\infty\frac{(r-z)+z}{(r-z)^2}dr=
\int\limits_{-\infty}^\infty\frac{z}{(r-z)^2}dr.\nonumber
\ee
Accordingly there holds
\be
\frac{1}{\pi}\left(\int\limits_{-\infty}^\infty\frac{z\psi(z)}{(r-z)^2}-\int\limits_{-1}^1\frac{r\psi(r)}{(r-z)^2}  dr \right)=E z\psi(z),\nonumber
\ee
to be compared with Eq. (4),  originally defining the eigenvalue problem for the $d=1$ Cauchy operator.

An immediate conclusion follows. If an even  (purely radial)  function  $\psi(z)$  is a solution of the $d=3$ eigenvalue problem,
then the  odd  function  $z\psi(z)$  actually is a solution of the $d=1$ eigenvalue problem.  Surely an odd function cannot be
 a ground state, but an excited state of the $d=1$ spectral
 problem. \\
The above reasoning can be inverted and thence by departing from the odd $d=1$ eigenfunction 1D ($z\psi(z)$, (where clearly  $\psi(z)$  is even)
we end up with  $\psi(z)$  as a legitimate eigenfunction of the $d=3$ spectral problem. Moreover,  both functions share the same eigenvalue.
\\

{\bf Remark 2:}  In Ref.  \cite{ZG}   we have  proposed  an  analytic expression  for the  approximate  excited eigenfunction
of the $d=1$ infinite Cauchy well:
\be
\psi_2(x)=-C\sin(\beta x)\sqrt{(1-x^2)\cos(\beta x)}, \ee
where C=1.99693 is a normalization constant in $d=1$, while the parameter $\beta $ has been optimized to take the value  $\beta=1760\pi/4096$.
Our discussion in the previous Remark 1, sets a transparent link between the first excited state in $d=1$ and the ground state in $d=3$.
Let us introduce
 \be
 \psi(r)=C\frac{\sin(\beta
r)\sqrt{(1-r^2)\cos(\beta r)}}{r},\qquad r=\sqrt{x^2+y^2+z^2}
\ee
as an admissible   analytic expression for the "natural" approximation of the ground state in $d=3$.
Here, the normalization constant needs to be evaluated in $d=3$   and equals (with an accuracy up to six decimal digits)
$C=0.796658$.

In Fig. 5 we display  comparatively the $d=3$  analytic curve  (red) (35) against the  approximate ground state
 $\psi _{(1,0)}^{(500)}$ (black).
 An agreement is striking. For more detailed discussion of the $d=1$ case, see e.g. Section II.D in Ref. \cite{ZG}.
\begin{figure}[h]
\begin{center}
\centering
\includegraphics[width=100mm,height=100mm]{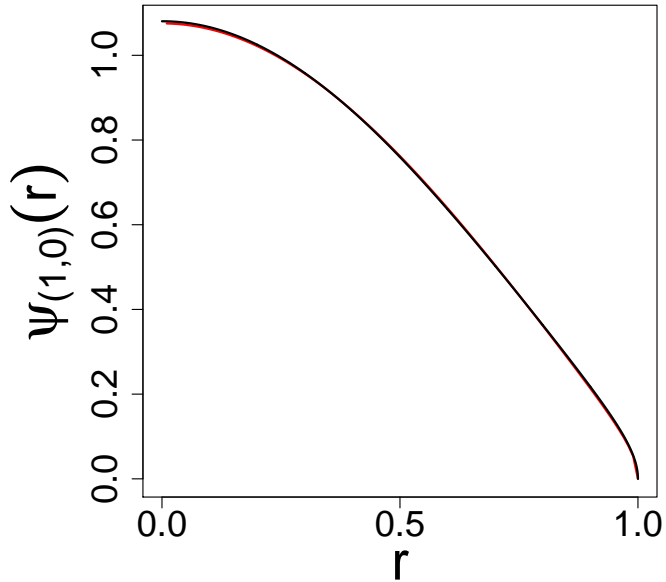}
\caption{Comparative display of two approximations of the ground state in $d=3$, the analytic  form (in red) and
$\psi _{(1,0)}^{(500)}$, see  e.g.  Fig. 1 (in black). The maximum of the analytic curve (red) is residually  shifted down
if compared with  the computed (black) approximate  outcome.}
\end{center}
\end{figure}

 {\bf Remark 3:}
   The previously discussed   $d=1$ versus $d=3$ interplay of Cauchy well spectral problems has its close analog in
     standard quantum mechanics, where the minus  Laplacian  replaces our   fractional operator).
      Let us consider  the $d=1$ symmetric infinite well on the open set  $(-1,1)$.  For all  $x\in(-1,1)$  we have
     the Schr\"{o}dinger  eigenvalue problem  (we set  $m=\hbar=1$) in the form
 \be
-\frac{1}{2}\frac{\partial^2 \psi}{\partial x^2}=E\psi  \nonumber
\ee
Its solutions have the standard form:  $\psi^p_n(x)=C\cos[(\pi/2+n\pi)x]$, $E^p_n=(\pi/2+n\pi)^2/2$, $n\in
\mathbb{N}$ (even, i.e. $n=2k$)  and  $\psi^n_n(x)=C\sin(n\pi x)$,
$E^n_n=(n\pi)^2/2$, $n\in \mathbb{N}_+$ (odd, $n=2k+1$).
Energy eigenvalues, in the notation encompassing  both families of eigenfunctions, read:
$E_n=n^2\pi^2/8$, where  $n\in \mathbb{N}_+$.
The ground state energy  $E_1=\pi^2/8$  corresponds to the even function, while the  first odd one  $C\sin(\pi x)$
refers to the excited state with  $E_2=\pi^2/2$. \\
The radial part of the $d=3$   Schr\"{o}dinger equation for $l=0$ takes the form
\be
-\frac{1}{2}\frac{\partial^2 R}{\partial r^2}-\frac{1}{r}\frac{\partial R}{\partial r}=E R. \nonumber
\ee
Both  $\cos(\gamma_{2n+1} r)/r$  and   $\sin(\gamma_{2n} r)/r$, where $\gamma_n=n\pi/2$,     are solutions of this equation.
However the blow-up property  of  $\cos(\gamma_{2n+1} r)/r$  as $r\to 0$  enforces discarding of that function  from the analysis.
Accordingly, solutions of the radial equation have the form  $\sin(\gamma_{2n} r)/r$. Clearly  $\sin(\pi r)/r$  (up to normalization)
stands for the ground state  with the eigenvalue  $E=\pi^2/2$.\\
In passing, we point out  \cite{G},  that other eigenfunctions for the $d=3$  infinite  spherical well  can be deduced by addressing
 the fully-fledged eigenvalue problem with $l \neq 0$.  Eigensolutions are given in terms of Bessel functions  and energy values read
  $E_{kl}=(u_{l,k}^2/2)$, where  $u_{l,k}$  are the Bessel function zeroes.  For each choice of
   $l= 0,1,2,...$  we  recover  the $l$-th eigenvalue series labelled by $k=1,2,...$. It is worthwhile to mention that for large $k$, the series
   look quite regular, in view of $u_{l,k}  \sim \pi (k + {\frac{l}{2}}) $.
    In particular, one can prove that
  for $l=0$  the Bessel function takes the form  $j_0(r)=\sin(r)/r$  which  clearly has zeroes at  $k\pi$. The corresponding
eigenvalues read $E_{k0}=(k\pi)^2/2$, $k\in \mathbb{N}_+$   and form the $l=0$-series of eigenvalues.   \\

\section{Orbitally nontrivial  eigenstates,  $l\geq 1$  series.}

\subsection{Prerequisites.}
In the previous section we have  relied on some $d=1$ intuitions in computing the ground state data for the infnite spherical well.
We find them useful in the search for  non-radial eigenfunctions,  albeit after some  preliminary discussion on how  the  rotational symmetry
of the problem may help in making computations easier.

Let us specify a point  $P$ as  the endpoint of the vector  $OP= \textbf{p}  =(x_1,x_2,x_3)\in D$.   By  executing a suitable three dimensional rotation,
  we may pass to a new coordinate system whose $Oz$ axis contains $P$.
Clearly, for such $P$,  the identity   $x_1^2+x_2^2= 0$  would imply   $x_1=x_2=0 $. Consistently,
 we may safely assume  $x_1^2+x_2^2\neq 0$ to hold true in general.

Our further considerations critically rely  on  a  proper change  (rotation)  of the
 reference frame in $\mathbb{R}^3$, under the assumption made.    The pertinent   frame of reference  change would
 result in  the rotation of coordinates
\be
y=R_y R_z x = \left(0,0, \sqrt{\iksy}\right)^T,
\ee
where we denote  $y=(y_1,y_2,y_3)^T$, $x=(x_1,x_2,x_3)^T$, ($T$ indicates that the vector is transposed)  while  $R_x$   and  $R_y$  are rotation matrices around  $OY$ and   $OZ$  respectively.
They read:
\be
R_z=\left(
      \begin{array}{ccc}
        \cos\phi & \sin\phi & 0 \\
        -\sin\phi & \cos\phi & 0 \\
        0 & 0 & 1 \\
      \end{array}
    \right),
\ee
\be
R_y=\left(
      \begin{array}{ccc}
        \cos\theta & 0 & -\sin\theta \\
        0 & 1 & 0 \\
        \sin\theta & 0 & \cos\theta \\
      \end{array}
    \right),
\ee
where
\begin{eqnarray}
\cos\phi=\frac{x_1}{\sqrt{x_1^2+x_2^2}},\qquad \sin\phi=\frac{x_2}{\sqrt{x_1^2+x_2^2}},\\
\cos\theta=\frac{x_3}{\sqrt{\iksy}},\qquad \sin\theta=\frac{\sqrt{x_1^2+x_2^2}}{\sqrt{\iksy}}.
\end{eqnarray}
Clearly, the inverse rotation matrix   gives rise to  $x=R^{-1}_z R^{-1}_y y$.
We denote  $S= R^{-1}_z R^{-1}_y$. Its explicit form is
\be
S=\left(
    \begin{array}{ccc}
      \frac{x_1 x_3}{\sqrt{x_1^2+x_2^2}\sqrt{\iksy}} & -\frac{x_2}{\sqrt{x_1^2+x_2^2}} & \frac{x_1}{\sqrt{\iksy}} \\
      \frac{x_2 x_3}{\sqrt{x_1^2+x_2^2}\sqrt{\iksy}} & \frac{x_1}{\sqrt{x_1^2+x_2^2}} & \frac{x_2}{\sqrt{\iksy}} \\
      -\frac{\sqrt{x_1^2+x_2^2}}{\sqrt{\iksy}} & 0 & \frac{x_3}{\sqrt{\iksy}} \\
    \end{array}
  \right).
\ee

In Section II we have given  $d=3$ Cauchy operator   a  somewhat formal but computationally convenient  integral form   $I_1-I_2$
(remember about our precautions concerning the close neighborhood of $(0,0,0)$).  We have:
\be
(-\Delta )^{1/2} \psi(x_1,x_2,x_3)=
\ee
$$
\frac{1}{\pi^2}\left(\int\limits_{\mathbb{R}^3}\frac{\psi(x_1,x_2,x_3)du}{((u_1-x_1)^2+(u_2-x_2)^2+(u_3-x_3)^2)^2}-
\int\limits_D\frac{\psi(u_1,u_2,u_3)du}{((u_1-x_1)^2+(u_2-x_2)^2+(u_3-x_3)^2)^2}\right),
$$
where $du=du_1du_2du_3$.

The integration procedure will be carried out as follows. We execute an  inverse  rotation of $u$ according to the previous recipe, e.g.
$u=Sv$  and employ  $x= Sy$ where $y=\left(0,0,\sqrt{\iksy }\right)^T$.
We note the  $D$  is rotation invariant (spherical well) and the modulus of the Jacobian of the transformation $S$ equals $1$.
 Substituting
\be
u_i=s_{i1}v_1+s_{i2}v_2+s_{i3}v_3,\qquad i=1,2,3,
\ee
where  $s_{ij}$  are matrix elements  of  $S$ we get
\begin{eqnarray}
\sqrt{u_1^2+u_2^2+u_3^2}\xrightarrow{S} \sqrt{v_1^2+v_2^2+v_3^2},\\
(u_1-x_1)^2+(u_2-x_2)^2+(u_3-x_3)^2\xrightarrow{S} v_1^2+v_2^2+ \left(v_3-\sqrt{\iksy}\right)^2 .\\
\end{eqnarray}
That is the starting point for our further analysis.

\subsection{$l=1$  series.}

We denote $\textbf{p}=(x_1,x_2,x_3)$  and    $p=\sqrt{\iksy}$. Our next assumption pertains to the  anticipated functional form of the
excited eigenstate with an orbital (angular)  input, i.e. being non-radial. We make a trial ansatz (note the  a priori insertion of
 orbital labels, to be justified in below):
\begin{eqnarray}
\psi_{(1,1,0)}(\textbf{p})=C\,  x_3\,  f(p),\\
f(p)=\sqrt{1-p^2}\sum\limits_{k=0}^\infty \beta_{2k}p^{2k},  \nonumber
\end{eqnarray}
where  $C$   is the normalization factor.  We assume furthermore that $\beta_0=1$.

We shall demonstrate that  Eq. (47)  indeed  determines  a proper functional  form of the first excited eigenfunction
and entails a computation of  its  fairly accurate approximations. Like in Section II, we shall execute integrations
 of the series expansion in  (47)  term after term.

First we shall prove that:
\be
(-\Delta )^{1/2} \left(u_3\sqrt{1-(u_1^2+u_2^2+u_3^2)}\right)(\textbf{p})=\frac{8}{3}x_3.
\ee
We need to evaluate (while taking care of divergent contributions two integral expressions $I_1(\textbf{p})$  and $I_2(\textbf{p})$, before
eventually subtracting them  and so  eliminating singular contributions.  We have:
\be
I_1(\textbf{p})=\frac{1}{\pi^2}\int\limits_{\mathbb{R}^3}\frac{x_3\sqrt{1-(\iksy)}}{((u_1-x_1)^2+(u_2-x_2)^2+(u_3-x_3)^2)^2}du=
\frac{x_3\sqrt{1-p^2}}{\pi^2}\int\limits_{\mathbb{R}^3}\frac{dv}{(v_1^2+v_2^2+(v_3-p)^2)^2}.
\ee

\begin{figure}[h]
\begin{center}
\centering
\includegraphics[width=50mm,height=50mm]{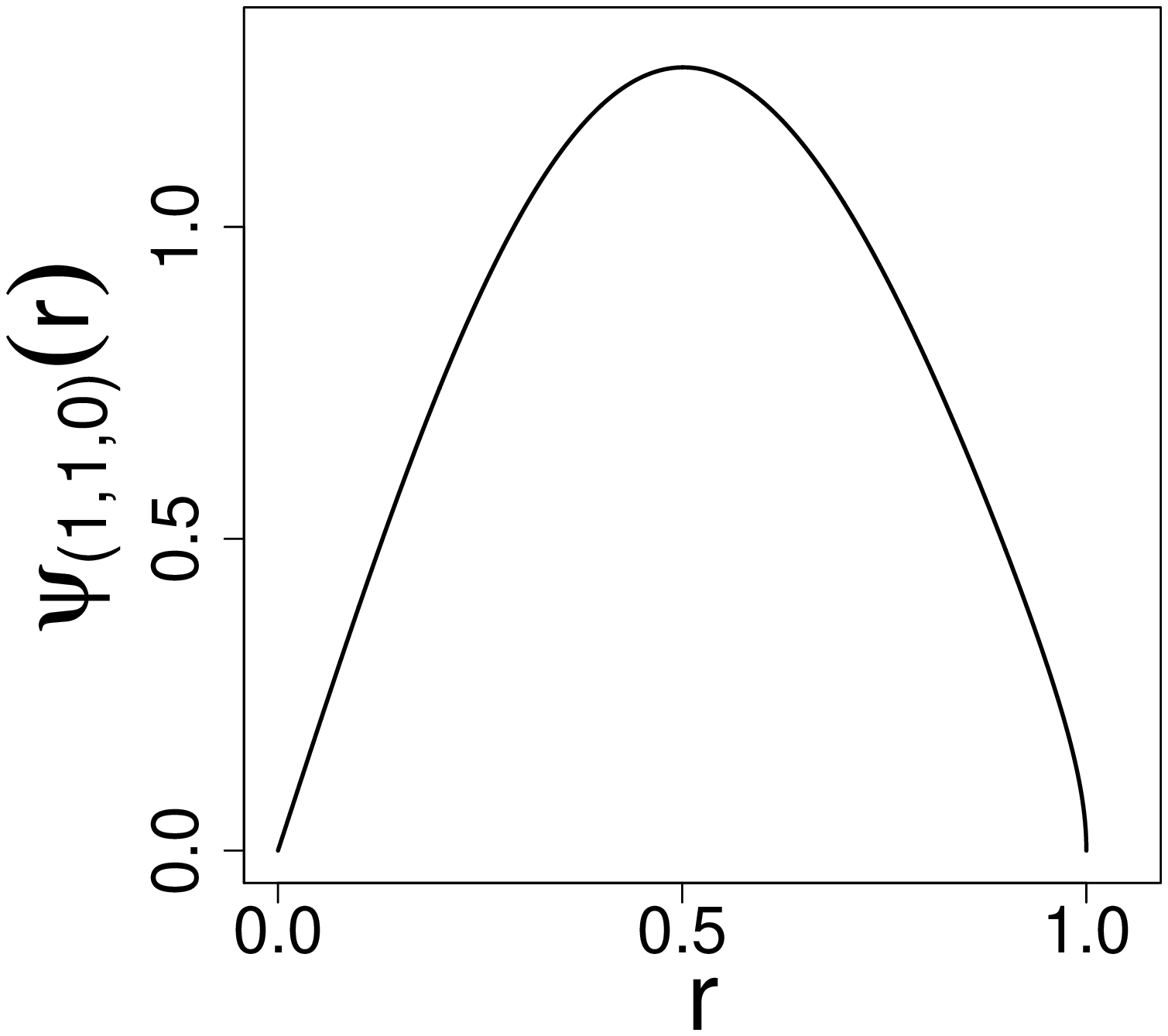}
\includegraphics[width=50mm,height=50mm]{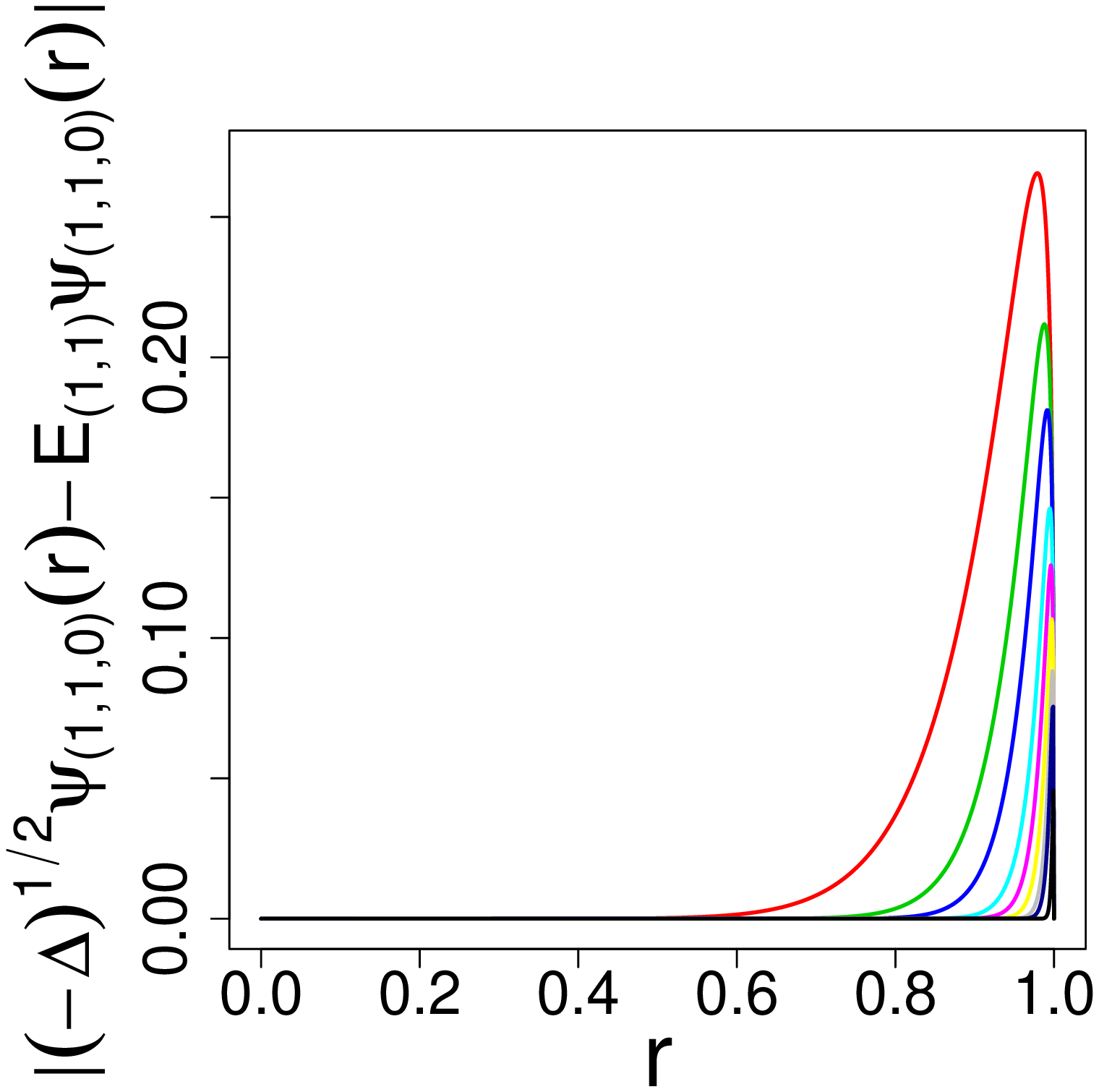}
\includegraphics[width=50mm,height=50mm]{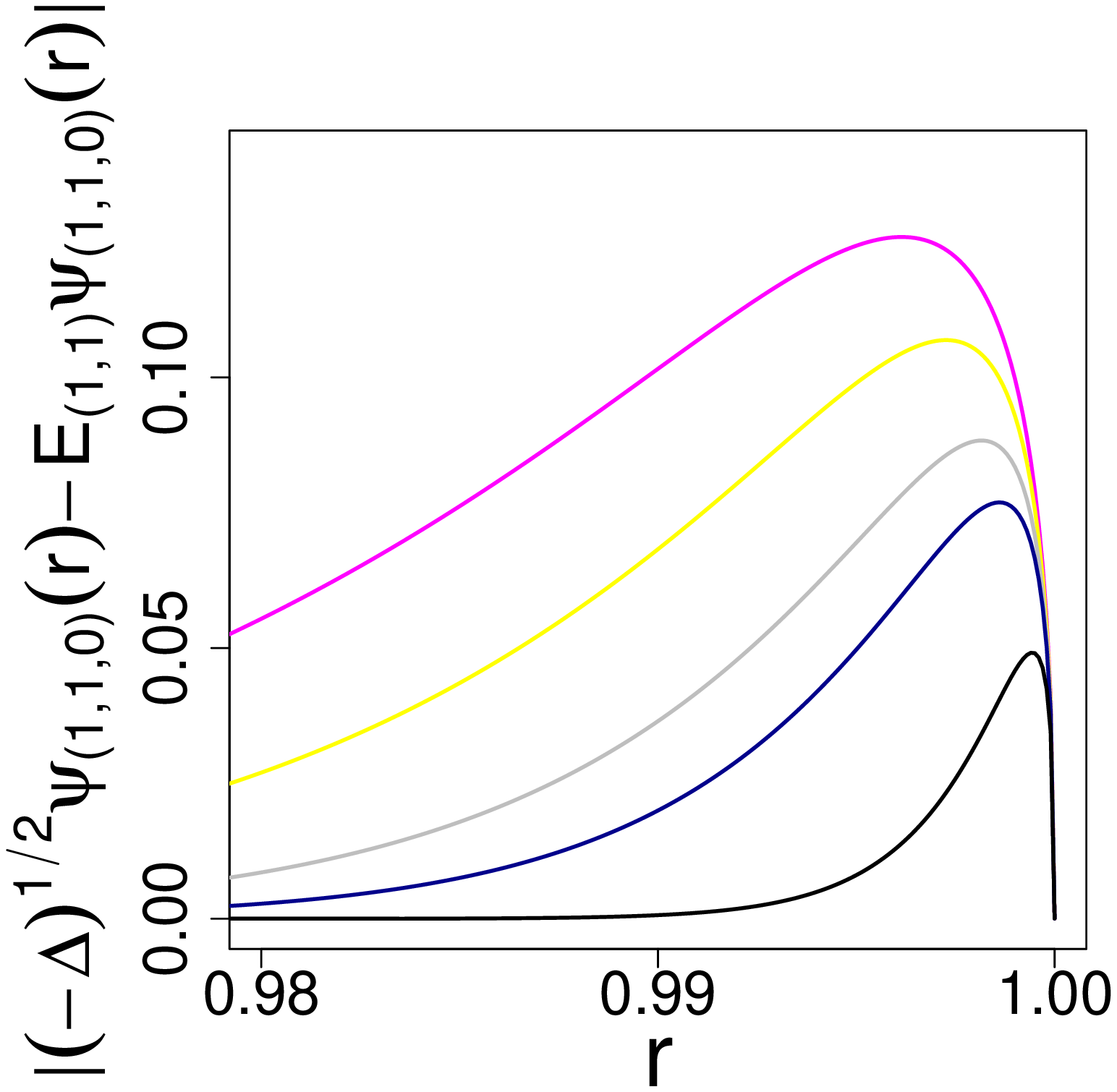}
\caption{Left panel: contour plot of  $\psi^{(500)}_{(1,1,0)}$  for  $\theta=0$.
Middle panel:  Comparative display of detuning  curves for $\psi_{(1,1,0)}$ at  $\theta=0$, polynomial
 approximations of degrees $10,20,30,50,70,100,150,200,500$. The maximum drops down with the growth of $2n$.
 The optimal (2n=500) curve is depicted in black. Right panel:  detuning for polynomial approximations
 of degrees  $70,100,150,200,500$.}
\end{center}
\end{figure}

\begin{figure}[h]
\begin{center}
\centering
\includegraphics[width=50mm,height=50mm]{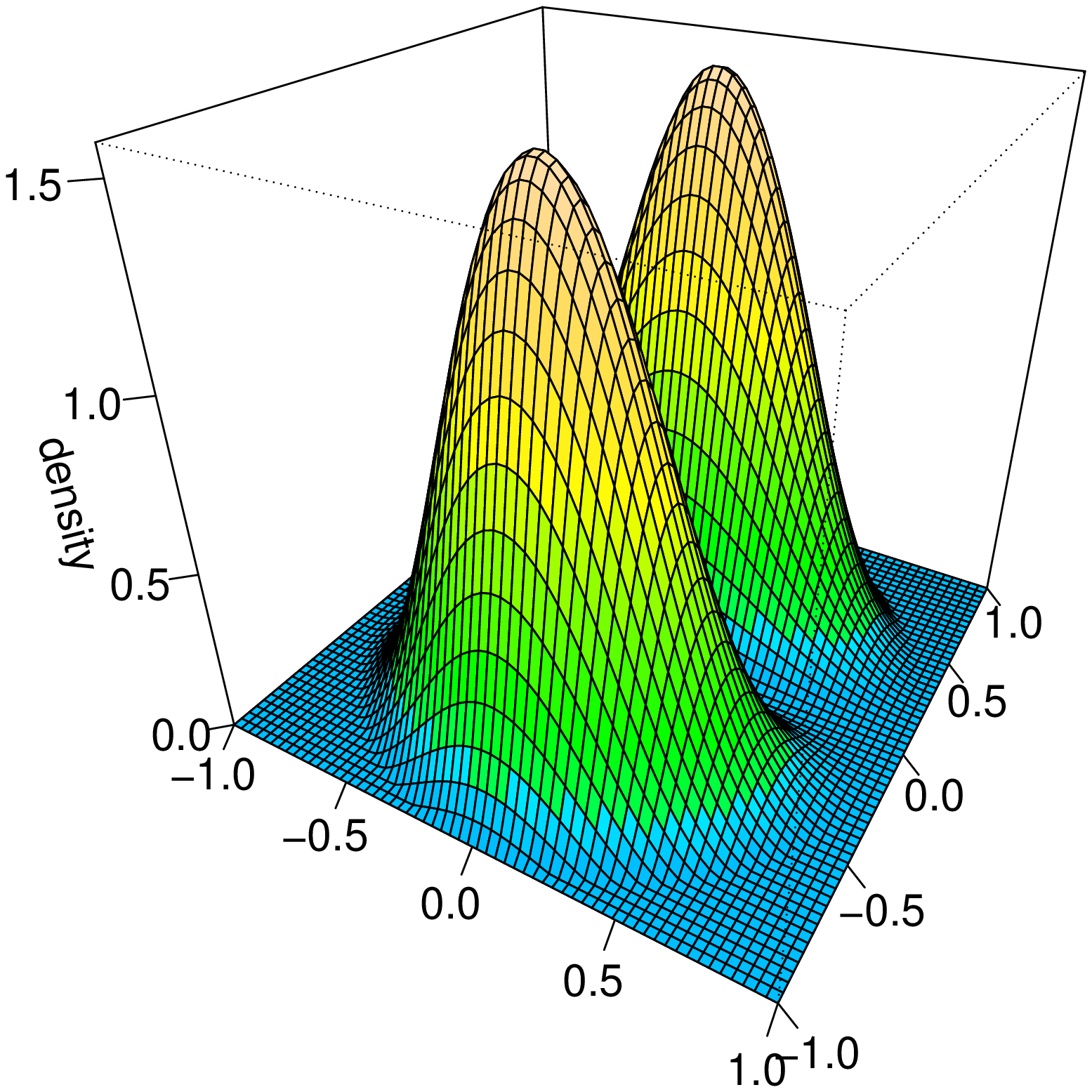}
\includegraphics[width=50mm,height=50mm]{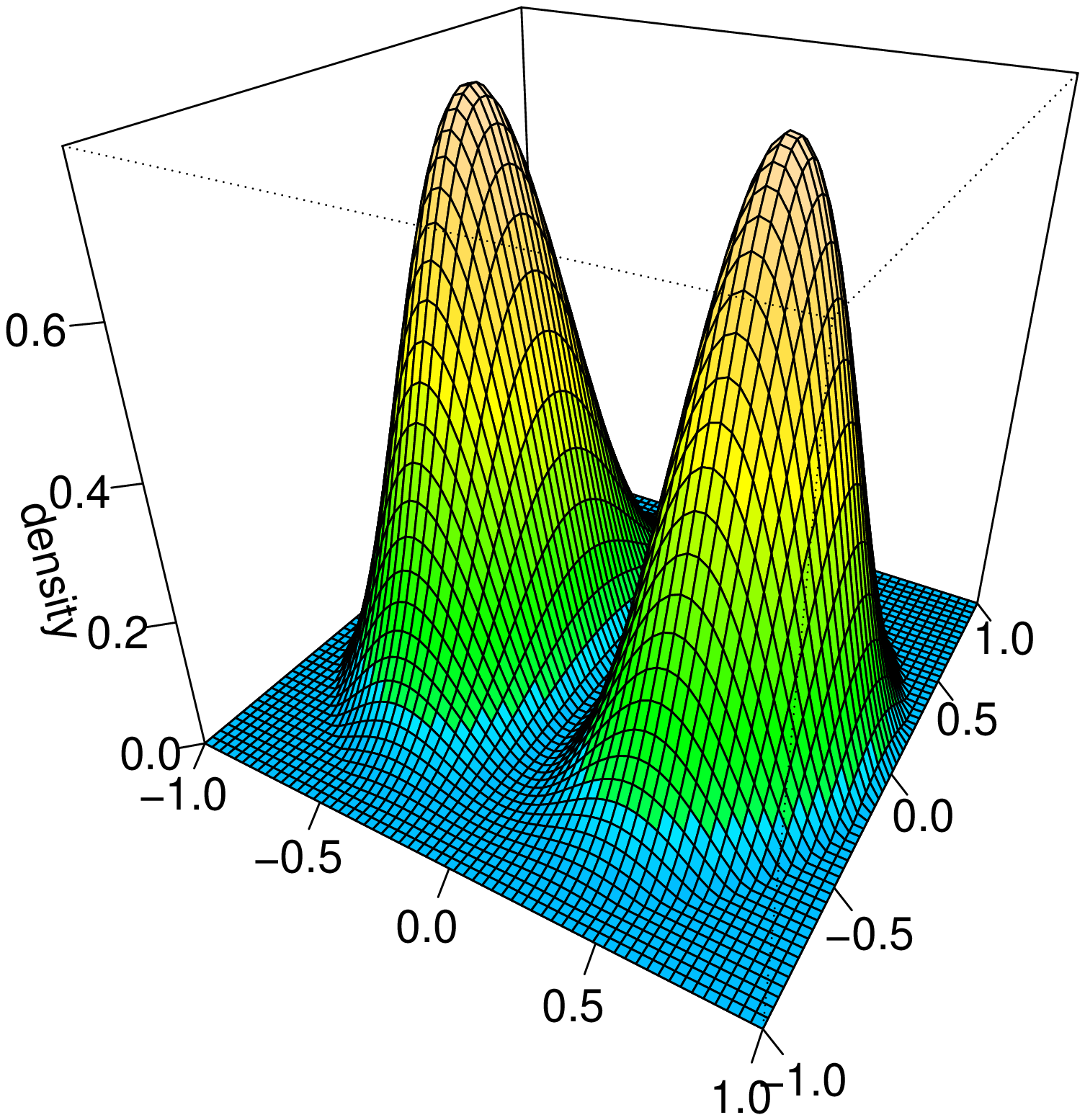}
\caption{Probability densities $|\psi^{(500)}_{(1,1,0)}(r,\phi,\theta)|^2$ (left) and  $|\psi^{(500)}_{(1,1,\pm
1)}(r,\phi,\theta)|^2$ (right) in polar   coordinates.}
\end{center}
\end{figure}
\begin{figure}[h]
\begin{center}
\centering
\includegraphics[width=50mm,height=50mm]{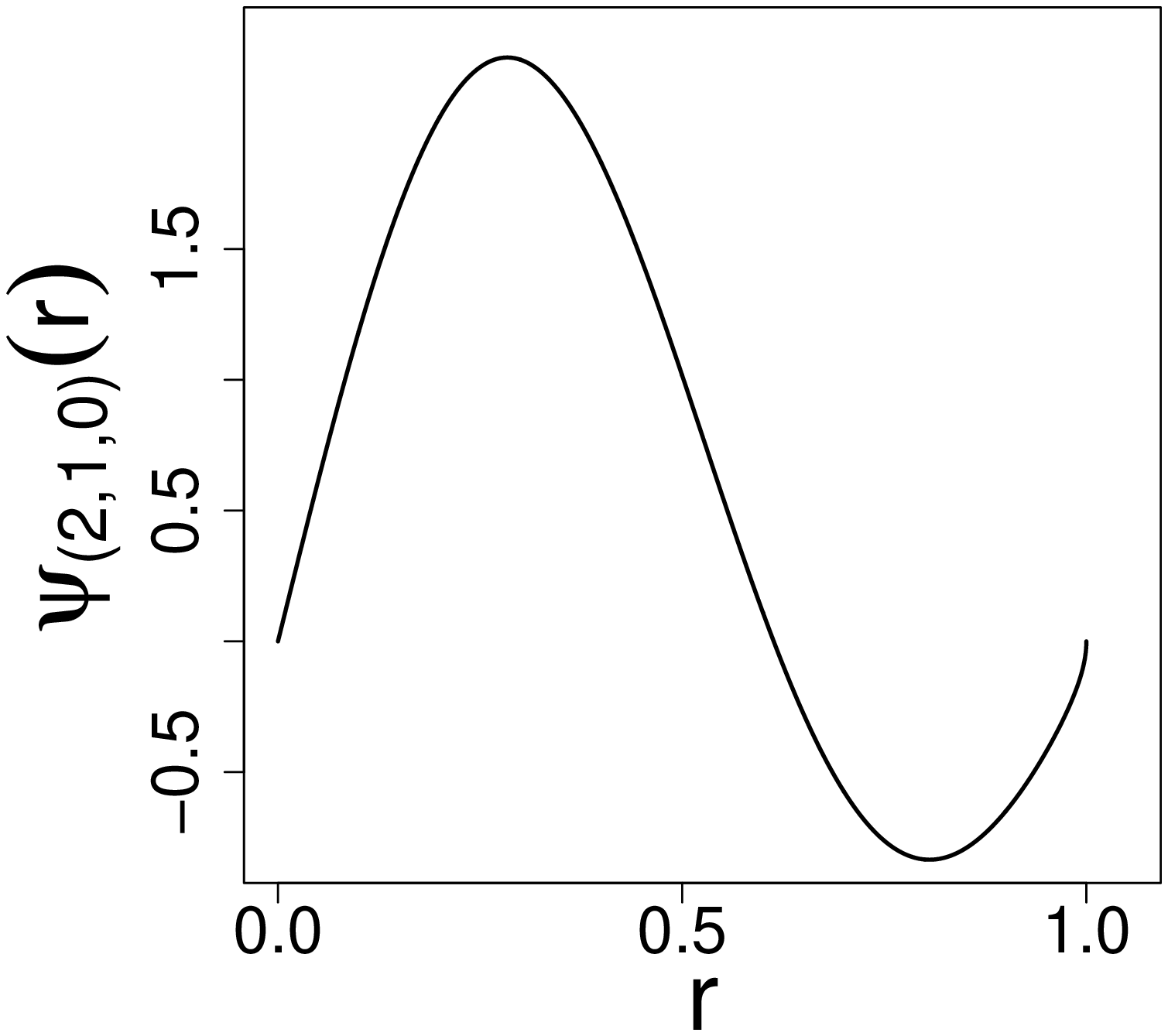}
\includegraphics[width=55mm,height=55mm]{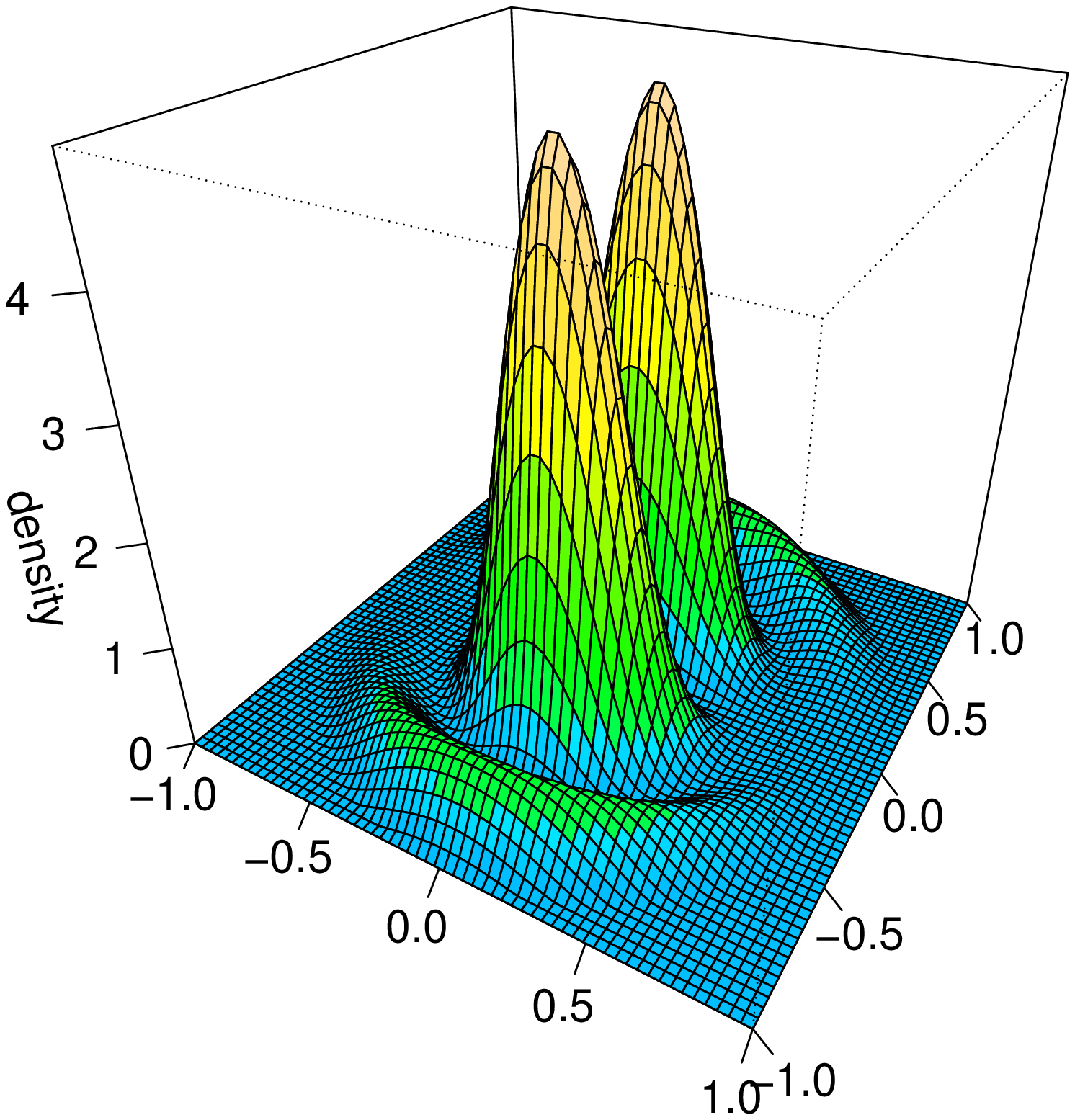}
\includegraphics[width=55mm,height=55mm]{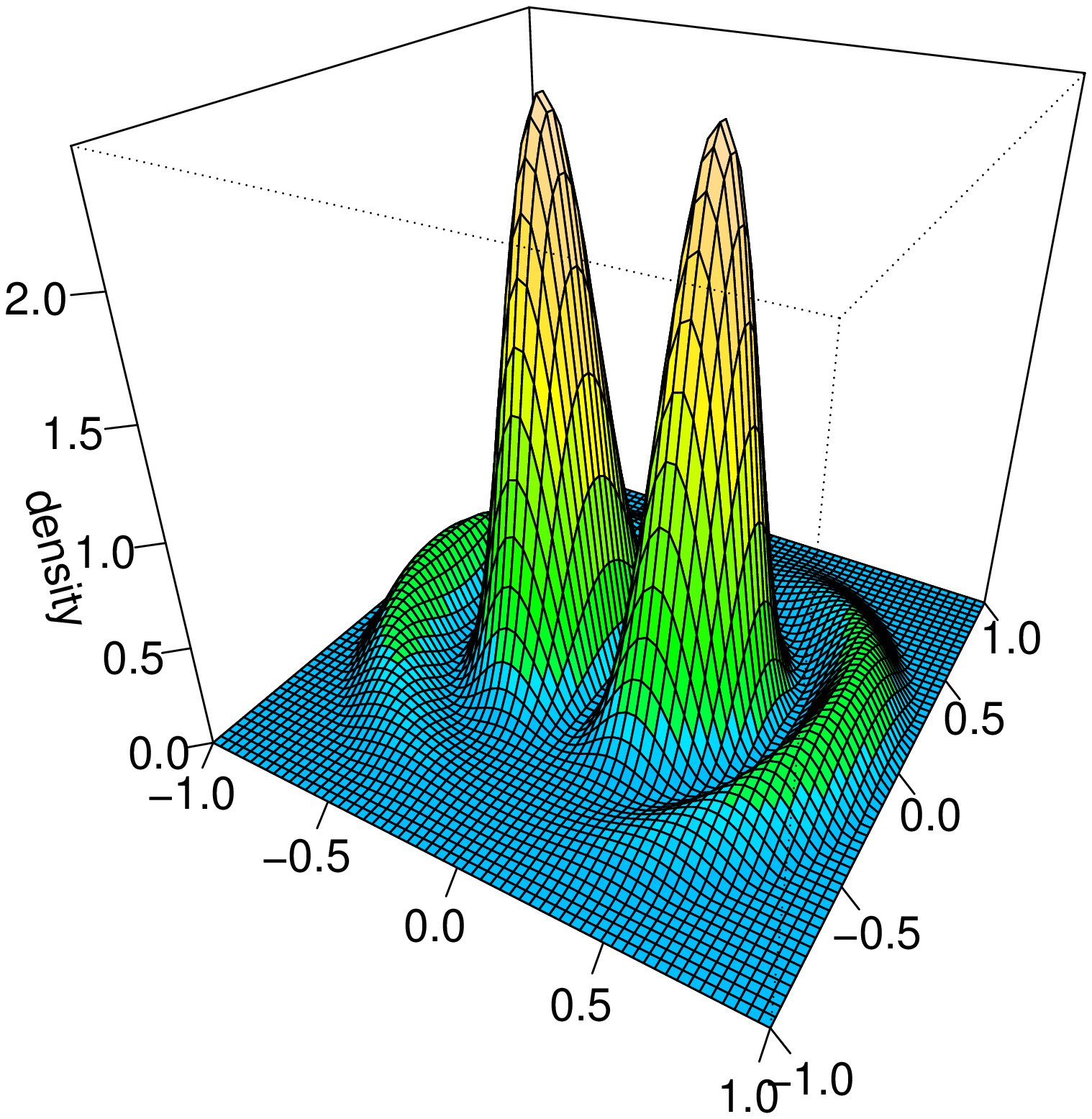}
\caption{Contour plot of  $\psi^{(500)}_{(2,1,0)}$  at   $\theta=0$ (left).  Probability densities
$|\psi^{(500)}_{(2,1,0)}(r,\phi,\theta)|^2$ (midddle) and   $|\psi^{(500)}_{(2,1,\pm 1)}(r,\phi,\theta)|^2$ (right) in polar coordinates.}
\end{center}
\end{figure}
\begin{figure}[h]
\begin{center}
\centering
\includegraphics[width=50mm,height=50mm]{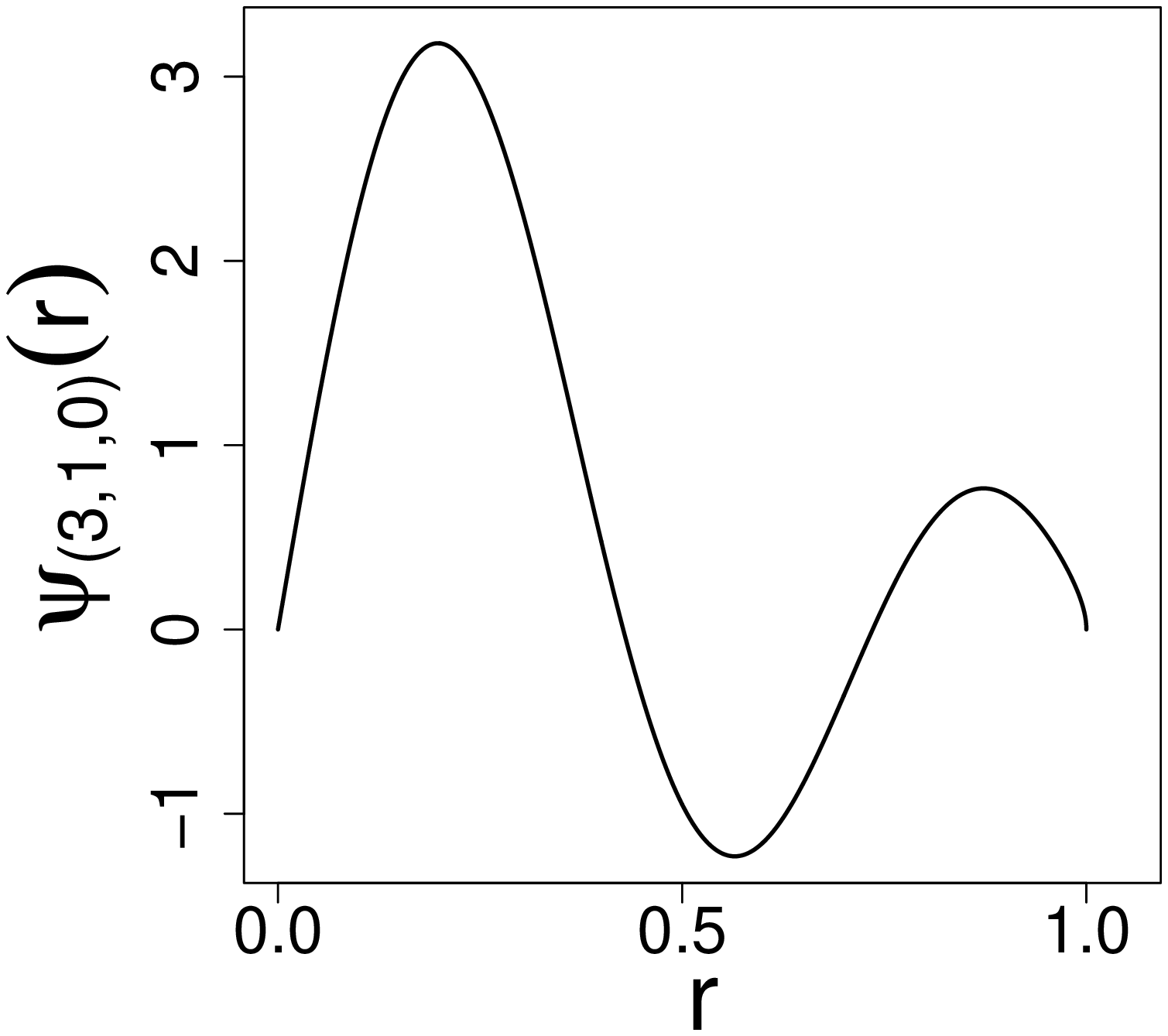}
\includegraphics[width=55mm,height=55mm]{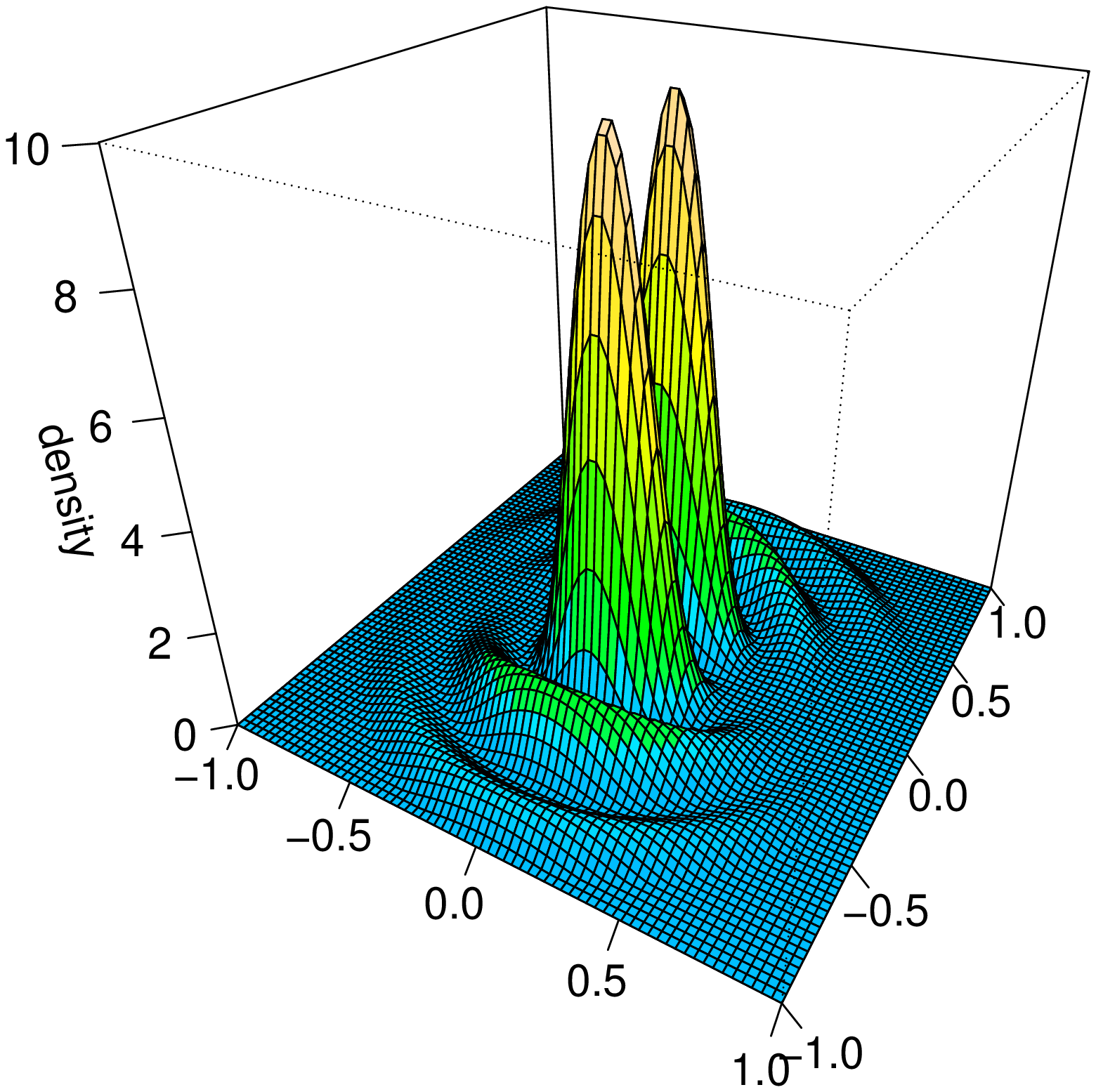}
\includegraphics[width=55mm,height=55mm]{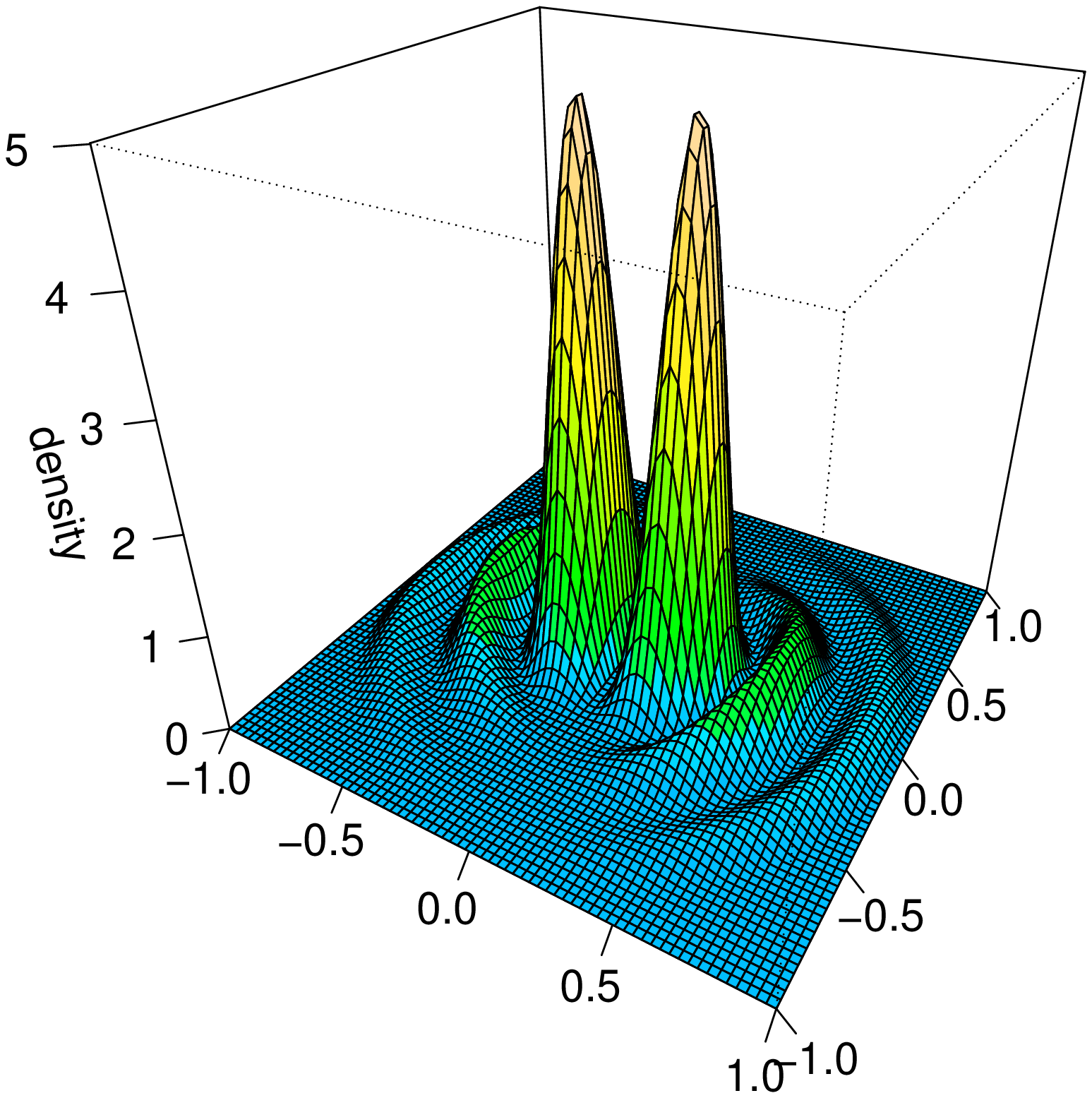}
\caption{Contour plot of $\psi^{(500)}_{(3,1,0)}$  at $\theta=0$.
 Probability densities $|\psi^{(500)}_{(3,1,0)}(r,\phi,\theta)|^2$ (middle) and   $|\psi^{(500)}_{(3,1,\pm 1)}(r,\phi,\theta)|^2$
 in polar coordinates.}
\end{center}
\end{figure}

After passing to spherical coordinates we get
\be
I_1(\textbf{p}) =\frac{x_3\sqrt{1-p^2}}{\pi^2}\calkan \frac{r^2\sin\theta}{(r^2-2r p\cos\theta+p^2)^2}=
\frac{2x_3\sqrt{1-p^2}}{\pi}\lim\limits_{\varepsilon\to 0}\frac{1}{\varepsilon}.
\ee
The integral entry $I_2(\textbf{p})$   reads
\be
I_2(\textbf{p})=\frac{1}{\pi^2}\int\limits_{D}\frac{u_3\sqrt{1-(\uki)}}{((u_1-x_1)^2+(u_2-x_2)^2+(u_3-x_3)^2)^2}du=
\frac{1}{\pi^2}\int\limits_{D}\frac{(s_{31}v_1+s_{32}v_2+s_{33}v_3)\sqrt{1-(\vki)}}{(v_1^2+v_2^2+(v_3-\sqrt{\iksy})^2)^2}dv,
\ee
We denote $v=\sqrt{\vki}$.   The  outcome  of the $\phi $-integration in the range $[0,2\pi ]$ is
\be
\int\limits_{D}\frac{v_1\sqrt{1-v^2}}{(v_1^2+v_2^2+(v_3-p)^2)^2}dv=
\calkap \frac{r^2\sin\theta\cdot r\cos\phi\sin\theta\sqrt{1-r^2}}{(r^2-2rp\cos\theta+p^2)^2}=0,
\ee
and quite analogously
\be
\int\limits_{D}\frac{v_2\sqrt{1-v^2}}{(v_1^2+v_2^2+(v_3-\sqrt{\iksy})^2)^2}dv=0.
\ee
Accordingly:
\be
I_2(\textbf{p}) =\frac{s_{33}}{\pi^2}\int\limits_{D}\frac{v_3\sqrt{1-v^2}}{(v_1^2+v_2^2+(v_3-\sqrt{\iksy})^2)^2}dv
=\frac{s_{33}}{\pi^2}\calkap \frac{r^2\sin\theta\cdot r\cos\theta\sqrt{1-r^2}}{(r^2-2rp\cos\theta+p^2)^2}.
\ee

To evaluate (48) few more steps are necessary. Let us notice that
\be
I_2(\textbf{p}) = {\frac{s_{33}}{2\pi   p^2}} (I_{21} + I_{22}),
\ee
where
\be
I_{21}=\int\limits_0^1 r(r^2+p^2)\sqrt{1-r^2}\left(\frac{1}{(r-p)^2}-\frac{1}{(r+p)^2}\right)dr,
\ee
\be
I_{22}=\int\limits_0^1 r\sqrt{1-r^2}\left(\ln(r-p)^2-\ln(r+p)^2\right)dr.
\ee
One may check the validity of the following indefinite integrals, \cite{GR}:
\begin{eqnarray}
\int \frac{r(r^2+p^2)\sqrt{1-r^2}}{(r\mp p)^2}dr=\frac{\sqrt{1-r^2}(r^3\pm p\pm 2r^2p\mp 18p^3+r(-1+9p^2))}{3(r\mp p)}\pm p(1-6p^2)
\arcsin(r)\nonumber\\
-\frac{2p^2(-2+3p^2)\ln\left|r\mp p\right|}{\sqrt{1-p^2}}+\frac{2p^2(-2+3p^2)\ln(1\mp r p+\sqrt{1-r^2}\sqrt{1-p^2})}{\sqrt{1-p^2}},
\end{eqnarray}
and
\begin{eqnarray}
\int r\sqrt{1-r^2}\ln(r\mp p)^2\,dr=\frac{1}{3}\left[\frac{1}{3}\sqrt{1-r^2}\left(8-2r^2\mp 3rp-6p^2\right)\pm p(-3+2p^2)\arcsin(r)\right.
\nonumber\\
\left.+2(1-p^2)^{3/2}\ln\left|r\mp p\right|-(1-r^2)^{3/2}\ln(r\mp p)^2-2(1-p^2)^{3/2}\ln(1\mp rp+\sqrt{1-r^2}\sqrt{1-p^2})\right].
\end{eqnarray}
Remembering about our precautions concerning singular terms  and   employing $s_{33}p=x_3$, we  ultimately arrive at
\be
I_2=-\frac{8}{3}x_3+\frac{x_3}{\pi}\lim\limits_{\varepsilon\to 0}\left(\frac{\sqrt{1-(p-\varepsilon)^2}}{\varepsilon}+\frac{\sqrt{1-(p+\varepsilon)^2}}{\varepsilon}\right).
\ee
While subtracting formal expressions we   note that all divergent terms cancel each other and consistently  there holds
\be
I_1-I_2=\frac{8}{3}x_3,
\ee
as anticipated.\\

Let  $u=\sqrt{\uki}$.  Analogously, albeit somewhat tediously, we  handle  subsequent  expansion  terms
in our formula  (47),  with an outcome valid for all $n$:
\be
(-\Delta )^{1/2} \left(u_3 u^{2n}\sqrt{1-u^2}\right)(x_1,x_2,x_3)=\left(4x_3\sum\limits_{k=0}^n c_{2k}\frac{(n+1-k)(n+2-k)}{(2n+3-2k)}p^{2n-2k}\right),
\ee
where  $c_{2k}$  are coefficients of the  Taylor expansion  of  $\sqrt{1-z^2}$, c.f. also Section II. \\

Upon inserting the trial function (47) to the eigenvalue equation (5),
we arrive at (c.f. also Section II.B and note that the resultant identity needs to hold true for all
 $x_3= p\cos \theta $):
\be
\sum_{k=0}^\infty\sum_{n=k}^\infty\beta_{2n} b_{k,n} p^{2n-2k}=\sum_{k=0}^\infty\sum_{n=0}^\infty E\beta_{2n} c_{2k}
p^{2k+2n},
\ee
where
\be
b_{k,n}=4\frac{(n+1-k)(n+2-k)}{(2n+3-2k)}c_{2k},
 \ee
and we have
\be
c_{2k}=\frac{(2k)!}{(1-2k)(k!)^2 4^k}.
\ee

 The analytic solution of the system of linear equations (63) is not in the reach. Therefore we reiterate to the very same truncation method (polynomial approximation) we have
 employed in Section II.B, and we follow steps  (i)-(iii)  there in.

The system of $2n+1$ equations for $2n+1$ unknowns   $E$ and  $\beta_{2k}$  (we recall that $\beta_0=1$, by assumption), corresponding   to the
polynomial approximation of the degree $2n$, has the form
\begin{eqnarray}
\sum\limits_{k=i}^n \beta_{2k}b_{k-i,k}=E\sum\limits_{k=0}^i \beta_{2k}c_{2(i-k)},\qquad i=0,1,\ldots,n-1,\nonumber\\
\sum\limits_{m=0}^n \left(\beta_{2m}\sum\limits_{k=0}^m b_{k,m}\right)=0.
\end{eqnarray}
The last identity is  an outcome of the boundary condition (iii) i.e.  $(-\Delta)^{1/2}\psi_{(1,1,0)}(r,\phi,\theta)$ at the boundary  $r=1$.

Wolfram Mathematica routines allow to   handle large systems of linear equations of the form  (66).  The computation allows to recover both the
approximate eigenfunction  $\psi^{(500)}_{(1,1,0)}(r,\phi,\theta)$  and the corresponding eigenvalue  $E^{(500)}_{(1,1)}=4.121332$.
It is worthwhile to mention that in Table 4 of Ref. \cite{D} the same eigenvalue has been independently computed with the
outcome (notation of \cite{D})   $\lambda_*=4.12131$.  The original motivation of Ref. \cite{D} was to  demonstrate that  the pertinent
$\lambda_*$ (first excited eigenvalue in $d=3$) is identical with the ground state eigenvalue of the spherical well problem in dimension $d+2=5$.

Analogous  considerations allow to prove that  three  trial  functions of the  form
\be
x_1f(p),\qquad x_2f(p), \qquad  x_3 f(p),
\ee
give rise to  real   (approximate) eigenfunctions  of the    spherical well problem,  sharing
 the eigenvalue $E_{(1,1)}$  and the  radial factor $f(p)$. \\

{\bf Remark 5:} Computations involve   respectively $x_1$ or $x_2$ instead of  $x_3$  in the integral expression $I_1$. Evaluation of the
integral expression  $I_2$  would look similarly. However, the change of variables (appropriate rotation of the intrinsic coordinate system)
would transform  $u_i$ ($i=1,2$)  to  $s_{i1}v_1+s_{i2}v_2+s_{i3}v_3$.  Integrals containing  $v_1$ i $v_2$  would  vanish identically.
We note that in the ultimate formulas one deals with    $s_{13}p=x_1$ and  $s_{23}p=x_2$.\\

Consequently, in case of $l=1$ to the label $k=1$ there correspond three linearly independent  real  eigenfunctions with a common
 radial part $f(p)$. It is  customary to pass to a complex system of eigenfunctions
\begin{eqnarray}
\psi_{(1,1,0)} (\textbf{p}) = C x_3 f(p)  = C' p  Y_1^0(\theta,\phi)  f(p) ,\\
\psi_{(1,1-1)}(\textbf{p}) =C(x_1-i\,x_2)f(p) = C' p Y_1^{-1}(\theta,\phi)  f(p),\\
\psi_{(1,1,1)}(\textbf{p})=C(x_1+i\,x_2)f(p) = C' p Y_1^1(\theta,\phi) f(p),\\
\end{eqnarray}
where  $\psi_{(1,1,m)}$, $m=-1,0,1$  can be given a familiar form of  linear combinations of spherical  harmonics (and solid harmonics in parallel),
 multiplied by the radial function $f(p)$, \cite{L,Lei}.\\

\subsection{$l=2$  series.}

\begin{figure}[h]
\begin{center}
\centering
\includegraphics[width=50mm,height=50mm]{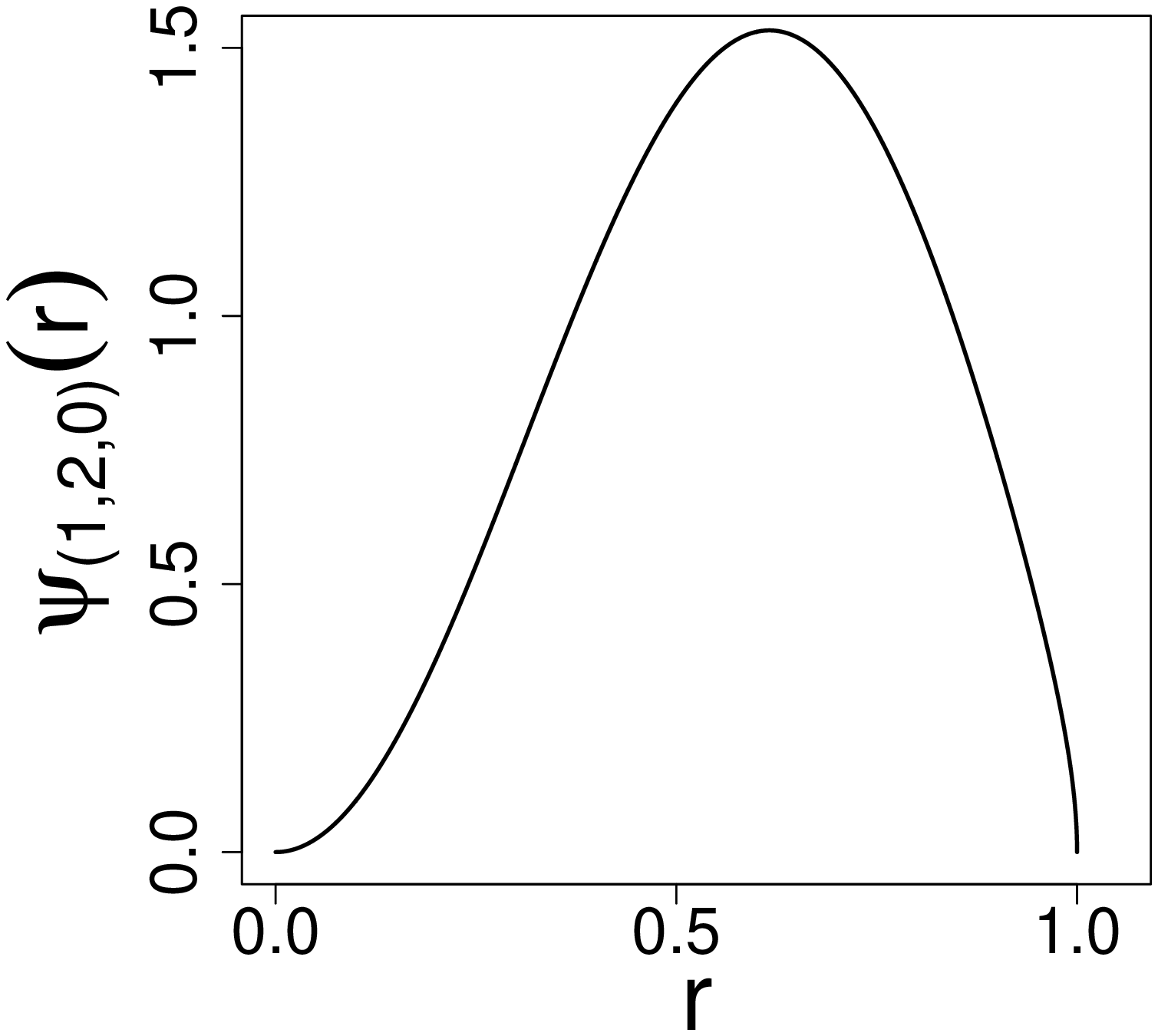}
\includegraphics[width=50mm,height=50mm]{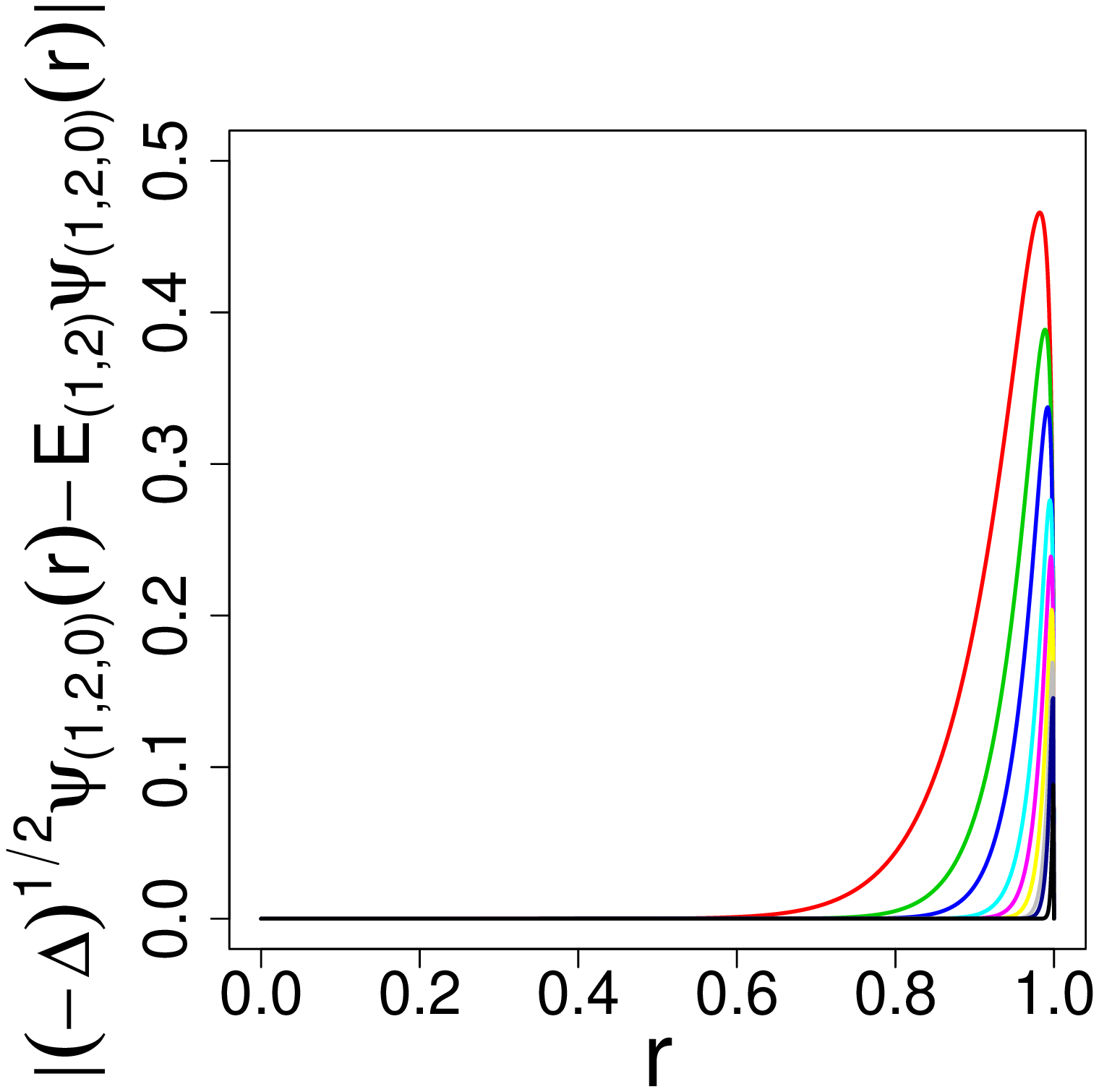}
\includegraphics[width=50mm,height=50mm]{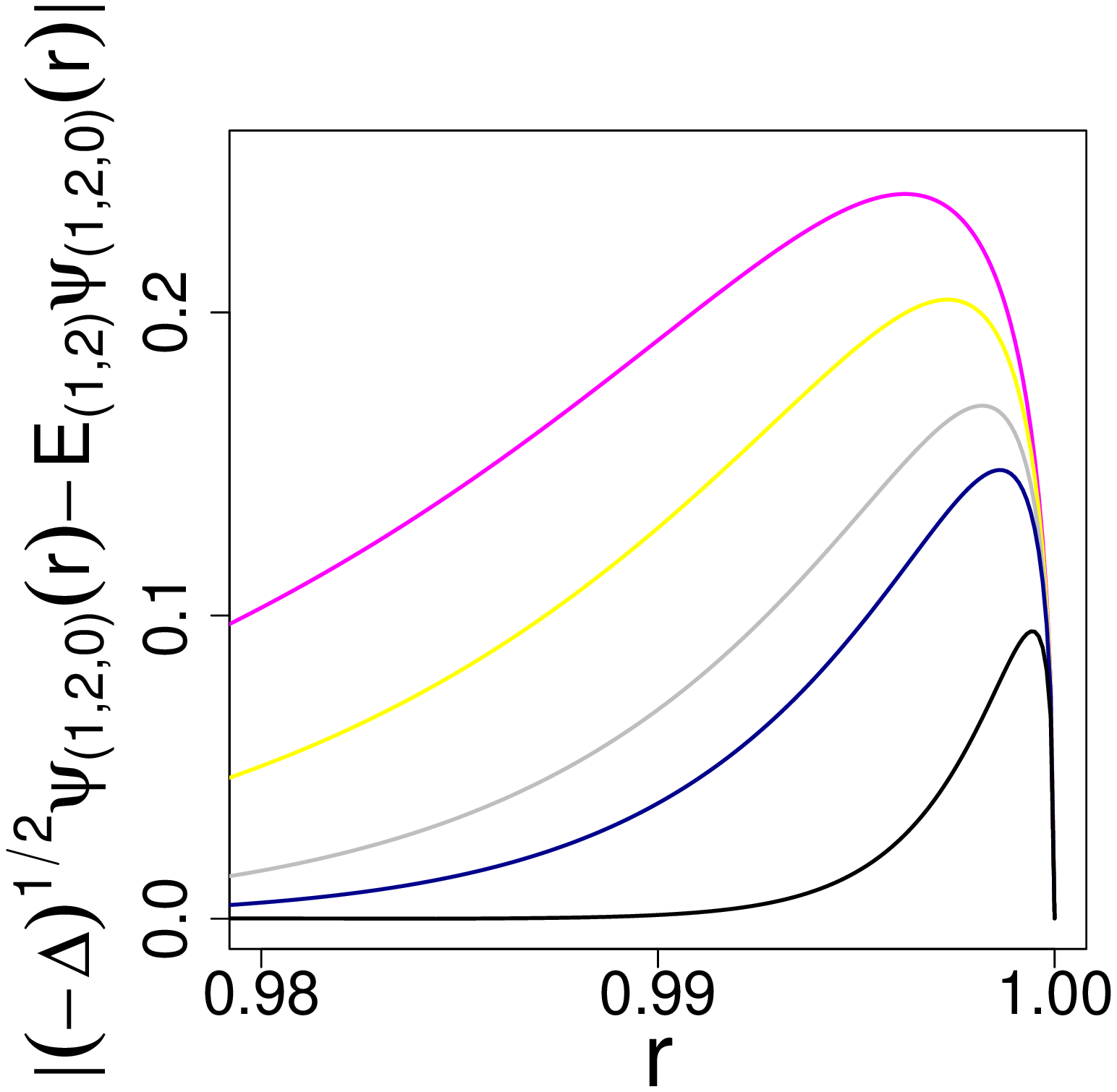}
\caption{Contour plot of  $\psi^{(500)}_{(1,2,0)}$   at $\theta=0$. The detuning for  $\psi^{(500)}_{(1,2,0)}$ at  $\theta=0$ for polynomial approximations
with degrees  $10,20,30,50,70,100,150,200,500$.   The maximum drops down with the growth of $2n$.
The $2n=500$ curve is depicted in black.  Right: detuning for degrees  $70,100,150,200,500$.}
\end{center}
\end{figure}
\begin{figure}[h]
\begin{center}
\centering
\includegraphics[width=50mm,height=50mm]{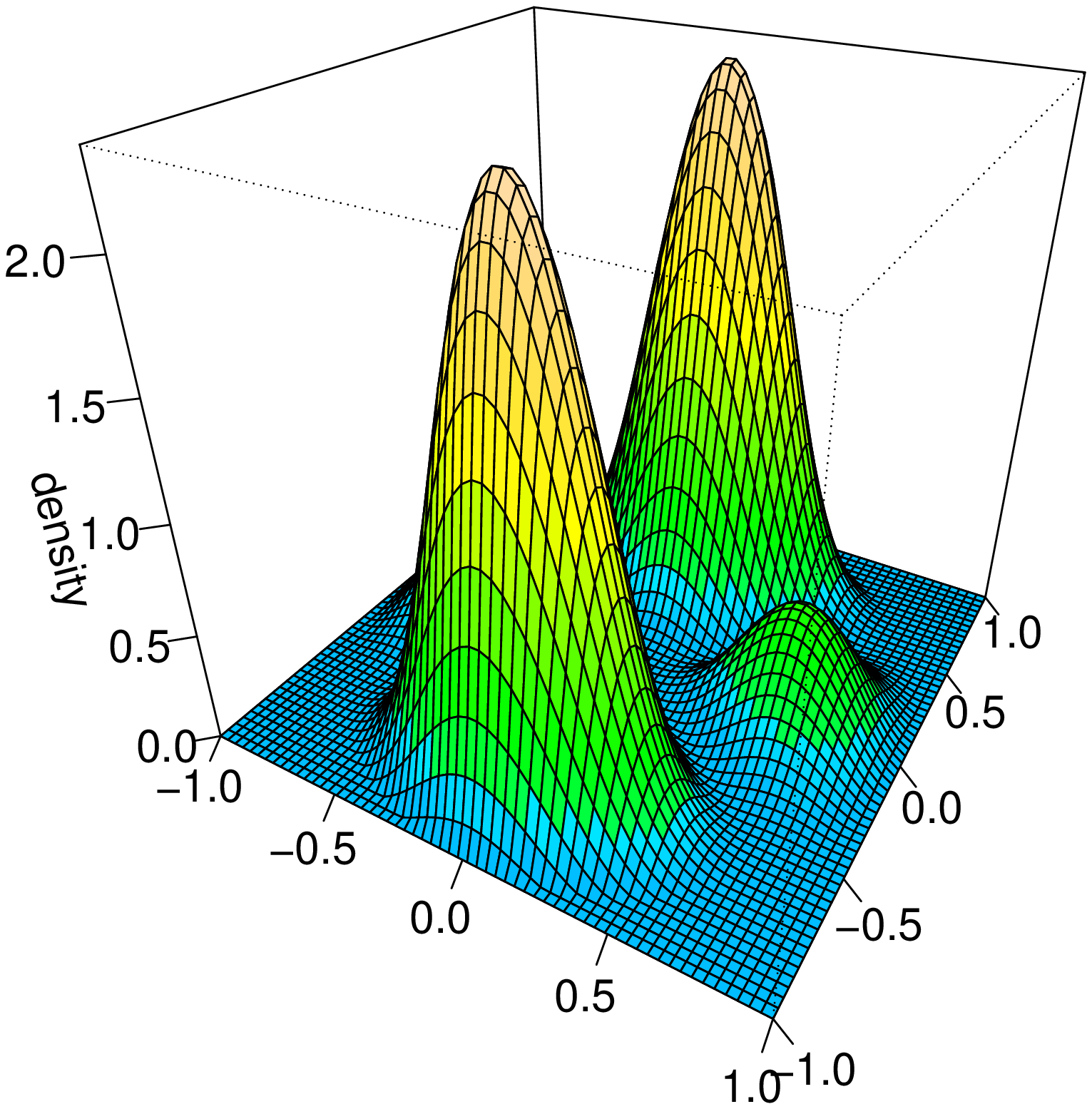}
\includegraphics[width=50mm,height=50mm]{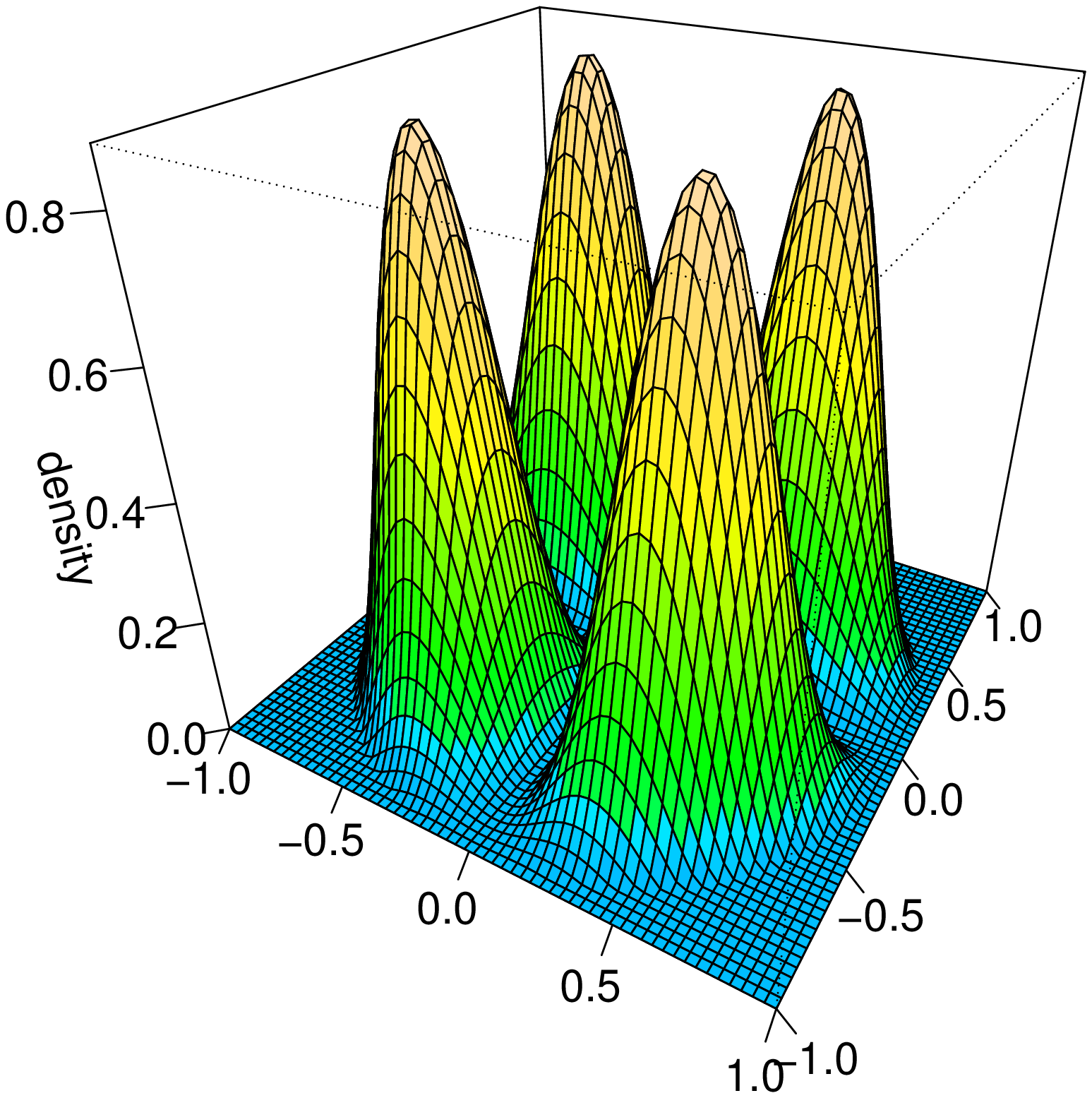}
\includegraphics[width=50mm,height=50mm]{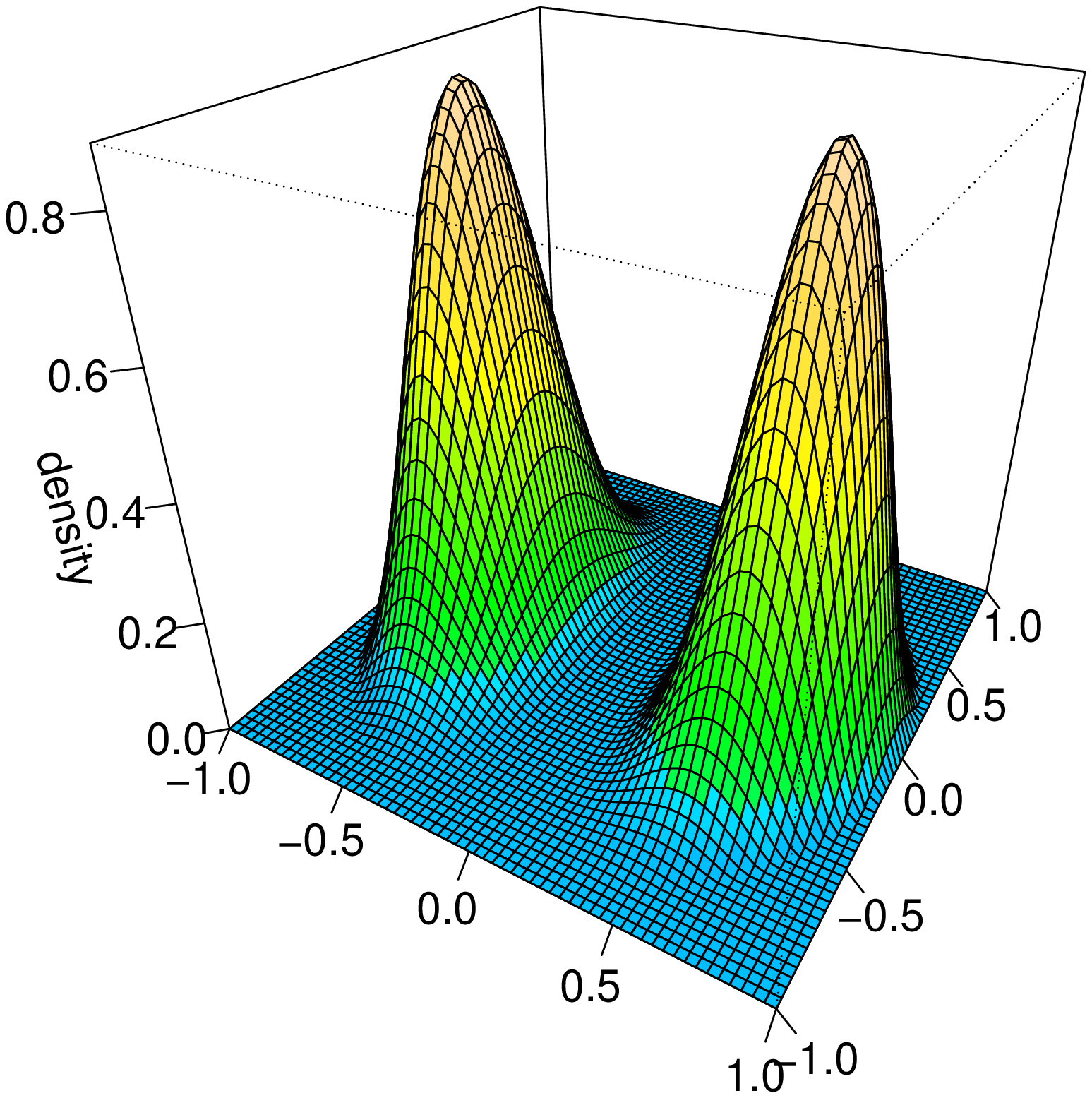}
\caption{Probability densities in polar coordinates (from left to right): $|\psi^{(500)}_{(1,2,0)}(r,\phi,\theta)|^2$,
$|\psi^{(500)}_{(1,2,\pm 1)}(r,\phi,\theta)|^2$  and $|\psi^{(500)}_{(1,2,\pm 2)}(r,\phi,\theta)|^2$ .}
\end{center}
\end{figure}

\begin{figure}[h]
\begin{center}
\centering
\includegraphics[width=50mm,height=50mm]{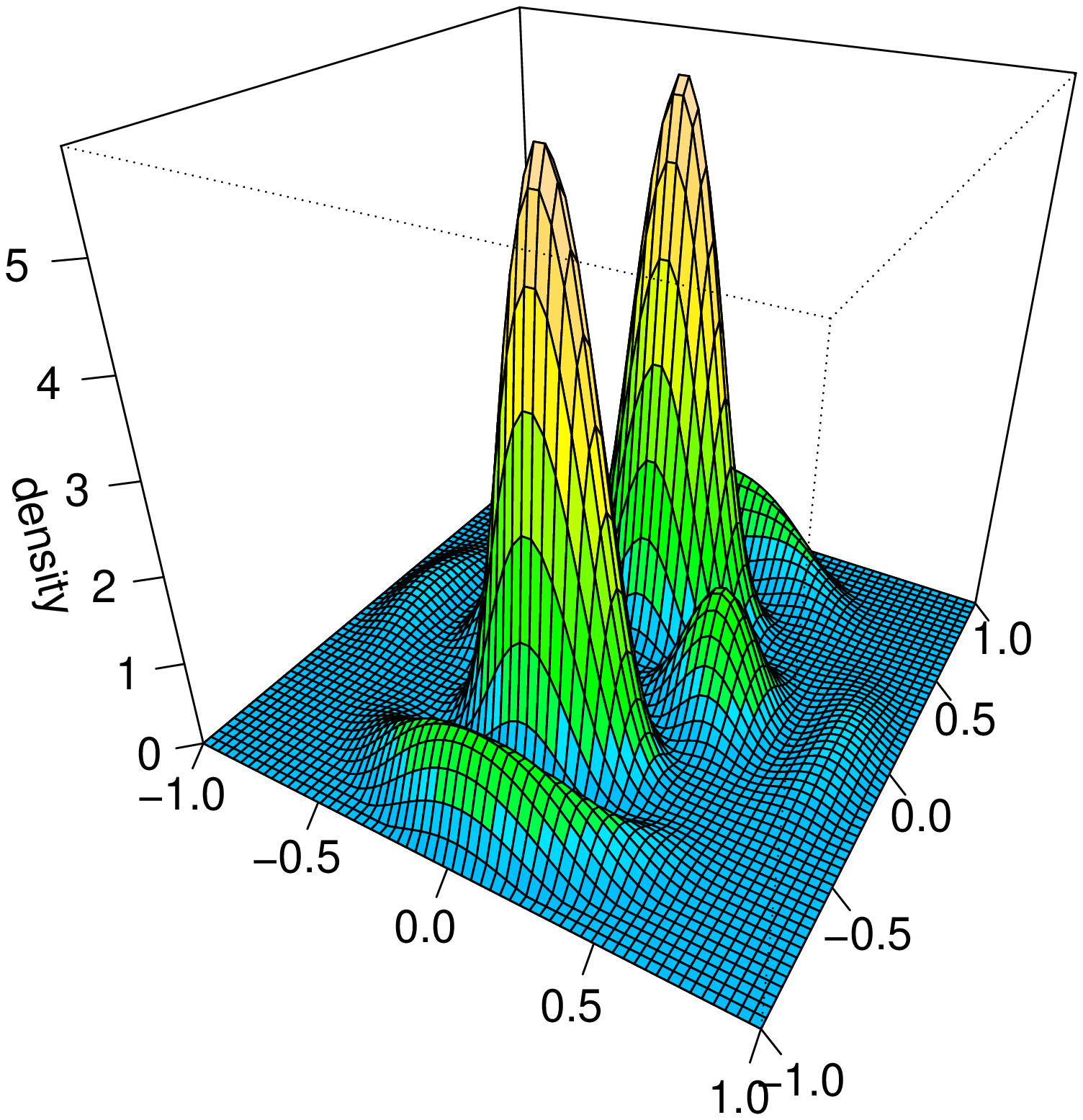}
\includegraphics[width=50mm,height=50mm]{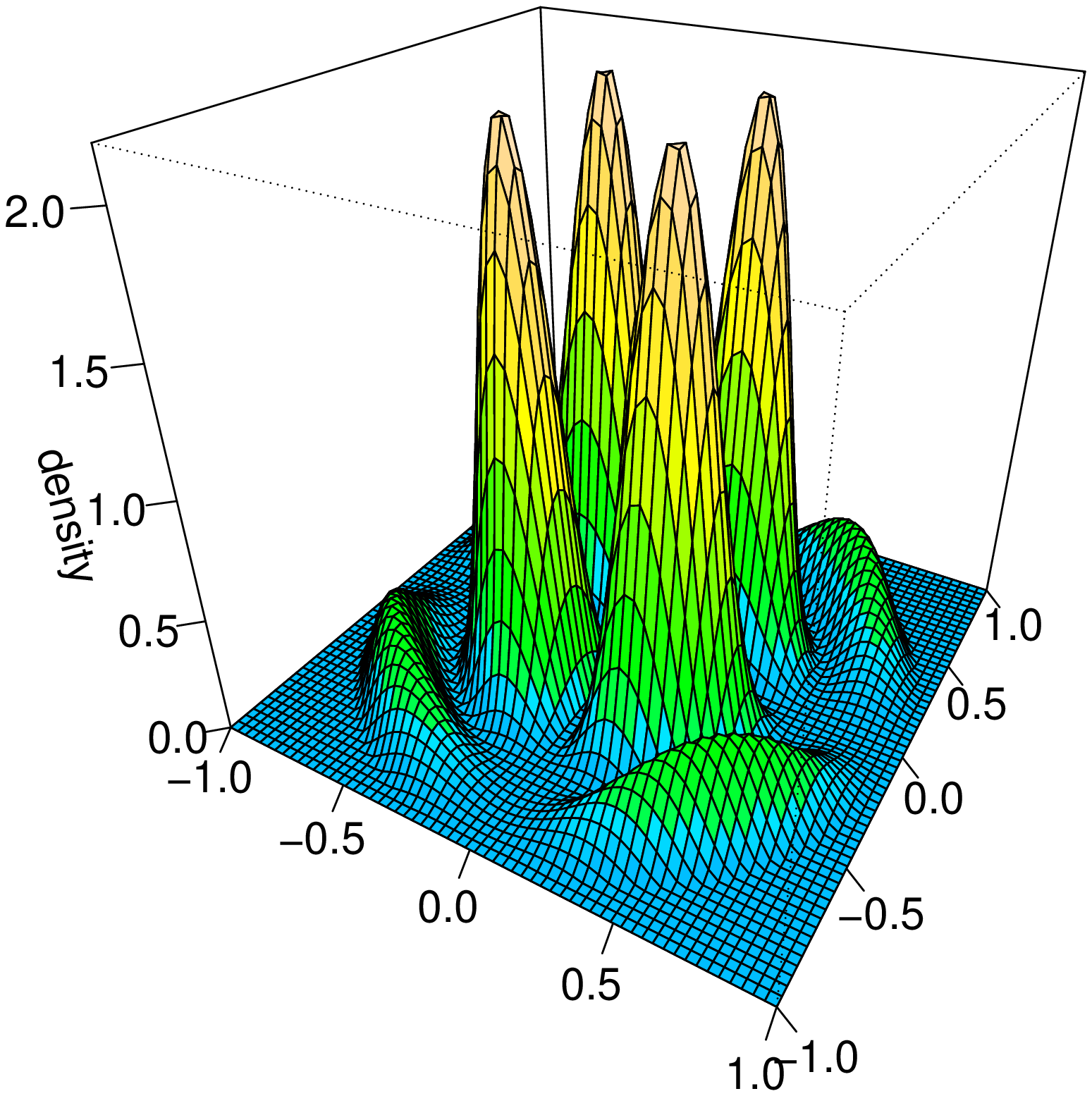}
\includegraphics[width=50mm,height=50mm]{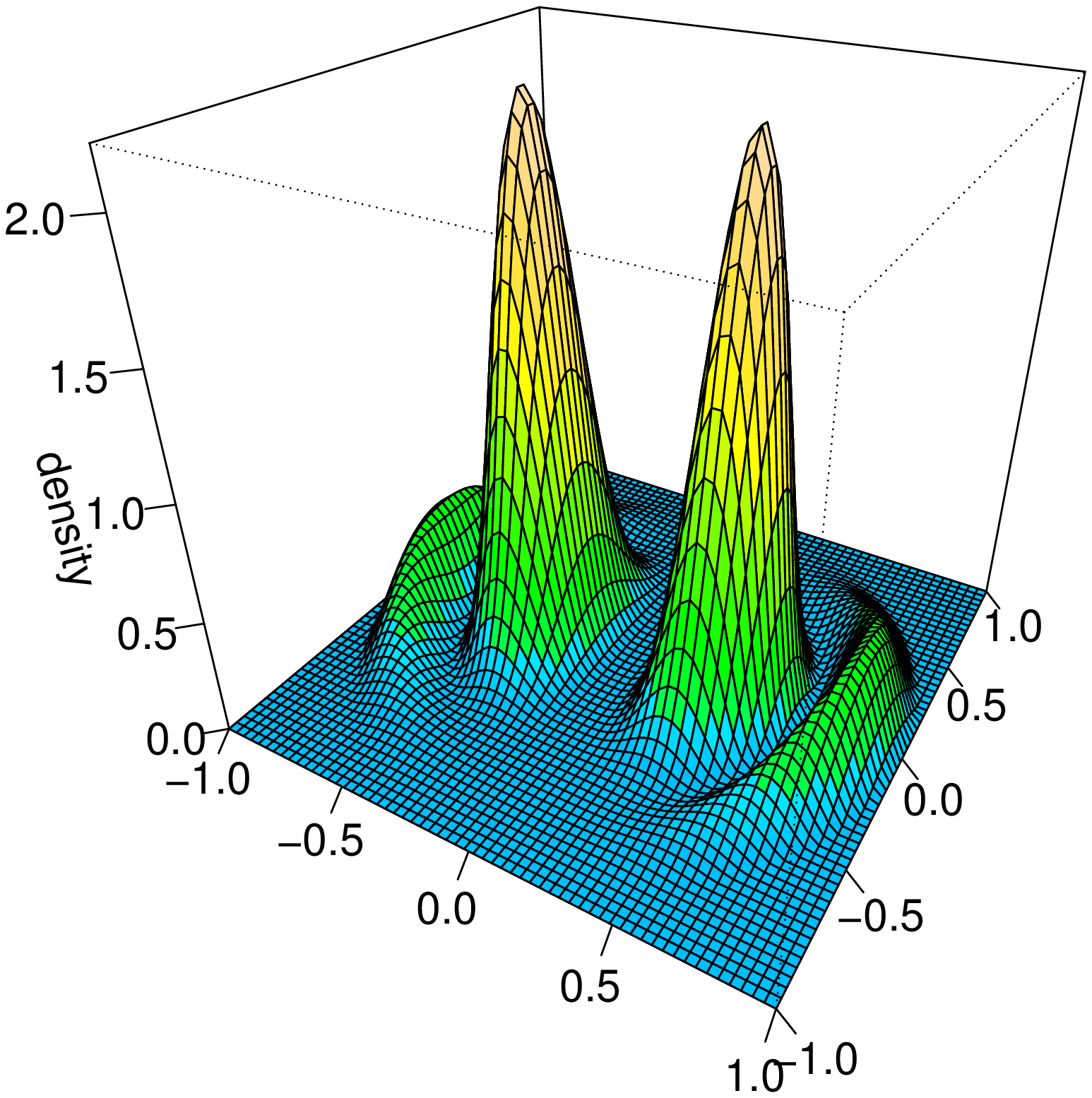}
\caption{ Probability densities (from left to right) $|\psi^{(500)}_{(2,2,0)}(r,\phi,\theta)|^2$,
 $|\psi^{(500)}_{(2,2,\pm 1)}(r,\phi,\theta)|^2$ and  $|\psi^{(500)}_{(2,2,\pm 2)}(r,\phi,\theta)|^2 $.}
\end{center}
\end{figure}

Our trial  choice for the next ($k=1,l=2$, $m=0$ being anticipated)  orbitally nontrivial bound state is
\be
\psi_{(1,2,0)}(x_1,x_2,x_3)=C\frac{3x_3^2-p^2}{2}f(p)
\ee
where
\be
f(p)=\sqrt{1-p^2}\sum\limits_{k=0}^\infty \gamma_{2k}p^{2k},\qquad \gamma_0=1.
\ee
We follow  the same methodology as before and skip detailed calculations.  However, for the reader's convenience
we present an outline of main steps  and reproduce  the
 ultimate outcomes.\\

 The integral expression $I_1$ takes the form
  \be
I_1=\frac{1}{\pi^2}\int\limits_{\mathbb{R}^3}\frac{3x_3^2-p^2}{2}\frac{\sqrt{1-p^2}}{((u_1-x_1)^2+(u_2-x_2)^2+(u_3-x_3)^2)^2}du=
\frac{(3x_3^2-p^2)}{\pi}\sqrt{1-p^2}\lim\limits_{\varepsilon\to 0} \frac{1}{\varepsilon},
\ee
while the  evaluation of $I_2$  is more intricate. We have
\be
I_2=\frac{1}{2\pi^2}\int\limits_{D}\frac{(3u_3^2-u^2)\sqrt{1-u^2}}{((u_1-x_1)^2+(u_2-x_2)^2+(u_3-x_3)^2)^2}du=
\frac{1}{2\pi^2}\int\limits_{D}\frac{[3(s_{31}v_1+s_{32}v_2+s_{33}v_3)^2-v^2]\sqrt{1-v^2}}{(v_1^2+v_2^2+(v_3-p)^2)^2}dv,
\ee
and a number of  integrals  need  to be evaluated explicitly.

 The final  outcome is
\be
(-\Delta )^{1/2} \left(\frac{3u_3^2-u^2}{2}\sqrt{1-u^2}\right)(x_1,x_2,x_3)=\frac{16}{5}\left(\frac{3x_3^2-p^2}{2}\right)=\frac{1}{2}\left[3\left(4-\frac{1}{3}-\frac{1}{5}\right)-4\right]c_0\left(\frac{3x_3^2-p^2}{2}\right),
\ee
Analogously we arrive at
\be
(-\Delta )^{1/2}  \left(\frac{3u_3^2-u^2}{2}u^2\sqrt{1-u^2}\right)(x_1,x_2,x_3)=\left(\frac{3x_3^2-p^2}{2}\right)\frac{1}{2}\left\{\left[3\left(4-\frac{1}{3}-\frac{1}{5}\right)-4\right]c_2
+\left[3\left(6-\frac{1}{5}-\frac{1}{7}\right)-6\right]c_0 p^2\right\},
\ee
\be
\begin{split}
 (-\Delta )^{1/2} \left(\frac{3u_3^2-u^2}{2}u^4\sqrt{1-u^2}\right)(x_1,x_2,x_3)&=\left(\frac{3x_3^2-p^2}{2}\right)\frac{1}{2}\left\{\left[3\left(4-\frac{1}{3}-\frac{1}{5}\right)-4\right]c_4\right.\\
&\left. +\left[3\left(6-\frac{1}{5}-\frac{1}{7}\right)-6\right]c_2 p^2+\left[3\left(8-\frac{1}{7}-\frac{1}{9}\right)-8\right]c_0 p^4\right\},
\end{split}
\end{equation}
where  $c_{2k}$ are Taylor series  expansion coefficients  for  $\sqrt{1-z^2}$.  We note that
\be
\frac{1}{2}\left[2(2n-2k+4)-3\left(\frac{1}{2n-2k+3}+\frac{1}{2n-2k+5}\right)\right]=
\frac{8(n-k+1)(n-k+2)(n-k+3)}{(2n-2k+3)(2n-2k+5)},
\ee
hence, the general formula (referring to the $2n$-th power of $u$)  takes the form
\be
(-\Delta )^{1/2}   \left(\frac{3u_3^2-u^2}{2}u^{2n}\sqrt{1-u^2}\right)(x_1,x_2,x_3)=\left(\frac{3x_3^2-p^2}{2}\right)8\sum\limits_{k=0}^n \frac{(n-k+1)(n-k+2)(n-k+3)}{(2n-2k+3)(2n-2k+5)} c_{2k}p^{2n-2k}.
\ee

Upon inserting  the trial function  $\psi_{(1,2,0)}$  (Eqs. (72) and (73)) to the eigenvalue equation,    we get
\be
\sum_{k=0}^\infty\sum_{n=k}^\infty g_{k,n}  \gamma_{2n} p^{2n-2k}=\sum_{k=0}^\infty\sum_{n=0}^\infty E\gamma_{2n} c_{2n}  p^{2k+2n}.
\ee
where
\be
g_{k,n} = c_{2n}\, \frac{8(n+1-k)(n+2-k)(n+3-k)}{(2n+3-2k)(2n+5-2k)}.
\ee

Like before,  we have no tools to solve  (81) analytically. Therefore we follow the approximation route of Section II.B, specifically
 steps (i) - (iii).  The polynomial approximation of the degree $2n$ results in the linear system of $2n+1$ equations for unknowns
 $E$   and  $\gamma_{2n}$  ($\gamma_0=1$ is presumed):
\begin{eqnarray}
\sum\limits_{k=i}^n \gamma_{2k}g_{k-i,k}=E\sum\limits_{k=0}^i \gamma_{2k}c_{2(i-k)},\qquad i=0,1,\ldots,n-1,\nonumber\\
\sum\limits_{m=0}^n \left(\gamma_{2m}\sum\limits_{k=0}^m g_{k,m}\right)=0.
\end{eqnarray}
The last identity is an outcome of the boundary condition (iii): $(-\Delta)^{1/2}\psi_{(1,2,0)}(r,\phi,\theta)$ at  $r=1$.

The system is amenable to Wolfram Mathematica routines and allows to compute all  $\gamma_{2k}$  and the  (approximate)
 eigenvalue $E_{(1,2)}=5.400079$    associated with $\psi ^{(500)}_{(1,2,0)}$.
 We can  can demonstrate that  five   real functions:
 \be
(x_1^2 -  x_2^2) f(p),  \quad     x_1 x_2 f(p),  \quad  x_1 x_3 f(p),  \quad  x_2 x_3 f(p), \quad (2x_3^2- x_1^2  - x_2^2)f(p),
 \ee
give rise to the system of  linearly independent approximate eigenfunctions, that    share the same  (approximate)  eigenvalue $E_{(1,2)}$.

By employing this  real  eigenfunctions quintet we can readily pass to their complex-valued relatives   which directly involve
 spherical harmonics (and solid harmonics as well).
Indeed, we have
\begin{eqnarray}
\psi_{(1,2,0)}(\textbf{p}) = C{\frac{3x_3^2-p^2}{2}} f(p)  = C' p^2 Y^0_2 f(p) ,\\
\psi_{(1,2,\pm 1)}(\textbf{p})=C  x_3(x_1\pm i x_2) f(p)= C' p^2 Y^{\pm 1}_2  f(p) ,\\
\psi_{(1,2,\pm 2)}(\textbf{p})=  C (x_1 \pm i x_2)^2 f(p) = C' p^2 Y^{\pm 2}_2 f(p) .\\
\end{eqnarray}

\begin{table}[h]
\begin{center}
\begin{tabular}{|l||c|c|c|c|c|c|}
\hline
\backslashbox{$l$}{$k$} & 1 & 2 & 3 & 4 & 5& 6  \\
\hline \hline
0 & 2.754769 & 5.892214 & 9.033009 & 12.174403& 15.316005 & 18.457716  \\
\hline
1 & 4.121332 & 7.342181 & 10.517287 &  13.677648 & 16.831345 & 19.981459  \\
\hline
2 & 5.400079 & 8.718436&   11.940889 & 15.129721  & 18.302539 &  21.466420    \\
\hline
3 & 6.630371 &  10.045716 & 13.320189 & 16.542195 & 19.738192 &  22.919240  \\
\hline
\end{tabular}
\end{center}
\caption{Spectral  $l$-series  for $l=0,1,2,3$ and  $1\leq k\leq 6$ . Approximate eigenvalues $E_{(k,l)}$   computed for polynomial
truncations of the  degree  $2n=500$.}
\end{table}

\begin{figure}[h]
\begin{center}
\centering
\includegraphics[width=50mm,height=50mm]{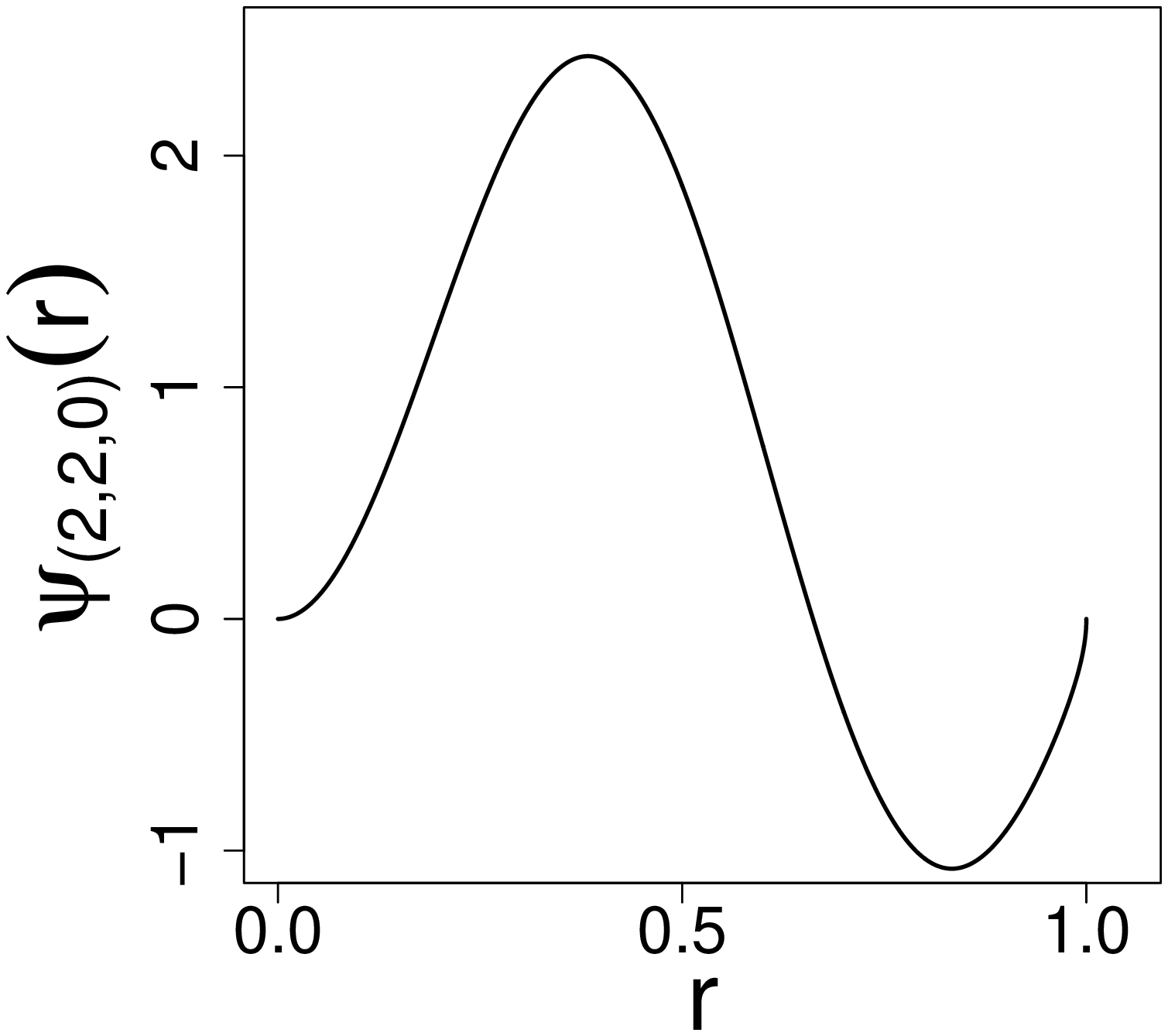}
\includegraphics[width=50mm,height=50mm]{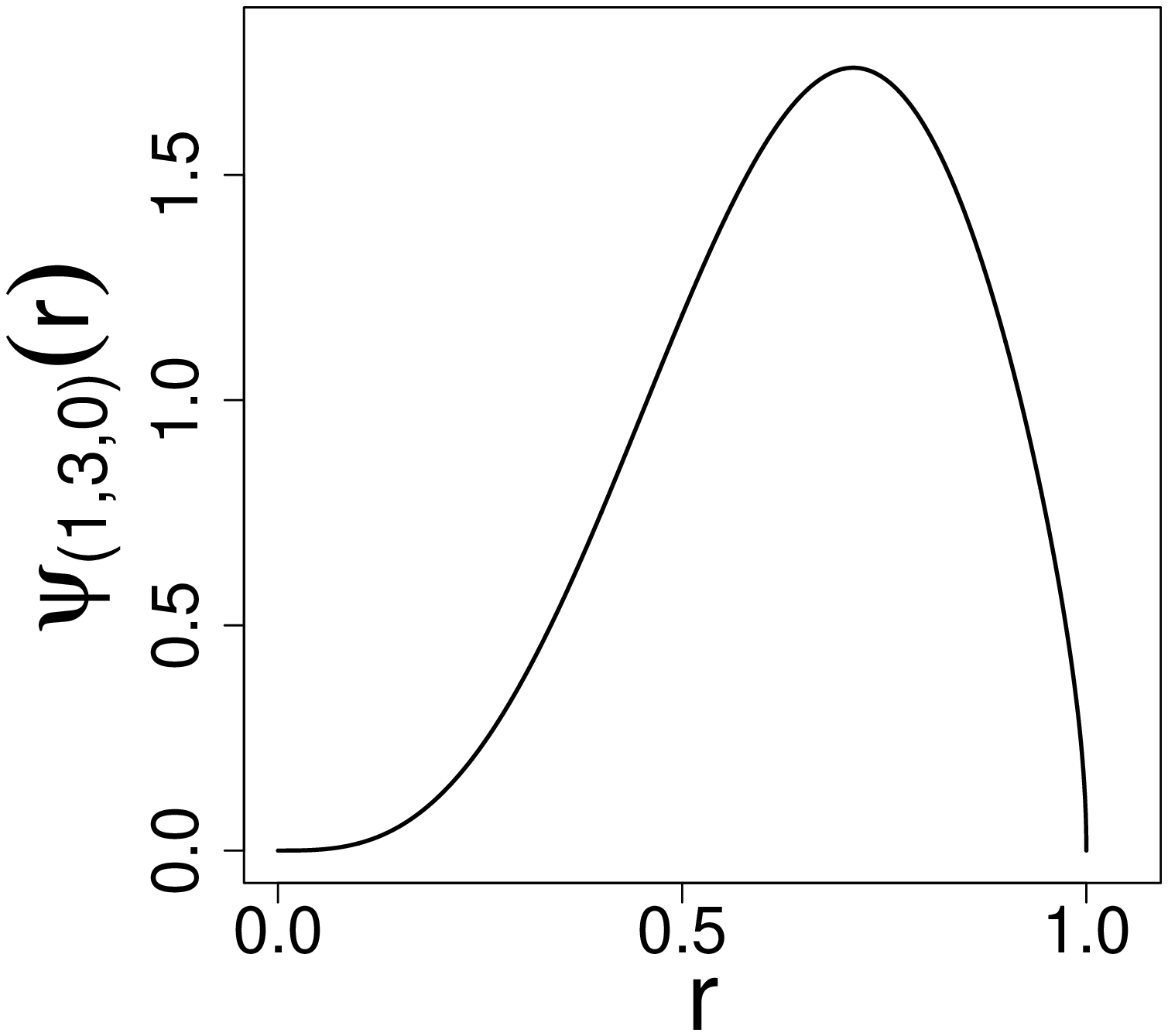}
\caption{Contour plots of  $\psi^{(500)}_{(2,2,0)}$ and  $\psi^{(500)}_{(1,3,0)}$  at  $\theta=0$.}
\end{center}
\end{figure}
\begin{figure}[h]
\begin{center}
\centering
\includegraphics[width=50mm,height=50mm]{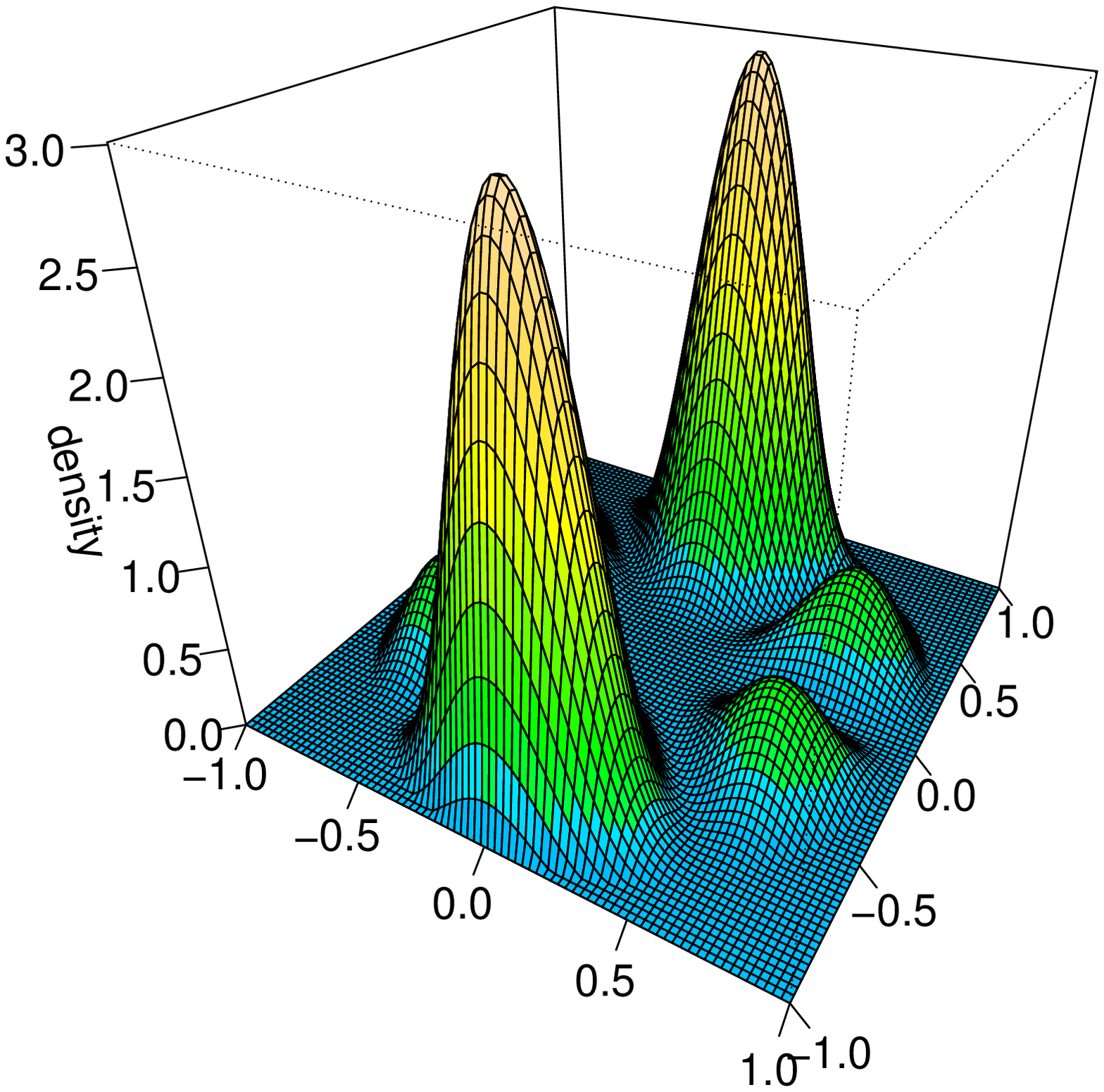}
\includegraphics[width=50mm,height=50mm]{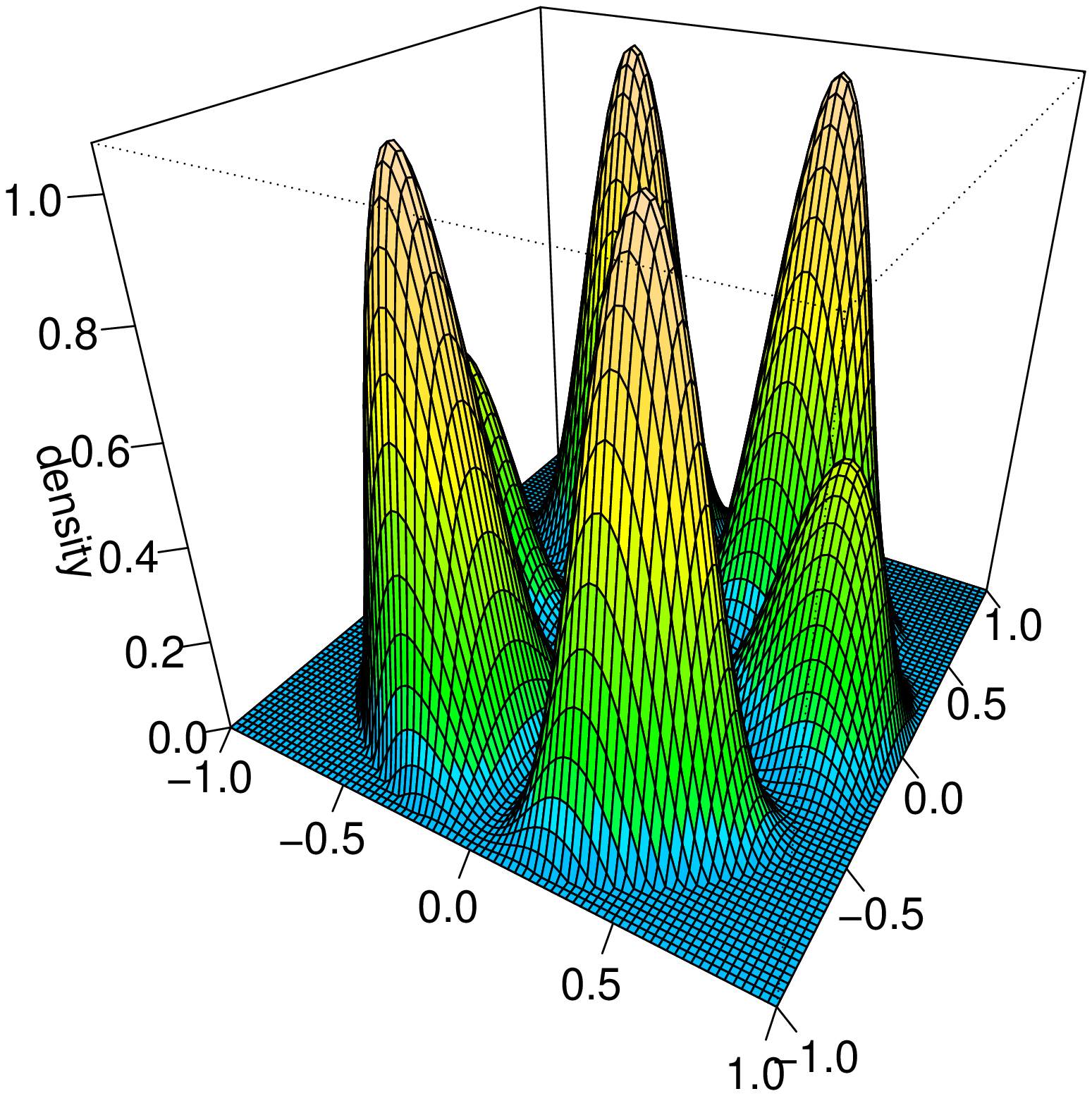}
\caption{Polar plots of probability densities $|\psi^{(500)}_{(1,3,0)}(r,\phi,\theta)|^2$ and
$|\psi^{(500)}_{(1,3,\pm 1)}(r,\phi,\theta)|^2.$ }
\end{center}
\end{figure}
\begin{figure}[h]
\begin{center}
\centering
\includegraphics[width=50mm,height=50mm]{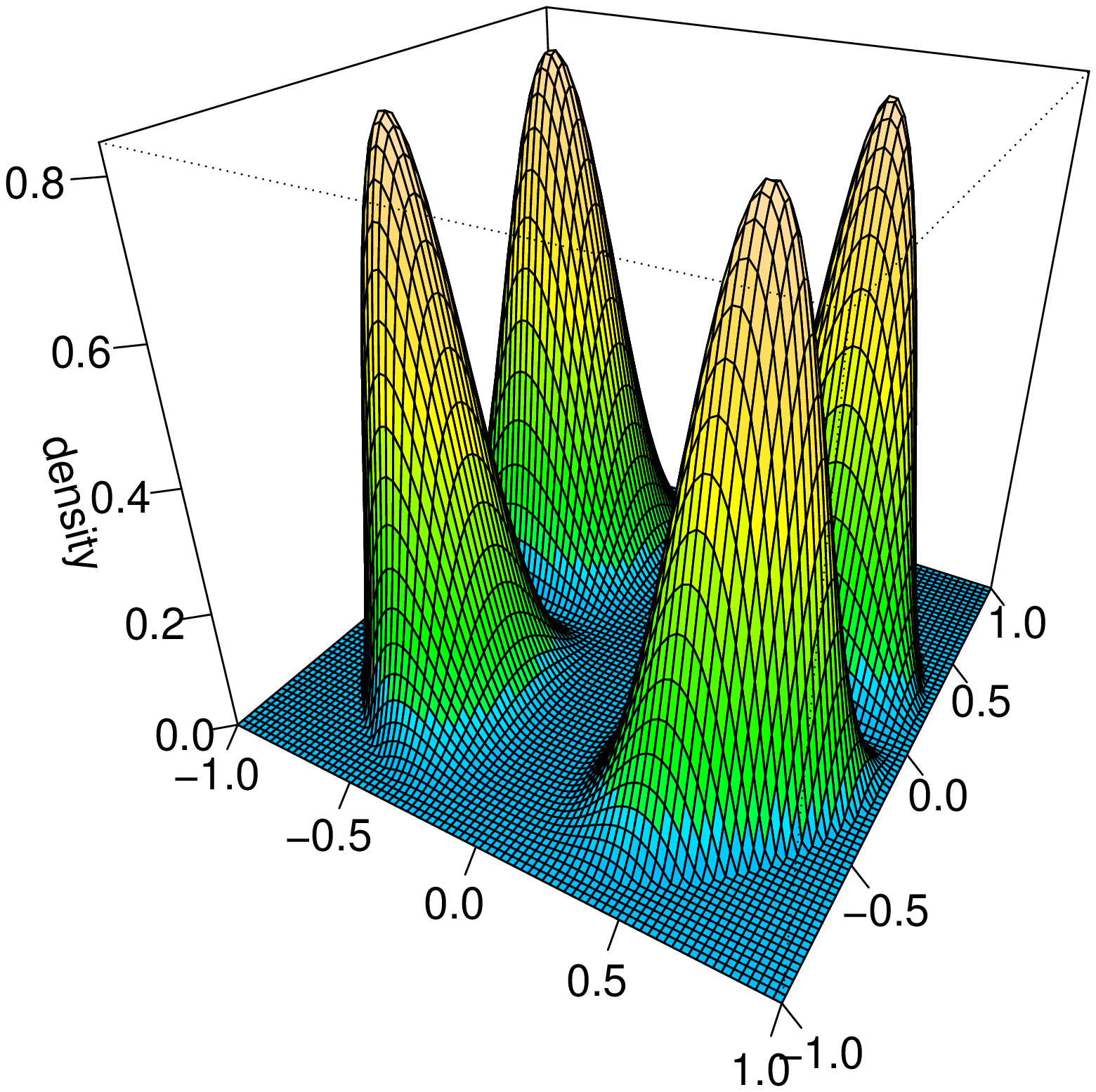}
\includegraphics[width=50mm,height=50mm]{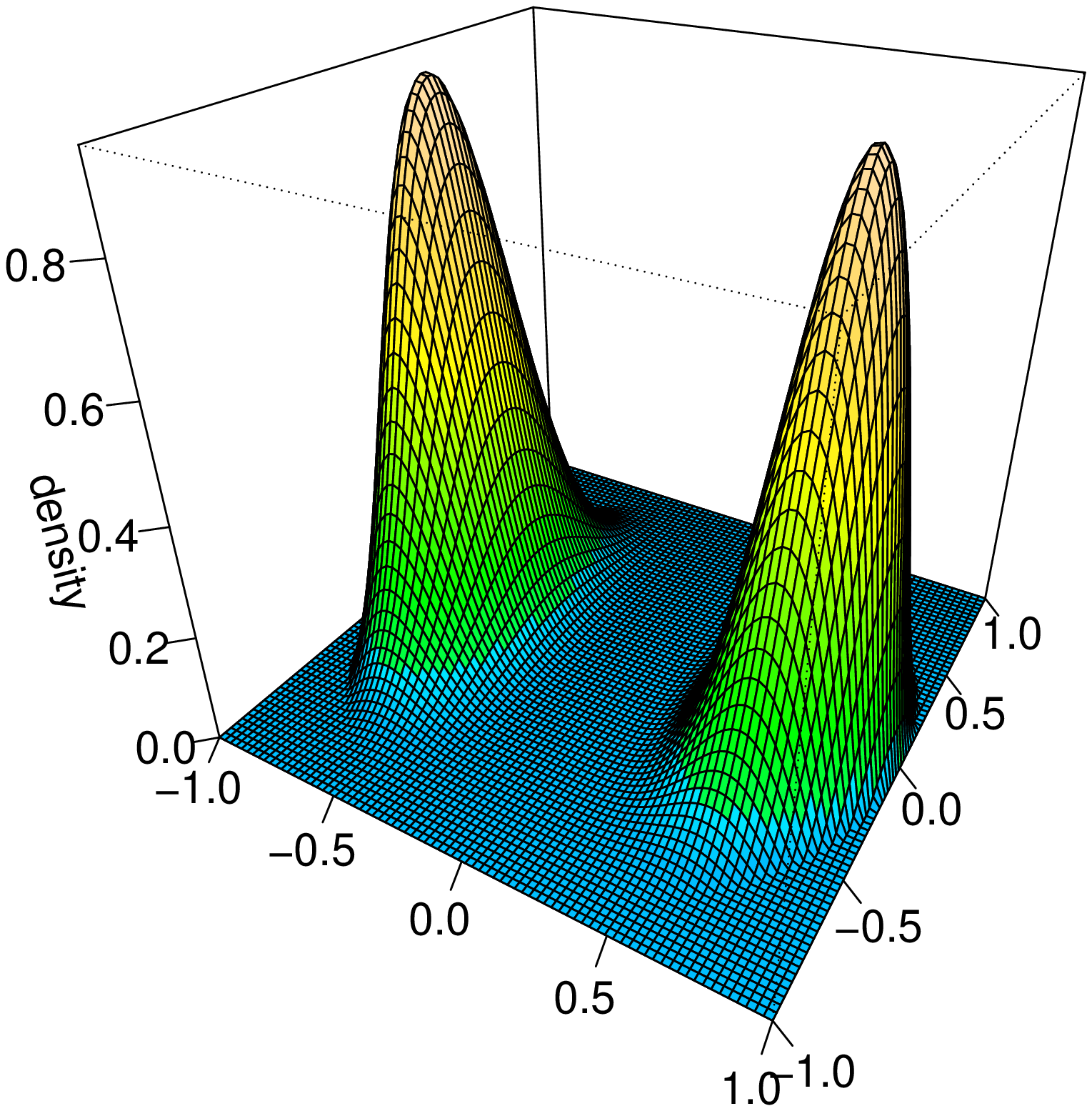}
\caption{Polar plots of probability densities  $|\psi^{(500}_{(1,3,\pm 2)}(r,\phi,\theta)|^2$ and
   $|\psi^{(500)}_{(1,3,\pm 3)}(r,\phi,\theta)|^2$ }
\end{center}
\end{figure}

\subsection{ Higher excited states,   $l\geq 3$ series.}

Let us notice that in the matrix re-writing of the eigenvalue problem (5)   for $l=0, 1, 2$   functions, we have encountered the
 generating matrices (27), (64), (82)
respectively, which we list  comparatively in a single formula:
\begin{eqnarray}
a_{k,n}^{(0)}= a_{k,n}=2(n+1-k)\,  c_{2k},\quad n\geqslant k, \\
a_{k,n}^{(1)}=b_{k,n}=4  {\frac{(n+1-k)(n+2-k)}{2n+3-2k}}\, c_{2k}   ,\quad n\geqslant k, \\
a_{k,n}^{(2)}= g_{k,n}=8{\frac{(n+1-k)(n+2-k)(n+3-k)}{(2n+3-2k)(2n+5-2k)}} \,  c_{2k},\quad n\geqslant k.\\
\end{eqnarray}

It is clear that we can proceed by induction and  take for granted that higher eigenfunctions of the Cauchy well  will be determined
by linear systems of equations  with generating matrices of the form
\be
a_{k,n}^{(l)} =  2^{l+1}\frac{\prod\limits_{s=1}^{l+1} (n+s-k)}{\prod\limits_{s=1}^{l}(2n+2s+1-2k)} \, \,  c_{2k},\quad n\geqslant k,\quad l\geq 0.
\ee
We have explicitly (by means of calculations)  checked the validity of the  formula  (93)  for the case of $l=3$.
The pertinent  calculations are skipped here. In particular, for $k=1$ and $l=3$  we have arrived at (approximate) eigenfunctions:

\be
\begin{split}
\psi_{1,3,0}(\textbf{r})&= C\left( \frac{5x^3_3 - 3x_3r^2}{2} \right) f(r)=  C\left(\frac{5\cos^3\theta-3\cos\theta}{2}\right)r^3 f(r)=
C' r^3 Y_3^0(\theta,\phi) f(r) ,\\
\psi_{1,3,\pm 1}(\textbf{r})&= C  (x_1 \pm x_2)(5x_3^2 - r^2) f(r)= C\sin\theta\left(5\cos^2\theta-1\right)e^{\pm i\phi}r^3 f(r)
= C' r^3Y_3^{\pm 1}(\theta,\phi) f(r) ,\\
\psi_{1,3,\pm 2}(\textbf{r})&= C  (x_1 \pm x_2)^2 x_3 f(r)=   C\sin^2\theta\cos\theta e^{\pm 2i\phi}r^3 f(r) = C' r^3 Y_3^{\pm 2}(\theta,\phi) f(r),\\
\psi_{1,3,\pm 3}(\textbf{r})&= C (x_1 \pm x_2)^3 f(r) =  C\sin^3\theta e^{\pm 3i\phi}r^3 f(r)= C' r^3 Y_3^{\pm 3}(\theta,\phi) f(r),
\end{split}
\end{equation}
with a common for all these  eigenfunctions  $f(r)$  factor of the form
\be
f(r)=\sqrt{1-r^2}\sum\limits_{n=0}^\infty \delta_{2n}r^{2n},\quad \delta_0=1.
\ee
One readily  recognizes both spherical harmonics $Y_3^m$   and solid harmonics  $r^3 Y_3^m$ with $m=0,\pm 1,\pm 2, \pm 3$
 in the presented formulas.
The  computed eigenvalue reads  $E_{(1,3)}=6.630371$.

 Since, in the polynomial approximation of the $2n$-th degree we have
 computed all expansion coefficients   $\delta _{2k}$, we  know precisely the functional form of  respective (approximate)  eigenfunctions.
 and that of resultant probability densities  $\psi_{1,3,m}(r,\phi,\theta)$  with   $m=0,\pm 1,\pm 2, \pm 3$. Those
 are depicted in Figs. (14) - (16). \\

We stress that the  universal   form    (93) of the generating matrix  $a_{k,n}^{(l)}$   opens the door to a direct computation
of approximate eigensolutions   (eigenvalues and eigenfunctions)  for the  $l$-the   spectral series of arbitrary length,
  by means of the Wolfram Mathematica routines. Thus, ultimately  we  are allowed to  skip detailed, sometimes tedious and demanding,
preliminary    calculations, whose outcome   would-be  the  specific (in view of a particular choice of $l$)
   matrix versions of the  spectral problem, like e.g.  those encoded in Eqs. (27), (64), (82).

The generic  functional  form of any trial    eigensolution of Eq. (5), corresponding to the eigenvalue $E$ in the $l$-th series,
  reads  as follows:
\begin{eqnarray}
\psi (\textbf{r}) = C\, r^l \, Y_l^m(\theta ,\phi )\,  f(r),\\
f(r)= \sqrt{1-r^2}\sum\limits_{n=0}^\infty \delta_{2n}r^{2n}.
\end{eqnarray}

 Actually, in the  polynomial approximation  of the $2n$-th degree  involving $f(r)$ of the form (95) or (97),
  we end up with  a universal (c.f. (93)) matrix eigenvalue problem, valid for any $l=0,1,2,...$,  from which one can deduce
    the corresponding  expansion  coefficients   $\delta_{2k}$, $k\leq n$, with $\delta_0=1$ being  presumed:
\begin{eqnarray}
\sum\limits_{k=i}^n \delta_{2k}  a_{k-i,k}^{(l)}=E\sum\limits_{k=0}^i \delta_{2k}c_{2(i-k)},\qquad i=0,1,\ldots,n-1,\nonumber\\
\sum\limits_{m=0}^n \left(\delta_{2m}\sum\limits_{k=0}^m a_{k,m}^{(l)}\right)=0.
\end{eqnarray}
The last identity is an outcome of the boundary condition $(-\Delta)^{1/2}\psi (\textbf{r})$ at  $r=1$, imposed on the trial function
 $\psi (\textbf{r})$,
Eq. (96), when truncated appropriately (polynomial approximation of the degree $2n$)..

A computer assisted computation, while  augmented by an eigenvalue sieve (we order the eigenvalues into  the non-decreasing series)
allows to  associate with each eigenvalue a corresponding eigenstate (or a family of them, in view of the   degeneracy of the spectrum).
The latter  are defined   (c.f. (95))  in terms of  directly  evaluated   coefficients   $\delta_{2k}$, $k\leq n$,  $\delta_0=1$ being  presumed.

\section{Outlook}

While attempting to solve the spectral problem for the  infinite  spherical well,  we have relied on  explicit calculations that show
 how the nonlocal ultrarelativistic operator acts on  properly chosen  trial functions in its domain.   We have
 employed  an efficient truncation method, which yields  approximate eigenvalues and eigenfunctions of the problem,
 with basically unlimited accuracy (depending on the degree $2n$ of the polynomial  truncation).

   We have identified  universal features of the method of solution, summarized in Eqs. (93), (96)-(98).
 The structure of the spherical well  spectrum  resembles that of the standard (Laplacian-induced)  quantum mechanical
 spherical well. Namely, the spectrum splits into  non-overlapping  eigenvalue and eigenfunction  families, each family being labeled
 by a corresponding orbital label  $l=0,1,2,...$.
 Links of the purely radial family of eigenstates   with spectral solutions of the  $d=1$ infinite well problem have been established.

 In connection with the addressed  ultrarelativistic  spherical  well problem, we refer to Ref. \cite{GS}
 for a broader background   and rationale for our analysis  of nonlocal operators.
  In the present paper we have contributed to seldom  investigated and still unexplored area,  where even simplest  spectral
  problems as yet have not received  full solutions, specifically those    exhibiting  nontrivial orbital features.

Like in the standard quantum mechanical reasoning, we regard  the infinite well  as a an approximation of a deep   finite
well.  Therefore, it is of interest to analyze ind detail the ultrarelativistic finite  spherical  well   (the $d=1$  case  has found
its solution, \cite{ZG}).  As well, quite  an ambitious research   goal  could be an analysis a  spatially
  random distribution ("gas") of finite ultrarelativistic  spherical wells,   embedded in a  spatially extended finite
  energy background.

\end{document}